\documentclass[fleqn,usenatbib]{mnras}

\usepackage{newtxtext,newtxmath}

\usepackage[T1]{fontenc}
\usepackage{bm}
\usepackage{cleveref}
\crefname{figure}{Fig.}{Figs.}
\crefname{table}{Table}{Tables}

\DeclareRobustCommand{\VAN}[3]{#2}
\let\VANthebibliography\thebibliography
\def\thebibliography{\DeclareRobustCommand{\VAN}[3]{##3}\VANthebibliography}

\usepackage{graphicx}	
\usepackage{amsmath}	
\usepackage{wasysym}    
\usepackage{xcolor}
\usepackage{xspace}
\usepackage{float}
\makeatletter 
  \patchcmd{\NAT@citex}
    {\@citea\NAT@hyper@{%
      \NAT@nmfmt{\NAT@nm}%
      \hyper@natlinkbreak{\NAT@aysep\NAT@spacechar}{\@citeb\@extra@b@citeb}%
      \NAT@date}}
    {\@citea\NAT@nmfmt{\NAT@nm}%
    \NAT@aysep\NAT@spacechar\NAT@hyper@{\NAT@date}}{}{}

  \patchcmd{\NAT@citex}
    {\@citea\NAT@hyper@{%
      \NAT@nmfmt{\NAT@nm}%
      \hyper@natlinkbreak{\NAT@spacechar\NAT@@open\if*#1*\else#1\NAT@spacechar\fi}%
        {\@citeb\@extra@b@citeb}%
      \NAT@date}}
    {\@citea\NAT@nmfmt{\NAT@nm}%
    \NAT@spacechar\NAT@@open\if*#1*\else#1\NAT@spacechar\fi\NAT@hyper@{\NAT@date}}
    {}{}
\makeatother

\newcommand\tb{\textcolor{black}}
\newcommand\as{\textcolor{black}}

\newcommand\thesan{\mbox{\textsc{thesan}}\xspace}

\title[High-$z$ AGN in cosmological simulations]{High-redshift AGN population in radiation-hydrodynamics simulations}

\author[T.-E. Bulichi et al.]{%
Teodora-Elena Bulichi,$^{1,2}$\thanks{E-mail: \href{mailto:teob1823@mit.edu}{teob1823@mit.edu}}
Oliver Zier,$^{3,1}$
Aaron Smith,$^{4}$
Mark Vogelsberger,$^{1,2,5}$
Anna-Christina Eilers,$^{1,2}$
\newauthor
Rahul Kannan,$^{6}$ 
Xuejian Shen,$^{1,2}$
Ewald Puchwein,$^{7}$
Enrico Garaldi\,$^{8}$
and
Josh Borrow\,$^{9}$
\\
$^{1}$Massachusetts Institute of Technology, Kavli Institute for Astrophysics and Space Research, 77 Massachusetts Avenue, Cambridge, MA 02139, USA \\
$^{2}$Department of Physics, Massachusetts Institute of Technology, 77 Massachusetts Avenue, Cambridge, MA 02139, USA\\
$^{3}$Center for Astrophysics $\vert$ Harvard $\&$ Smithsonian, 60 Garden Street, Cambridge, MA 02138, USA\\
$^{4}$Department of Physics, The University of Texas at Dallas, Richardson, TX 75080, USA\\
$^{5}$The NSF AI Institute for Artificial Intelligence and Fundamental Interactions, Massachusetts Institute of Technology, Cambridge, MA 02139, USA \\
$^{6}$Department of Physics and Astronomy, York University, 4700 Keele Street, Toronto, ON M3J 1P3, Canada\\
$^{7}$Leibniz-Institut für Astrophysik Potsdam, An der Sternwarte 16, D-14482 Potsdam, Germany\\
$^{8}$Kavli Institute for the Physics and Mathematics of the Universe, The University of Tokyo, 5-1-5 Kashiwanoha, Kashiwa, 277-8583, Chiba, Japan\\
$^{9}$Department of Physics and Astronomy, University of Pennsylvania, 209 South 33rd Street, Philadelphia, PA 19104, USA
}

\date{Accepted 2025 October 10. Received 2025 October 10; in original form 2025 July 15}

\pubyear{2025}

\begin{document}
\label{firstpage}
\pagerange{\pageref{firstpage}--\pageref{lastpage}}
\maketitle

\begin{abstract}
High-redshift active galactic nuclei (AGN) have long been recognized as key probes of early black hole growth and galaxy evolution. However, modeling this population remains difficult due to the wide range of luminosities and black hole masses involved, and the high computational costs of capturing the hydrodynamic response of gas and evolving radiation fields on-the-fly. In this study, we present a new suite of simulations based on the IllustrisTNG galaxy formation framework, enhanced with on-the-fly radiative transfer, to examine AGN at high redshift ($z \gtrsim 5$) in a protocluster environment extracted from the MillenniumTNG simulation. We focus on the co-evolution of black holes and their host galaxies, as well as the radiative impact on surrounding intergalactic gas. The model predicts that black holes form in overdense regions and lie below the local black hole-stellar mass relation, with stellar mass assembly preceding significant black hole accretion. Ionizing photons are primarily produced by stars, which shape the morphology of ionized regions and drive reionization. Given the restrictive black hole growth in the original IllustrisTNG model, we reduce the radiative efficiency from 0.2 to 0.1, resulting in higher accretion rates for massive black holes, more bursty growth, and earlier AGN-driven quenching. However, the resulting AGN remain significantly fainter than observed high-redshift quasars. As such, to incorporate this missing population, we introduce a quasar boosted model, in which we artificially boost the AGN luminosity. This results in strong effects on the surrounding gas, most notably a proximity effect, and large contributions to He ionization.

\end{abstract}


\section{Introduction}

Hydrodynamical simulations have become a key tool for modeling galaxy formation in a cosmological framework. By incorporating complex, non-linear physical processes and bridging the gap between galactic and cosmological scales through sub-grid prescriptions, simulations have enabled meaningful comparisons with a wide range of observational key features of galaxy populations, across cosmic time 
(see \citealt{Vogelsberger2020} for a review).

In particular, active galactic nuclei (AGN), fueled by accretion onto supermassive black holes (SMBHs), have proved to be a crucial component of hydrodynamical simulations, as one of the most efficient star formation (SF) regulator across cosmic time (e.g., \citealt{Springel2005b, Fabian2012, Cicone2014, KingPounds2015,Bower2017, Scharre2024}). The current theoretical paradigm implies that SMBH growth occurs in massive dark matter halos (e.g., \citealt{Springel2005,Costa2014,Latif2022}). This picture is generally supported by observations, via e.g., the observed clustering of high\tb{-}redshift quasars (e.g., \citealt{Shen2007,GarciaVergara2022}), as well as the high number of binary quasars at $z > 4$ (e.g., \citealt{Hennawi2006, McGreer2016, Yue2021}, see also \citealt{Overzier2016} for a review). 
In simulations, SMBHs are generally prescribed to reside in overdense environments,  typically employing a halo-dependent seeding mechanism by placing black hole seeds in the most massive halos (e.g., \citealt{Springel2005b,Vogelsberger2013,Habouzit2020}). 

AGN feedback is also believed to be the main driver of the co-evolutionary growth path of SMBHs and their host galaxies. There is now a wealth of local observational evidence that links the masses of SMBHs with the properties of their host galaxies, such as stellar masses ($M_\mathrm{BH}$--$M_\star$ relation) and velocity dispersions ($M_\mathrm{BH} - \sigma $ relation; see e.g., \citealt{KormendyHo2013} for a review). In most cases, large-scale cosmological simulations have been calibrated to match these observational constraints at $z = 0$. At high redshift, however, the AGN population remains highly unconstrained in simulations (see e.g., \citealt{Habouzit2020,Habouzit2022} for a review)\tb{, but recently, increasing efforts have emerged to explore the high redshift counterparts (e.g., \citealt{Bennett2024,Bhowmick2024})}. 

From an observational perspective, the advent of the \textit{James Webb Space Telescope} (\textit{JWST}) has pushed the boundaries of our understanding of massive black hole growth and evolution in the early Universe. Looking at the environments of quasars, e.g., \cite{Wang2024} found a massive protocluster around a luminous quasar at $z = 6.63$, and e.g., \cite{Eilers2024} observed a wider range of overdensities around quasars, 
indicating that high redshift quasars reside in overdense regions, and are generally expected to track overdensities. 
Moreover, \textit{JWST} allows for image decomposition to disentangle between the emission from the quasar and the one from the host galaxy, in order to study the properties of the quasars' host galaxies and their co-evolution. For example, \cite{Ding2023} reported the detection of two quasar host galaxies at $z > 6$, with the corresponding $M_\mathrm{BH}$--$M_\star$ relation agreeing with the local relation reported in \cite{KormendyHo2013}. However, \tb{other studies (e.g., \citealt{Stone2023,Harikane2023,Kokorev2023,Uebler2023,Maiolino2024,Yue2023} )} found that the inferred masses of the SMBHs in their \textit{JWST} studies lie above the local \cite{KormendyHo2013} $M_\mathrm{BH}$--$M_\star$ relation, suggesting an over-massive population with respect to the host galaxy stellar mass (see also \citealt{Scoggins2023, Pacucci2023,Natarajan2024}), albeit selection biases may be at play \citep{Li2024}. Thus, it is unclear at what redshift the observed correlation between the SMBH's and host galaxy's mass starts taking shape. 

With \textit{JWST} it is also now possible to observe a new population of high-redshift AGN down to the bolometric luminosity range \tb{$L_\mathrm{bol} \approx 10^{43}-10^{46}\,\mathrm{erg~s^{-1}}$ (e.g., \citealt{Furtak2023, Kocevski2023, Kokorev2023, Onoue2023, Maiolino2023, Matthee2023, Taylor2024,Juodvzbalis2025})}. These \tb{AGN} are usually identified from the existence of a broad component in the H$\alpha$ or H$\beta$ line, and, in some cases, display small discrepancies in abundance, for a given mass range, with previous theoretical predictions (e.g., \citealt{Habouzit2024}). Lastly, the unprecedented capabilities of \textit{JWST}'s mid-infrared instrument (MIRI) are expected to reveal the high-redshift, obscured AGN, currently largely undetected at $z \gtrsim 4$ due to the absence of UV/optical observations and the inability of X-ray surveys to detect Compton-thick AGN \citep{Ueda2014,Ananna2019}. This will also provide constraints on the mass assembly of early BHs (i.e., the black hole accretion history BHAD), in the highly unexplored regime $z \gtrsim 5$, via e.g., the MIRI Early Obscured-AGN Wide Survey (MEOW; Leung et al., in prep).

Under the ``Soltan argument'' \citep{Soltan1982}, the growth of SMBH is directly linked with the emission of quasar light. The most luminous quasars emerge as early as $z \gtrsim 6$, and attain luminosities as high as $10^{47}-10^{49}$ erg~s$^{-1}$ (e.g., \citealt{Mortlock2011, Jiang2016, Matsuoka2018,Wang2019, Banados2023}, for reviews see e.g., \citealt{SmithSMBH2019,Inayoshi2020,FanBanadosSimcoe2023}). Modeling the emission of light from AGN and effects on surrounding gas is challenging in simulations, and generally relies on post-processing techniques due to the high computational costs. The post-processing technique has been proved to allow a wide range of parameters to be studied, such as the sizes of quasar proximity zones (e.g., \citealt{Davies2020,ChenGnedin2021,Zhou2023}), or ionized bubbles (e.g., \citealt{Asthana2024}). However, the post-processing methods do not account for gas dynamics and the evolution of the background radiation field. Hence, they are not able to reproduce the hydrodynamic response of IGM to the inhomogeneous photoheating, nor model in detail the galaxy population and escape of ionizing radiation.

It is now possible however to model the radiation effects on-the-fly, using for example the \thesan radiation-magnetohydrodynamic simulations \citep{Kannan2022,Garaldi2022,Smith2022,Garaldi2023}. \thesan simultaneously models both the large-scale statistical properties of the intergalactic medium during reionization and the detailed characteristics of the galaxies responsible for it. The simulations utilize the efficient radiation hydrodynamics solver {\small AREPO-RT} \citep{Kannan2019,Zier2024}, which accurately captures the interaction between ionizing photons and gas. These simulations align with a wealth of observational data, including the stellar-to-halo-mass relation, galaxy stellar mass function, star formation rate density, and the mass–metallicity relation, at high redshift. They have also proven effective across a wide range of scientific applications, including studies of ionizing escape fractions \citep{Yeh2023}, the sizes of ionized bubbles \citep{Neyer2024, Jamieson2024}, and galaxy sizes during the epoch of reionization \citep{Shen2024}; as well as \textsc{thesan-zoom} for more detailed follow-up comparisons with zoom-in simulations more accurately capturing the multi-phase interstellar medium \citep[e.g.,][]{Kannan2025,  Zier2025, Zier2025-PopIII, McClymont2025a,McClymont2025b,Shen2025,Wang2025}.

In this study, we make use of the IllustrisTNG galaxy formation model \citep{Pillepich2018a,Springel2018,Nelson2018,Marinacci2018,Weinberger2018}, together with an updated version of the self-consistent radiative transfer treatment of the \thesan simulations. We select a massive halo from the parent MillenniumTNG simulation box \citep{pakmor2023millenniumtng}, in order to study the high-$z$ AGN population, and effects on surrounding gas due to the radiation emerging from AGN and stars. The rest of the paper is structured as follows: Sec.~\ref{sect:methods} describes the methods employed by this study (galaxy formation model, radiation modeling, and different variations of free parameters); Sec.~\ref{sect:BH_pop} discusses the properties of the BH population, and their host galaxies; Sec.~\ref{sect:rad_mod} describes the radiation modeling and impact on gas properties and Sec.~\ref{sect:summary} summarises the main findings of this work. 

Throughout this work, we adopt the \citet{Planck2016} cosmological parameters: $H_0 = 100 h$ km\,s$^{-1}$\, Mpc$^{-1}$ with $h = 0.6774$, $\Omega_\mathrm{m} = 0.3089$, $\Omega_\Lambda = 0.6911$, $\Omega_\mathrm{b} = 0.0486$, $\sigma_8 = 0.8159$, and $n_s = 0.9667$.

\section{Methods}
\label{sect:methods}

In this section, we describe the simulation set-up, including the galaxy formation model and the self-consistent radiative transfer prescriptions. We also introduce different variations of the model that will be used throughout the paper to analyze the effects of radiative efficiency, radiation modeling, and cooling prescriptions on AGN and galaxy populations (Sec.~\ref{sect:BH_pop}), as well as the response of gas to radiation (Sec.~\ref{sect:rad_mod}).

\subsection{Simulation set-up}
All simulations presented in this paper are performed with the massively parallel {\small AREPO} code \citep{Springel2010, Pakmor2016, Weinberger2020}. {\small AREPO} solves the Euler equations on an unstructured, moving, Voronoi mesh using a second-order accurate finite-volume scheme. The mesh-generating points move approximately with the local fluid velocity, rendering the method Galilean invariant and maintaining an approximately constant mass per cell. This condition is further enforced by splitting or removing cells whose masses deviate by more than a factor of 2 or 0.5, respectively, from a predefined target mass.
Gravitational interactions between collisionless dark matter, star particles, and gas cells are computed using the hybrid TreePM method \citep{Bagla2022}, which combines a hierarchical octree \citep{Barnes1986} for short-range forces with a particle-mesh approach \citep{Aarseth2003} for long-range forces. Fixed gravitational softening lengths are used for dark matter and star particles, while the softening length for gas cells scales with their effective size.
{\small AREPO} can generate halo catalogs on-the-fly using the friends-of-friends (FOF) algorithm \citep{Davis1985}, which groups particles based on spatial proximity. Bound substructures within these FOF groups are further identified with the SUBFIND-HBT algorithm \citep{springel2001populating, Gadget4}.

In this paper, we perform cosmological zoom-in simulations using the IllustrisTNG galaxy formation model (see Sec.~\ref{sect:TNG}) coupled with radiation transport (see Sec.~\ref{sect:RT}), aiming to study the growth of AGN and their impact on the surrounding environment. We select the most massive halo at redshift $z = 5$ from the MTNG740 simulation as our target halo, which is part of the MillenniumTNG (MTNG) project \citep{pakmor2023millenniumtng}. Given the larger box size of MTNG, this simulations suite is expected to host a higher abundance of massive BHs than smaller IllustrisTNG simulations, due to the halo dependent BH seeding prescription (see Sec.~\ref{sect:TNG}), collected over a significantly larger volume.

The parent simulation, MTNG740, was carried out with {\small AREPO} and the IllustrisTNG model, but without radiation transport. It evolves a periodic box of side length 500\,$h^{-1}$\,cMpc with $4320^3$ equal-mass dark matter particles and approximately $4320^3$ gas cells, corresponding to mass resolutions of $2 \times 10^8 \,\mathrm{M}_\odot$ for dark matter and $3 \times 10^7 \,\mathrm{M}_\odot$ for baryons. The gravitational softening lengths are fixed at 3.7 ckpc for dark matter and stellar particles, and the gas softening length is adaptive with a minimum of 0.37 ckpc. Initial conditions were generated at redshift $z = 63$ using second-order Lagrangian perturbation theory within NGENIC, which is incorporated in the {\small GADGET-4} code \citep{Gadget4}.

The chosen cluster attained a halo mass of $M_{200} \sim 10^{13}\,\mathrm{M_\odot}$, in the original parent box, with the galaxy stellar mass $M_\mathrm{\star} \sim 2 \times 10^{11}\,\mathrm{M_\odot}$, hosting a SMBH with $M_\mathrm{BH} \sim 2.5 \times 10^8 \,\mathrm{M_\odot}$, at $z=5$. We choose a relatively large spherical high-resolution region with radius $R=60\,$cMpc around the target halo to ensure that the region captures meaningful observables, such as ionized bubbles from Ly$\alpha$ emitters, while also mitigating biases from the high-density environment surrounding the central galaxy and its SMBH. 
The initial conditions are created using a novel zoomed-initial condition code (Puchwein et al., in prep) which allows arbitrary shapes for different resolution levels. For the main part of the paper, we adopt the same resolution within the high-resolution region as in the parent volume, degrading the resolution outside of this region.
In \cref{app:resolution}, we assess the convergence of the black hole properties by re-simulating the central region, within a 10 cMpc radius from the center of the target halo, at eight times higher mass resolution.

\subsection{Galaxy formation model}
\label{sect:TNG}
All simulations presented in this work employ the same galaxy formation model as used in MTNG, based on the fiducial IllustrisTNG model, but omits magnetic fields and does not track individual metal species.
Given that our simulations are only evolved to $z = 5$, the omission of magnetic fields is justified, as they are not expected to be saturated at the resolutions we employ \citep{Pakmor2024}. Nonetheless, we note that \citet{pakmor2023millenniumtng} found that magnetic fields can impact the stellar mass of galaxies by $z=0$.

The model includes radiative cooling from metal lines \citep{Vogelsberger2013}, employing tabulated cooling rates that depend on metallicity, density, temperature, and redshift \citep{Smith2008, Wiersma2009}.
In simulations that do not include radiation transport (``EqCool-UVB'' in Table~\ref{tab:prop}), we instead model equilibrium cooling from primordial elements \citep{Cen1992, Katz1996} under the influence of a spatially uniform, redshift-dependent UV background \citep{Faucher2009}, with a density-based self-shielding correction applied according to \cite{Rahmati2013}. For all the other runs, we perform primordial cooling via a non equilibrium thermochemistry module (see Sec.~\ref{sect:RT}).

The resolution in MTNG is not sufficient to fully capture a multi-phase interstellar medium (ISM). To model the unresolved structure, we adopt the effective equation of state (eEOS) introduced by \citet{SpringelHernquist2003}, which assumes that gas cells with densities above $n_\mathrm{H} > 0.11\,\mathrm{cm}^{-3}$ exist in pressure equilibrium between a cold and a hot phase.
We interpolate between this eEOS and an isothermal equation of state at $10^4$ K, with the isothermal component contributing 70\% to the combined EOS. This approach imposes a tight relation between gas density and temperature, and gas is not allowed to cool below the eEOS (see also Sec.~\ref{sect:diff_thesan}). Gas with $n_\mathrm{H} > 0.11\,\mathrm{cm}^{-3}$ can temporarily exceed the eEOS temperature due to processes such as shock heating or AGN feedback, but it typically cools back onto the eEOS within a few time steps.
Star formation occurs stochastically from gas following the eEOS. To mitigate the computational cost associated with very dense gas, we adopt a boosted star formation efficiency above a density threshold of $n_\mathrm{H} \sim 24.4\,\mathrm{cm}^{-3}$, following the approach introduced in \citet{Nelson2019TNG50} and further discussed in \citet{Burger2025}.

Stellar feedback contributes indirectly to the pressure in the eEOS, but we also employ the wind particle model from \citet{Pillepich2018a} to describe galactic outflows. In this model, star-forming gas cells are stochastically converted into wind particles that carry mass, metals, and energy out of dense regions into the surrounding halo.
Metal enrichment from both supernovae and asymptotic giant branch (AGB) stars is also modeled following the IllustrisTNG model.

\begin{table*}
    \centering
  
    \begin{tabular}{|l|c|c|c|c|c|c|c|l|}
        \hline
        Name & $R_\mathrm{box}$   & Zoom  & AGN radiation? & $f_\mathrm{esc,\star}$   & $\epsilon_\mathrm{rad}$   & Cooling &UVB & Description\\ 
        &[cMpc]&  factor &  [Y/N] &  [$\%$]  &  &   & [Y/N]\\ \hline 
      &   &    &  &   &  &     &  & Fiducial model. Radiation is modeled  \\
        RT-fid &  60 & 1    & Y  & 25  & 0.1 & Non-eq    & N &self consistently, on-the-fly, including all sources: \\
       &   &    &  &   &  &     &  & stars (with the fiducial escape fraction $f_\mathrm{esc,\star} = 25\%$) \\
        &   &    &  &   &  &     &  & and AGN, including an obscuration scheme\\ \hline
      RT-$\epsilon_\mathrm{rad} = 0.2$ &  60 & 1    & Y  & 25  & 0.2 & Non-eq    & N & Same as fiducial model, except different \\ 
      &   &    &  &   &  &     &  & radiative efficiency: $\epsilon_\mathrm{rad} = 0.2$\\ \hline
      &   &    &  &   &  &     &  & Stars only. Radiation from stars is modeled \\
      RT-Stars &  60 & 1    & N  & 25  & 0.1 & Non-eq    & N &  self consistently, whilst the photons from  \\
      &   &    &  &   &  &     &  & AGN are not allowed to escape \\ \hline
      &   &    &  &   &  &     &  & AGN only. Radiation from AGN is modeled \\
      RT-AGN &  60 & 1    & Y  & 0  & 0.1 & Non-eq    & N &  self consistently, whilst the photons from\\ 
      &   &    &  &   &  &     &  & stars are not allowed to escape \\ \hline
      
       &   &    &  &   &  &     &  & No radiation. Cooling is performed via the \\ 
      noRad &  60 & 1    & N  & 0  & 0.1 & Non-eq    & N &  non-equilibrium thermochemistry module,  but \\ 
      &   &    &  &   &  &     &  &  photons from stars and AGN are not allowed to \\
      &   &    &  &   &  &     &  &  escape, and the UV background is turned off \\\hline
      &   &    &  &   &  &     &  & TNG model. Radiation is modeled  \\ 
      EqCool-UVB &  60 & 1    & -  & -  & 0.1 & Eq    & Y &  as a spatially uniform UV background \\
      &   &    &  &   &  &     &  &  It also assumes equilibrium cooling \\\hline
      &   &    &  &   &  &     &  &  Quasar boosted model. Same as fiducial  \\
      RT-Quasar  &  60 & 1    & x5, x10, x20 & 25  & 0.1 & Non-eq    & N & model, but AGN luminosity is artificially \\
      (x5, x10, x20)&   &    &  &   &  &     &  & boosted by a factor of 5, 10, 20 respectively \\
      &   &    &  &   &  &     &  &  and AGN obscuration effects are removed \\\hline

    \end{tabular}
    
    \caption{From left to right the columns indicate: name of the run, radius of our spherical Lagrangian region, zoom factor (where a zoom factor of one corresponds to the resolution of the original parent MillenniumTNG simulation), escape fraction for stars, radiative efficiency, cooling mechanism (non-equilibrium thermochemistry or equilibrium), presence of a spatially uniform external UV background, and a short description of each run.}
    \label{tab:prop}
\end{table*}

\begin{figure*}
    \includegraphics[width=1\textwidth]{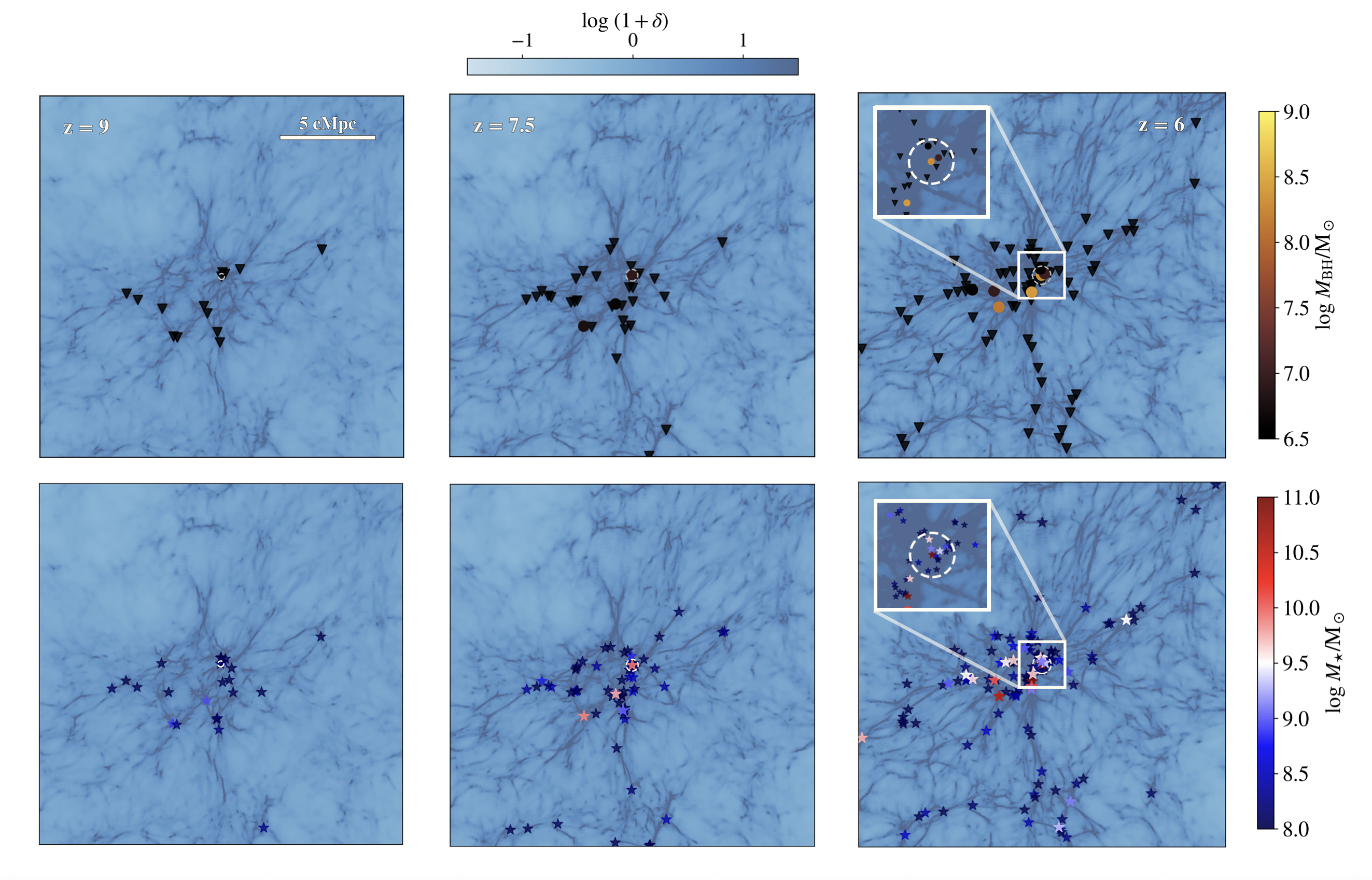}
    \vspace{-0.5cm}
    \caption{\textbf{Visual illustration of the mass assembly of black holes (\textit{top row}), and galaxy stellar mass (\textit{bottom row})} \textit{Background:} Gas distribution, color-coded by the corresponding gas overdensity $\delta_\mathrm{g} = \rho_\mathrm{g}/\bar{\rho}_\mathrm{g} - 1$. Each panel shows a projection slice measuring 20 cMpc across and 10 cMpc in depth, centered on the most massive halo in the simulation. We show three different redshifts: $z = 9$ (\textit{left column}), $z = 7.5$ (\textit{middle column}) and $z = 6$ (\textit{right column}). The right column also includes a 2.5 cMpc square zoom-in around the center of the slice. In the top row, black holes are depicted as circles color-coded by mass, with seed BHs shown as black triangles. In the bottom row, stellar masses of galaxies are indicated by star symbols, also color-coded by mass. The top row illustrates that the first BHs are seeded around $z \sim 9$ and tend to trace the dense, filamentary structures due to the halo-dependent seeding model (see Sec.~\ref{sect:TNG}). By $z = 6$, the most massive BH has grown to $M_\mathrm{BH} \sim 10^{8.5} \, \mathrm{M_\odot}$ and resides in the central protocluster. The bottom row highlights that galaxies assemble their stellar mass earlier and more rapidly than BHs, with $M_\star$ typically exceeding the mass of the central BH by at least two orders of magnitude.}

    \label{fig:env_proj}
\end{figure*} 

The treatment of AGN is fully described in \citet{Weinberger2017, Weinberger2018}. Black hole particles are seeded in halos identified by the FOF algorithm once the halo mass exceeds a threshold of $5 \times 10^{10}\, h^{-1}\mathrm{M}_\odot$, provided the halo does not already host a BH. Each seed is assigned an initial mass of $8 \times 10^5 h^{-1} \mathrm{M}_\odot $.
Black holes grow by gas accretion, following the Bondi–Hoyle formalism \citep{Bondi1944,Bondi1952}: 
\begin{equation} 
\dot{M}_\text{Bondi} = \frac{4 \pi G^2 M_\text{BH}^2 \rho}{c_s^3} \, , 
\label{eq:bondiRate}
\end{equation}
where $G$ is the gravitational constant, $M_\mathrm{BH}$ is the BH mass, $\rho$ is the ambient gas density, and $c_s$ is the local sound speed. Ambient quantities are estimated using a standard SPH-like kernel over the 48 nearest gas cells, matching the MillenniumTNG simulation.
Accretion is capped by the Eddington rate: 
\begin{equation} 
\dot{M}_\text{Edd} = \frac{4 \pi G M_\text{BH} m_p}{\epsilon_\mathrm{rad} \sigma_{\rm T}c} \, , 
\label{eq:eddingtonRate}
\end{equation} 
where $m_p$ is the proton mass, $c$ is the speed of light, $\sigma_{\rm T}$ is the Thomson cross-section, and $\epsilon_\mathrm{rad}$ is the radiative efficiency. While the TNG model typically adopts $\epsilon_\mathrm{rad} = 0.2$, we implement a lower value of $\epsilon_\mathrm{rad} = 0.1$ to enable more efficient growth of massive black holes (see Sec.~\ref{sect:erad}), except for a control run (RT-$\epsilon_\mathrm{rad}=0.2$, see Table~\ref{tab:prop}).
The final accretion rate is taken to be the minimum of the Bondi and Eddington rates: 
\begin{equation}
    \dot{M}_\text{BH} = \min\left( \dot{M}_\text{Bondi} , \dot{M}_\text{Edd}  \right) \, .
    \end{equation}
In addition to gas accretion, BHs also grow through mergers with other BHs \citep{Weinberger2017}.

Our resolution is insufficient to self-consistently follow the orbital decay of BHs toward the potential minimum via dynamical friction. Instead, we artificially reposition BH particles to the local minimum of the gravitational potential.
Feedback from BHs is implemented via two distinct modes. In the high-accretion regime, the thermal (or quasar) mode continuously injects thermal energy into the surrounding gas with a rate given by 
\begin{equation}
\dot{E}_{\rm therm} = \epsilon_{f,\text{high}} \epsilon_{\rm rad}\, \dot{M}_\text{BH} \,c^2 \, ,
\label{eq:etherm}
\end{equation}
with feedback efficiency $ \epsilon_{f,\text{high}} = 0.1$.
In the low-accretion regime, feedback energy is deposited in a pulsed fashion as pure kinetic feedback, with an energy injection rate of
\begin{equation}
\dot{E}_{\rm kin} = \epsilon_{f,\text{kin}}\, \dot{M} \,c^2 \, ,
\end{equation}
where the efficiency parameter is defined as 
\begin{equation}
\epsilon_{f,\text{kin}} = \min\left( \frac{\rho}{0.05 \,\rho_\text{SFthresh}} , 0.2\right) \, ,
\end{equation}
and $\rho_\text{SFthresh}$ is the star formation threshold density. 
The transition between the two modes is determined by the accretion rate. Specifically, thermal feedback activates when the accretion rate in units of the Eddington limit exceeds 
\begin{equation}
\chi = \min\left[ 0.002 \left(\frac{M_\text{BH}}{10^8 \,\text{M}_\odot} \right)^2, 0.1\right] \, .
\end{equation}

The density dependence of $\epsilon_{f,\text{kin}}$ serves as a regulatory mechanism that reduces the coupling efficiency of kinetic AGN feedback in low-density environments. Similarly, in the thermal mode, the TNG model includes a pressure-based correction following \citet{Vogelsberger2013}. When the external pressure falls below a reference threshold, the accretion rate is suppressed by a factor $\left(P_{\rm ext}/P_{\rm ref}\right)^2$ where $P_{\rm ext}$ is the kernel-weighted pressure of the gas surrounding the BH, and $P_{\rm ref}$ is a fixed reference pressure (see equation 23 of \citealt{Vogelsberger2013}).

The black hole mass–stellar mass relation at $z = 0$ was one of the key constraints used in calibrating the IllustrisTNG model \citep{Pillepich2018a, Weinberger2017}, and the simulation reproduces this relation in good agreement with observational data \citep[e.g.,][]{Bhowmick2020, Terrazas2020, Piotrowska2022}. However, the properties of high-redshift AGN remain largely unconstrained by current observations. Whilst it is possible to re-calibrate the model to match high-redshift constraints, in this study we will mostly focus on the interplay between the existing IllustrisTNG model and the self-consistent, on-the-fly radiation treatment. We introduce a new model in Sec.~\ref{sect:quasar_model}, aiming to reproduce the rare quasars observed in the early Universe, which we will expand on in a subsequent paper (Bulichi et al., in prep).

\subsection{Coupling to radiation transport}
\label{sect:RT}
In contrast to the IllustrisTNG simulations, we also model the evolution of the radiation field using the GPU-accelerated {\small AREPO-RT} extension \citep{Kannan2019, Zier2024}. {\small AREPO-RT} solves four additional hyperbolic conservation equations corresponding to the zeroth and first moments of the radiation field -- i.e., the photon number density and photon flux -- using the M1 closure relation \citep{Levermore1984, DubrocaFeugeas1999}.
To achieve second-order convergence, spatial gradients are computed using a least-squares fit method \citep{Pakmor2016}, followed by piecewise linear interpolation. The RT equations are coupled to the gas via a non-equilibrium thermochemistry model for hydrogen and helium (for details, see Appendix B of \citealt{Kannan2019}).
Cooling and heating processes are modeled using non-equilibrium abundances for the primordial species $\ion{H}{I}$, $\ion{H}{II}$, $\ion{He}{I}$, $\ion{He}{II}$, and $\ion{He}{III}$, along with equilibrium metal-line cooling and Compton cooling from the cosmic microwave background. The metal-line cooling rates are computed in the same way as in the IllustrisTNG model, assuming a uniform UV background (see Sec.~\ref{sect:TNG}). Although this is formally inconsistent with the self-consistently evolved radiation field, it has only minor effects (e.g., on ionization states; see discussion in \citealt{Kannan2022}).
The thermochemical network is integrated using a semi-implicit solver with subcycling, following the implementation described in \citet{Zier2024}, which is particularly efficient on GPUs. We model radiation in three frequency bins with boundaries at $[13.6, 24.6, 54.4, \infty)$ eV.

Radiation transport imposes a stringent Courant condition on the time step, scaling inversely with the speed of light. To improve computational efficiency, we adopt the reduced speed of light approximation, replacing the physical speed of light with $0.2c$. This value has been shown to be sufficient for achieving a converged reionization history in the \thesan simulations \citep{Kannan2022}.
Unlike \thesan (see also the next section), we do not evolve the full thermochemical network for gas above the eEOS threshold density of $\approx 0.11\,\mathrm{cm}^{-3}$. Instead, we apply only radiation transport and standard equilibrium cooling, as done in IllustrisTNG. This approach leads to improved numerical stability, including for BH properties.

Both stars and AGN serve as sources of radiation in our simulations. The spectral energy distribution (SED) of stellar populations is derived from the Binary Population and Spectral Synthesis (BPASS) models \citep{Eldridge2017}, assuming a \citet{Chabrier2003} initial mass function (IMF). To account for the unresolved absorption of Lyman continuum (LyC) photons within star-forming regions, the \thesan simulations introduced an escape fraction parameter for stellar birth clouds, adopting a value of $f_\mathrm{esc, \star} = 0.37$.
In our model, this escape fraction is interpreted as the fraction of LyC photons escaping from the eEOS imposed on dense gas. As such, we adopt a slightly lower default value of $f_\mathrm{esc, \star} = 0.25$. As demonstrated in Sec.~\ref{sect:ioniz}, this choice results in a reionization history consistent with observational constraints across the entire zoom-in region. We do not implement an escape fraction for the AGN radiation, but instead we account for obscuration effects via the power-law model introduced by \citet{Hopkins2008}, using the same parameter values as in \citet{Vogelsberger2013}:
\begin{equation*}
\frac{L_{\rm bol}^{\rm AGN, obs}}{L_{\rm bol}^{\rm AGN}} = \omega_{\rm 1}\left(\frac{L_{\rm bol}^{\rm AGN}}{10^{46}\rm{erg~s^{-1}}}\right)^{\omega_{\rm 2}} \, ,
\end{equation*}
with $\omega_{\rm 1}=0.3$ and $\omega_{\rm 2}=0.07$. The total bolometric luminosity of AGN is $L_{\rm bol}^{\rm AGN, tot} = \epsilon_\mathrm{rad} \dot{M} c^2$, out of which a fraction $\epsilon_{f,\text{kin}}$ is injected as thermal energy in the thermal (quasar) kinetic mode (eq.~\ref{eq:etherm}). \as{Following the implementation of AGN radiation in IllustrisTNG, which is based on the advection-dominated accretion flow (ADAF) disk model, we assume a transition to a radiatively inefficient accretion below $0.002\dot{M}_\text{Edd}$, for which we do not inject any radiation.} The injected power $L_{\rm bol}^{\rm AGN, obs}$ is distributed in the different frequency bins using the spectral shape prescribed by the parametrization of \citet{Lusso2015}.

\subsection{Differences with respect to the \thesan model}
\label{sect:diff_thesan}
Although our model is similar to the original setup from the \thesan project, there are several important differences, particularly in the treatment of black holes. As mentioned in the previous section, we treat all gas above the star formation threshold (i.e., governed by the eEOS) as transparent, and we apply equilibrium cooling to this gas. In contrast, \thesan employed non-equilibrium cooling with absorption for all gas cells, which enabled cooling below the equation of state. 

In \thesan, thermochemistry becomes decoupled from the eEOS during sub-cycled timesteps.
For large timesteps, this allows gas to become partially neutral, begin absorbing ionizing photons, and simultaneously lose pressure support through cooling. In the limit of infinitesimally small timesteps, this approach asymptotically reproduces a transparent eEOS, as adopted in our model.

A key consequence of this difference is seen in BH accretion. In \thesan, the Bondi accretion rate (Eq~\ref{eq:bondiRate}) was computed using the intermediate temperature of gas after it had cooled below the eEOS. Since eEOS gas is typically dense and cools efficiently, this artificially boosts BH growth, due to the reduced sound speed. This leads to a significant departure from the original IllustrisTNG calibrated BH model. Additionally, an undetected coding mistake in \thesan replaced the speed of light constant 
$c$ with the reduced value of $0.2c$ in both feedback routines and in the Eddington accretion rate (Eq.~\ref{eq:eddingtonRate}), further enhancing black hole masses relative to those in the standard IllustrisTNG model. 

Our implementation adheres more faithfully to the IllustrisTNG model, in the context of BH growth and feedback. As we show in Sec.~\ref{sect:rad_effects},~\ref{sect:MBH-Ms} and~\ref{sect:BHAD}, it produces galaxy and BH properties that closely match those from TNG, increasing our confidence that, like TNG, our model yields a physically plausible low-redshift galaxy and BH population. In the original \thesan simulations, the increased BH growth leads to an earlier start of the kinetic feedback and efficient quenching of galaxies at high redshift \citep{Chittenden2025}. This effect does not compromise the main science goals of \thesan, as the galaxies stellar mass assembly remains in good agreement with the high-redshift counterparts of TNG (see Appendix A of \citealt{Garaldi2022}). However, it does cause the BHs to be more massive than expected from the original TNG prescriptions, and start quenching their host galaxies much earlier~\citep{Shen2024}.

\subsection{Overview of simulations}

In this study, we aim to investigate the effects of radiation modeling on the growth of galaxies and BHs, as well as radiation effects on shaping gas properties. We therefore perform several runs, varying the stellar and AGN contributions, the cooling prescriptions and radiative efficiency $\epsilon_\mathrm{rad}$ (see Table~\ref{tab:prop}). All the runs were performed with the zoom-in region size of $R = 60 \,\mathrm{cMpc}$, and the same resolution as MTNG (i.e., zoom factor of one). We explore resolution convergence within a smaller region in Appendix~\ref{app:resolution}. 

Our fiducial model (RT-fid) employs an escape fraction of $f_\mathrm{esc,\star} = 0.25$ for stars, as well as radiation from AGN, using the obscuration prescription described in Sec.~\ref{sect:TNG}. We note that the value for $f_\mathrm{esc,\star}$ was chosen to reproduce a realistic reionization history (see Appendix~\ref{app:CalibrationEsc}). In order to disentangle the individual effects of stars and AGN, we perform two runs with the radiation from AGN turned off (i.e., stars only, RT-Stars), and with the radiation from stars turned off (i.e., AGN only, RT-AGN). We also investigate the case when both the radiation from stars and AGN are turned off (noRad), which still employs non-equilibrium cooling, but does not allow any ionizing photons to escape the sources. Similarly, we also include a run where radiation is not modeled self-consistently from the sources, but this time following the IllustrisTNG model: equilibrium cooling, and a spatially uniform UV background (see Sec.~\ref{sect:TNG}) -- EqCool-UVB. This allows us to identify any deviations from the TNG model caused by the on-the-fly radiative transfer prescriptions (see Sec.~\ref{sect:rad_effects},~\ref{sect:MBH-Ms} and~\ref{sect:BHAD}). All of these runs assume a radiative efficiency of $\epsilon_\mathrm{rad} = 0.1$, but we investigate the effects of changing this value from the fiducial TNG value of $\epsilon_\mathrm{rad} = 0.2$ in RT-$\epsilon_\mathrm{rad}=0.2$, keeping all the other parameters the same as in the fiducial model (RT-fid). 

Lastly, in order to explore the effects of unobscured quasars, which our BH growth prescriptions do not cover within the parameters explored in this study (Sec.~\ref{sect:TNG}), we introduce a boosted-model that mimics the effects of a central quasar (RT-Quasar). Since the maximum luminosity attained by the AGN in the fiducial run is about $L_\mathrm{bol} \approx 10^{46} \, \mathrm{erg~s^{-1}}$ (see Sec.~\ref{sect:erad} and Fig.~\ref{fig:QLF}), and further diminished by the obscuration prescription (Sec.~\ref{sect:TNG}), we do not expect the AGN to have a major contribution to the gas properties (e.g., formation of ionized bubbles). As such, the RT-Quasar run does not take obscuration into account, and we explore three luminosity boosts: by a factor of 5, 10 and 20, such that the maximum luminosity attained at $z \approx 6$ is $L_\mathrm{bol} \approx 5 \times 10^{46} \, \mathrm{erg~s^{-1}}$, $L_\mathrm{bol} \approx 10^{47} \, \mathrm{erg~s^{-1}}$, and $L_\mathrm{bol} \approx 2 \times 10^{47} \, \mathrm{erg~s^{-1}}$, respectively. These values are comparable to that of rare, observed quasars, and the set-up allows us to investigate properties of such objects, which will be reported in a subsequent paper (Bulichi et al., in prep).

\section{Black hole and galaxy growth}
\label{sect:BH_pop}
In this section, we explore the mass assembly of BHs along with their host galaxies. We first show a qualitative overview of black hole seeding and growth, in relation to the underlying
galaxy population. We then explore the change in the fiducial TNG radiative efficiency on BH growth, as well as the redshift evolution of the AGN luminosity function. Lastly, we focus on the co-evolution between galaxies and their central SMBHs, under the different models explained in Table~\ref{tab:prop}, in relation to recent \textit{JWST} results. 

\subsection{Overview}
\label{sect:env}

We illustrate the growth of the BHs in our box, as well as the growth of all the galaxies in Fig.~\ref{fig:env_proj}. As explained in Sec.~\ref{sect:TNG}, the simulated SMBHs are prescribed to trace overdensities, since the BH growth prescription is directly connected to the host halo. As a result, they are also hosted by massive galaxies. Our model predicts that the galaxies grow first, and grow faster, reaching orders of magnitude higher stellar masses than the BH masses at their centers (see also Sec.~\ref{sect:MBH-Ms} and \ref{sect:BHAD}), across all redshifts. 

Fig.~\ref{fig:env_proj} shows that the first BHs are seeded at around $z \approx 9$ ($M_\mathrm{seed} \sim 10^6 \, \mathrm{M_\odot}$), in galaxies with $M_\mathrm{\star} \sim 10^8\,\mathrm{M_\odot}$, along filamentary structures (i.e., overdense regions), due to the halo-dependent seeding model (see Sec.~\ref{sect:TNG}).
The Bondi--Hoyle accretion remains mostly inefficient between $z = 9$ and $z = 7.5$, with the first seeded BHs growing by less than 0.5 dex in this time, and mainly by means of mergers rather than accretion (see Fig.~\ref{fig:centralBH_z}). The galaxies, on the other hand, assemble faster, with the central galaxy growing by about two orders of magnitude in stellar mass in the same redshift interval. At $z = 6$, the most massive BH attains a mass of $M_\mathrm{BH} \sim 10^{8.5}\, \mathrm{M_\odot}$, and its host galaxy $M_\mathrm{\star} \sim 10^{11} \, \mathrm{M_\odot}$. They are located in the center of the box as part of a protocluster (zoomed-in in the sub-panels of Fig.~\ref{fig:env_proj}), with a halo virial mass $M_\mathrm{200} \approx 4 \times 10^{12} \, \mathrm{M_\odot}$. 

Lastly, we assume a radiative efficiency of $\epsilon_\mathrm{rad} = 0.1$ in Fig.~\ref{fig:env_proj}, and a fiducial RT set-up ($f_\mathrm{esc,\star} = 25 \, \%$ and AGN radiation; see Table.~\ref{tab:prop}), but we will explore the effects of the radiative efficiency and RT treatment on BH and galaxy growth in the subsequent subsections.

\subsection{Radiative efficiency}
\label{sect:erad}

\begin{figure}
    \includegraphics[width=\columnwidth]{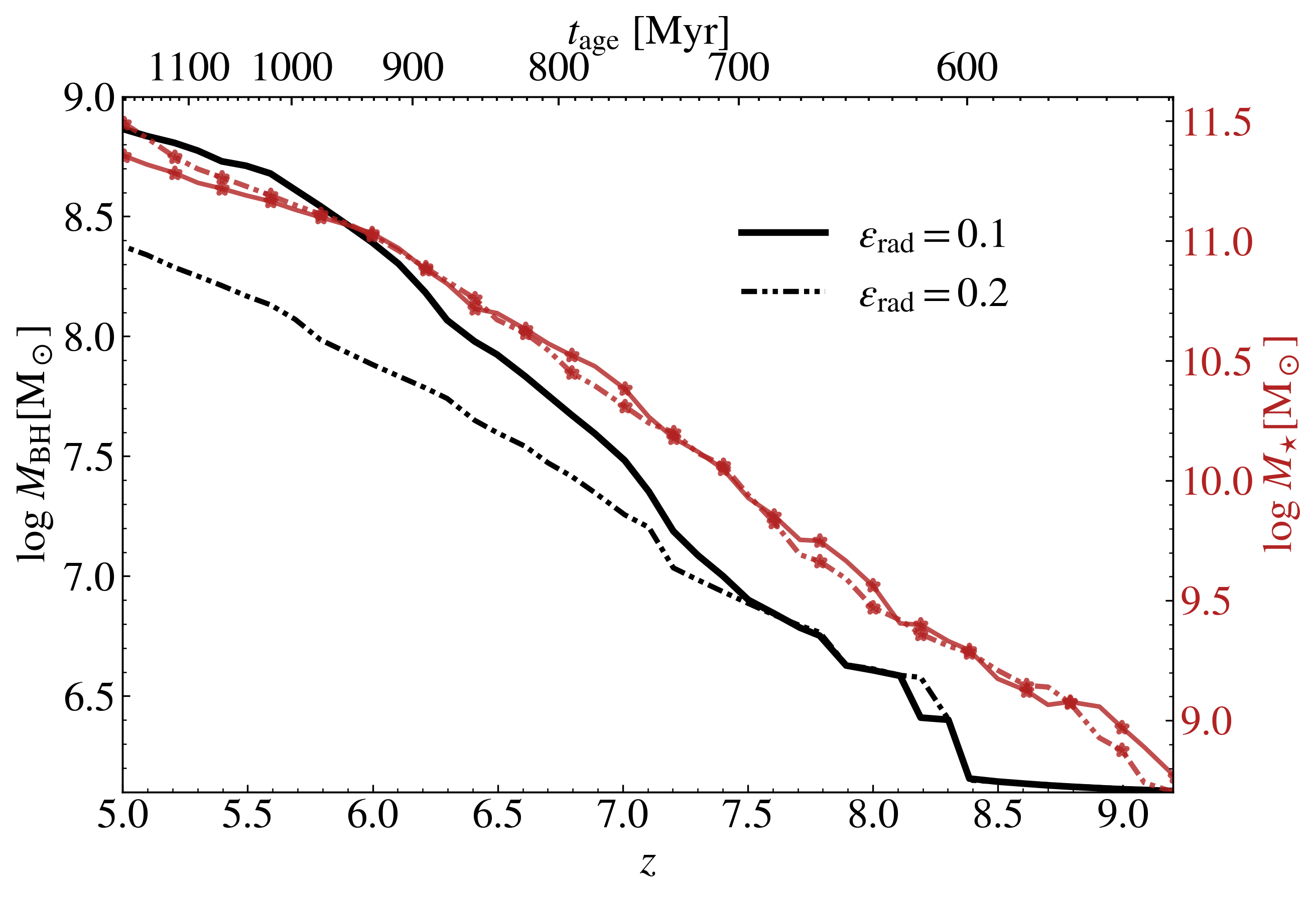}
    \caption{\textbf{Growth of the central SMBH (black) and its host galaxy (red) over time.} Solid and dotted-dashed lines represent $\epsilon_{\rm{rad}} = 0.1$ and $\epsilon_{\rm{rad}} = 0.2$, respectively. In terms of black hole growth, at very high redshift ($z \gtrsim 7.5$, and low BH mass, $M_\mathrm{BH} \lesssim 10^7\,\mathrm{M_\odot}$), the Bondi--Hoyle accretion is highly inefficient, and BH growth is dominated by mergers. Once $M_\mathrm{BH}$ attains $M_\mathrm{BH} \gtrsim 10^7\,\mathrm{M_\odot}$, the 
     differences between the two models ($\epsilon_\mathrm{rad} = 0.1$ and $\epsilon_\mathrm{rad} = 0.2$) grow over time, as the Bondi--Hoyle accretion rate scales as $\propto M_\mathrm{BH}^2$. The growth of the host galaxy is largely insensitive to the choice of $\epsilon_{\rm{rad}}$, apart from minor stochastic variations and the onset of quenching due to AGN kinetic feedback, which becomes effective at $z \lesssim 5.7$ in the $\epsilon_{\rm{rad}} = 0.1$ run. As a result, the co-evolutionary trajectories of the BH and its host galaxy diverge between the two models, a trend explored further in Fig.~\ref{fig:MBH-Ms}.} 
    \label{fig:centralBH_z}

\end{figure} 

\begin{figure}
    \includegraphics[width=\columnwidth]{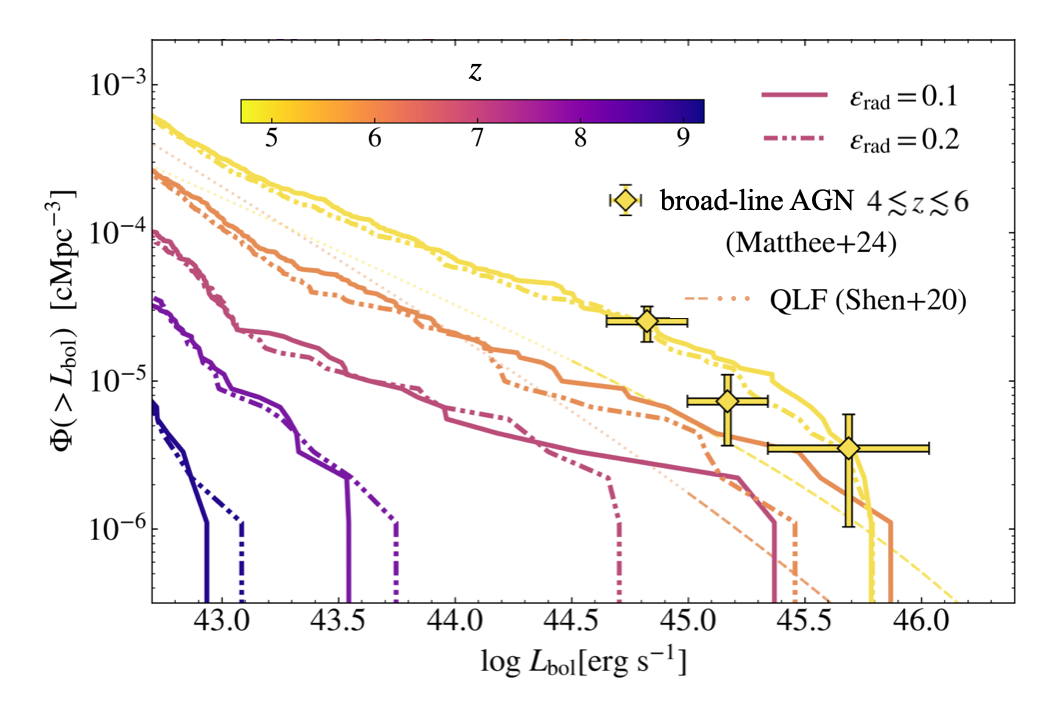}
    \caption{\textbf{Effects of the radiative efficiency value on the AGN luminosities.} The lines (\textit{solid}: fiducial model, $\epsilon_\mathrm{rad} = 0.1$; \textit{dotted-dashed}: $\epsilon_\mathrm{rad} = 0.2$) show the number density of AGN exceeding a given bolometric luminosity threshold, as indicated on the x-axis ($L_\mathrm{bol} = \epsilon_\mathrm{rad} \dot{M} c^2$, without accounting for attenuation from obscuration effects). The lines are color-coded by redshift, $z=9\rightarrow5$, in increments of $\Delta z =1$. At $z = 9$ and $z = 8$, when accretion is still highly inefficient and nearly identical in both simulations (see Fig.~\ref{fig:centralBH_z}), the AGN in the $\epsilon_\mathrm{rad} = 0.2$ run are approximately twice as luminous as those in the $\epsilon_\mathrm{rad} = 0.1$ run -- an expected result given the direct scaling with the radiative efficiency. Once the BH masses become high enough for accretion to become efficient ($z = 7$ and $z = 6$), $\epsilon_\mathrm{rad} = 0.1$ produces more luminous AGN, due to the significantly higher accretion rates. By $z = 5$, however, AGN kinetic feedback has already been triggered in the $\epsilon_\mathrm{rad} = 0.1$ simulation, suppressing BH growth for the massive BHs and leading to reduced accretion rates and consequently, lower AGN luminosities. We overplot observational constraints from \tb{broad-line AGN} \citep{Matthee2023} at $z \tb{ \approx} 5$, \tb{and the quasar luminosity function \citep{Shen2020} at $z \approx 6$ and $z\approx 5$ (dashed lines, and extrapolations shown with dotted lines),} finding good agreement with our simulations predictions.} 
    \label{fig:QLF}

\end{figure} 

Given our aim to reproduce closely the high\tb{-}redshift AGN population, we enhance the black hole growth by lowering the radiative efficiency from the fiducial TNG value $\epsilon_\mathrm{rad} = 0.2$ to $\epsilon_\mathrm{rad} = 0.1$. The value of the radiative efficiency is maintained constant over time, and we show its effects on the growth track of the central, most massive BH in our box, and its corresponding host galaxy in Fig.~\ref{fig:centralBH_z}. Given the scaling of the Bondi--Hoyle accretion with the square of the BH mass (Eq.~\ref{eq:bondiRate}), the differences between the BH growth tracks shortly after seeding at $z \approx 9$ (Fig.~\ref{fig:env_proj}), are minimal. In fact, accretion is mostly inefficient until the black hole mass reaches $M_\mathrm{BH} \sim 10^7\,\mathrm{M_\odot}$, at $z \approx 7.5$, and the growth in both cases is dominated by mergers. Once the black hole mass attains $M_\mathrm{BH} \sim 10^7\,\mathrm{M_\odot}$, the differences between the two radiative efficiencies become more pronounced with time, with the BH attaining a  $\sim 0.5\, \mathrm{dex}$ higher mass for $\epsilon_\mathrm{rad} = 0.1$ by the end of the simulation ($z = 5$). Additionally, given the high BH mass reached in the $\epsilon_\mathrm{rad} = 0.1$ run, the AGN kinetic feedback mode (Sec.~\ref{sect:TNG}) turns on at around $z \approx 5.7$ and quenches the subsequent accretion onto the BH, as well as the host galaxy growth, as seen in Fig.~\ref{fig:centralBH_z}.

Figure~\ref{fig:centralBH_z} also shows the stellar mass of the host galaxy as a function of redshift, for both $\epsilon_\mathrm{rad} =0.1$ and $\epsilon_\mathrm{rad} =0.2$. At $z \gtrsim 5.7$, the mass assembly of the host is unaffected by the value of $\epsilon_\mathrm{rad}$ (only influencing accretion onto the BH), modulo small stochastic effects. Below $z \approx 5.7$, as explained previously, the AGN kinetic feedback turns on and slows down the star formation of the host galaxy, resulting in lower stellar masses for the $\epsilon_\mathrm{rad} = 0.1$ run. \tb{We note that these findings are closely tied to the assumptions built into the IllustrisTNG model, in particular its relatively weak AGN thermal feedback. In models where feedback is implemented through direct momentum injection (e.g., \textsc{MISTRAL}; \citealt{Farcy2025}), we would expect the host galaxy’s stellar mass to be more strongly affected.} 

As such, changing the radiative efficiency has a considerable effect on the properties of the central BH, 
albeit the changes only become visible once $M_\mathrm{BH}$ reaches $M_\mathrm{BH} \sim 10^7\,\mathrm{M_\odot}$, about 10 times more massive than the seed mass. We also note that the mass where the accretion becomes more important than mergers is resolution dependent (see e.g., Appendix B of \citealt{Weinberger2018}), as the Bondi--Hoyle accretion is also highly sensitive to the gas density (eq.~\ref{eq:bondiRate}). The fact that \tb{in our model} the stellar mass of the host is mostly unaffected by the value of $\epsilon_\mathrm{rad}$ indicates two different co-evolutionary paths between the black hole and the host (see also Fig.~\ref{fig:MBH-Ms}), but in both cases the ratio between the two remains $M_\mathrm{BH}/M_\mathrm{\star} \lesssim 0.003$, falling below the local $M_\mathrm{BH}-M_\mathrm{\star}$ relation (\citealt{KormendyHo2013}). 

Looking at the properties of the AGN population, using a smaller value for the radiative efficiency aids in producing more luminous AGN at $z \sim 7.5 \rightarrow 6$, as depicted in Fig.~\ref{fig:QLF}. Figure~\ref{fig:QLF} shows the cumulative luminosity function (i.e., AGN number density as a function of bolometric luminosity), for both $\epsilon_\mathrm{rad} = 0.2$ and $\epsilon_\mathrm{rad} = 0.1$ at five different redshifts: $z = 9,\, 8,\, 7,\, 6,\, 5$. The bolometric luminosity in this case represents the intrinsic/unobscured bolometric luminosity: $L_\mathrm{bol} = \epsilon_\mathrm{rad} \dot{M}c^2$. At $z = 9$ and $z = 8$, given that the BH accretion is very similar between the two runs (Fig.~\ref{fig:centralBH_z}), the $\epsilon_\mathrm{rad} = 0.2$ results in more luminous AGN by a factor of $\sim 2$, given that $L_\mathrm{bol} \propto \epsilon_\mathrm{rad}$ in this regime. At $z = 7$ and $z = 6$, the high-mass end of the AGN population attains high enough masses such that the accretion is significantly higher for $\epsilon_\mathrm{rad} = 0.1$ than $\epsilon_\mathrm{rad} = 0.2$ (Fig.~\ref{fig:centralBH_z}), resulting in a more luminous end of the luminosity function. As before, the fainter AGN ($M_\mathrm{BH} \lesssim 10^7 \, \mathrm{M_\odot}$, $L_\mathrm{bol} \lesssim 10^{44}\,\mathrm{erg~s^{-1}}$) do not show strong differences between the two runs. At $z = 5$, given that the accretion onto the most massive BH has quenched in the $\epsilon_\mathrm{rad} = 0.1$, the luminosity drops to a lower value compared to $z = 6$, and the differences in the luminosity function between the two radiative efficiencies become minimal, as the competition between accretion rate, and the factor of two from the ratio of $\epsilon_\mathrm{rad}$ smooths out the differences. 

Lastly, we note that our model does not reproduce rare, luminous quasars with sustained high luminosities $L_\mathrm{bol} \sim 10^{47}\,\mathrm{erg~s^{-1}}$, at any of the snapshots output (redshift cadence $\Delta z = 0.1$), due to the limited size of the original MTNG box and restrictive BH growth. In particular, we remark that the Bondi-Hoyle model assumes the accretion flow is adiabatic, spherically symmetric, steady and unperturbed, which can underpredict the accretion rate in the low resolution regime (e.g., \citealt{Hopkins2011,Gaspari2013,Beckmann2018,Kho2025}). To this end, we will explore a quasar boosted model later in the paper, to analyze the effects of such sources not captured within the BH masses and accretion rates ranges of this study (see Sec.~\ref{sect:effects_gas}). For lower luminosities, however, \tb{we find a good agreement between our simulated AGN population and the observed abundances of broad-line AGN at $4 \lesssim z \lesssim 6$, and luminosities $L_\mathrm{bol} \sim 10^{45}\,\mathrm{erg~s^{-1}}$, reported in \cite{Matthee2023}. We note, however, that \cite{Matthee2023} includes only broad-line AGN, whereas our simulations encompass the entire AGN population, i.e., both broad- and narrow-line sources. This makes it difficult to interpret the comparison in the absence of observational constraints on the abundance of narrow-line AGN. Such constraints could in fact reveal a discrepancy in simulations in reproducing the high-redshift SMBH population (see also \citealt{Habouzit2024})}.

\tb{We also include the quasar luminosity function constraints from \cite{Shen2020}, reported at $z \approx 6$ for sources with $L_\mathrm{bol} \gtrsim 10^{45}\,\mathrm{erg\,s^{-1}}$ and at $z \approx 5$ for AGN with $L_\mathrm{bol} \gtrsim 10^{44.5}\,\mathrm{erg\,s^{-1}}$. In the regions of overlap, our results show good agreement with \cite{Shen2020}, with minor discrepancies likely arising because the \cite{Shen2020} fits miss the obscured, lower-luminosity sources. We also note that the faint-end extrapolation in \cite{Shen2020} was highly uncertain prior to \textit{JWST}.}

\begin{figure*}
    \includegraphics[width=\textwidth]{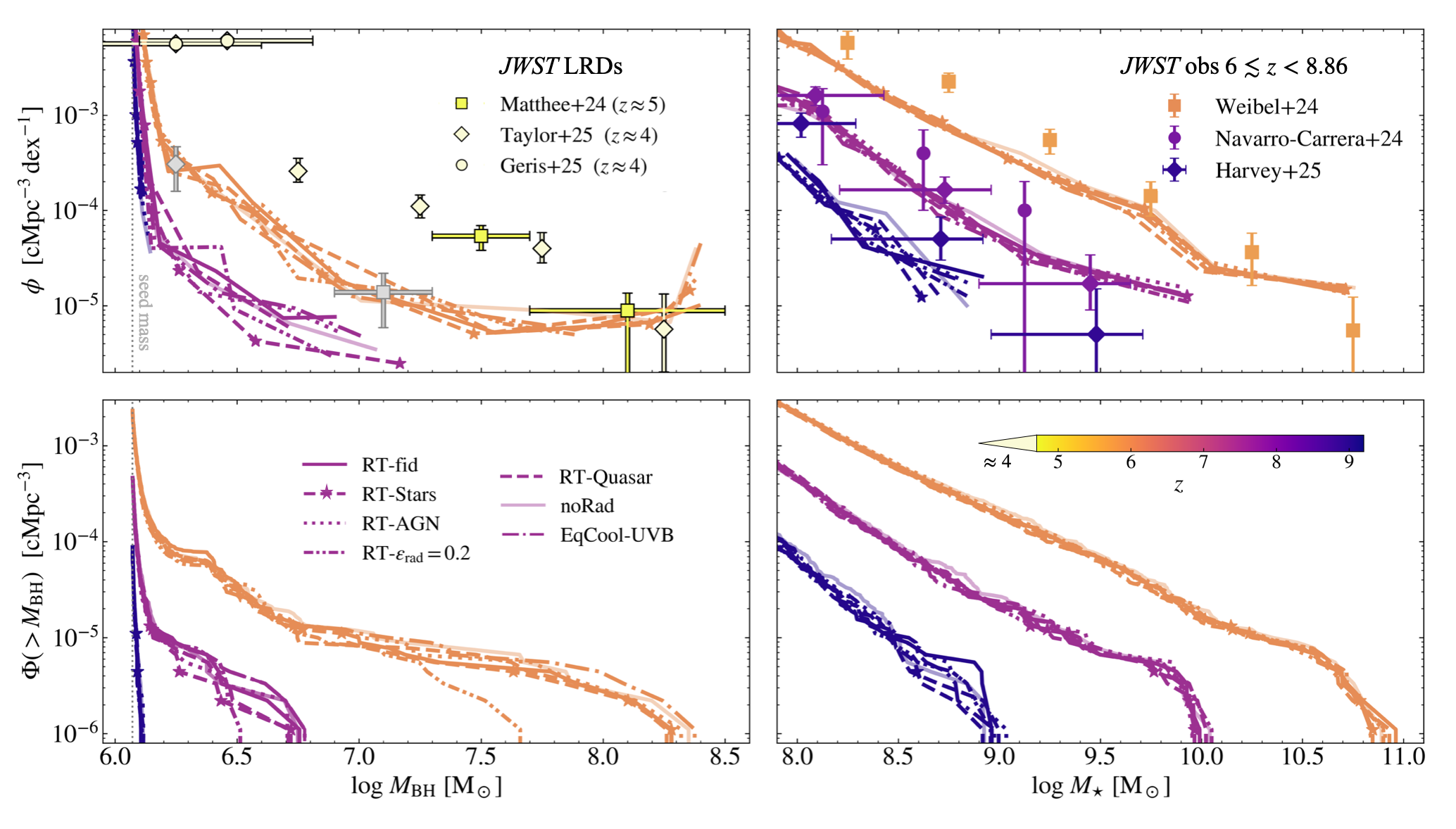}
    \caption{\textbf{Black hole mass function (\textit{left column}) and stellar mass function (\textit{right column}), at three different redshifts: $z = 9$, $z = 7.5$ and $z = 6$.} We show all radiation models, explained in Table~\ref{tab:prop}, and for an easier distinction between them, we include the cumulative black hole and stellar mass functions in the bottom row. As seen in Fig.~\ref{fig:env_proj}, the first BHs are seeded at around $z \approx 9$ (seed mass is indicated by the gray dotted line), and grow slower than their host galaxies. The radiation models show good overall convergence in both the black hole and galaxy populations, with small differences due to feedback and stochastic effects. The main difference arises in the $\epsilon_\mathrm{rad} = 0.2$ run, which shows a suppressed high-mass end of the black hole mass function at $z = 7.5$ and $z = 6$, as expected (see Sec.~\ref{sect:erad}). The upturn at the high-mass end of the black hole mass function at $z = 6$ is driven by the dense environments surrounding the most massive BHs (see Fig.~\ref{fig:env_proj}), which boost accretion rates, along with the strong mass dependence of the Bondi--Hoyle accretion model. However, this trend will be suppressed by $z \approx 5.5$, as AGN kinetic feedback becomes active. \tb{We compare our predictions with observational constraints from LRDs BHMFs (top-left panel, in yellow, with gray points representing data not corrected for completeness), and stellar mass functions from \textit{JWST} studies (top-right panel, color-coded by redshift). Although the LRDs BHMF constraints are derived at lower redshifts than our simulations, they suggest very rapid black hole growth between $z = 6$ and $z \approx 4$, which may be difficult to reproduce with current models. We find overall good agreement with the observational constraints on the stellar mass function.}}
    \label{fig:BHMF6}
\end{figure*}

\subsection{Black hole and stellar mass function}
\label{sect:rad_effects}

Figure~\ref{fig:BHMF6} explores the black hole and stellar mass functions for all BHs, and galaxies in our box, at three redshifts: $z = 9$, $z = 7.5$ and $z = 6$ (same redshifts as in Fig.~\ref{fig:env_proj}), for the radiation models described in Table~\ref{tab:prop}. In order to distinguish the small differences between the models, we include the cumulative BH and stellar mass functions also, in the bottom panel of Fig.~\ref{fig:BHMF6}. As discussed in Sec.~\ref{sect:env}, the BHs are seeded around $z \approx 9$, and experience a slower growth compared to their hosts, with the Bondi--Hoyle accretion being highly inefficient for $M_\mathrm{BH} \lesssim 10^7\,\mathrm{M_\odot}$ (see also Sec.~\ref{sect:erad}). In all cases, the mass functions agree between the different radiation models explored, as expected since the radiative feedback does not impact BH accretion, nor SFR (Sec.~\ref{sect:TNG}). The most noticeable difference is at the high-mass end of the black hole mass function (BHMF), at $z = 6$, where a radiative value of $\epsilon_\mathrm{rad} = 0.2$ produces less massive BHs, by about 0.5 dex, than the $\epsilon_\mathrm{rad} = 0.1$ counterparts (see also Fig.~\ref{fig:centralBH_z}). Interestingly, the high-mass end of the BHMF also displays a small upturn, which we attribute to the more efficient BH growth in this regime, for $\epsilon_{\mathrm{rad}} =0.1$. 

\tb{We also include observational constraints from the so-called ``little red dots'' (LRDs) studies at $z \approx 4$ \citep{Taylor2024, Geris2025} and $z \approx 5$ \citep{Matthee2023}. Although a direct comparison with our simulations is not possible due to the redshift mismatch, the observations imply that our simulated black holes would need to undergo very rapid growth to be consistent. Moreover, since LRDs likely represent only a subset of the full AGN population, this underscores the potential difficulty for current models, at this resolution, to reproduce the observed abundance of AGN (see also \citealt{Habouzit2024}).}

The stellar mass function shows excellent convergence between all models and is unaffected by the radiation modeling, as well as the value of the radiative efficiency, $\epsilon_\mathrm{rad}$. By $z = 6$, the most massive galaxies grow to $M_\mathrm{\star} \sim 10^{11} \, \mathrm{M_\odot}$. \tb{We overplot observational constraints from \cite{Weibel2024, Navarro-Carrera2024, Harvey2024} and find overall agreement within the uncertainties, though our predictions tend to fall consistently below the observational estimates. This offset has been previously explored in the literature, with several possible explanations proposed (e.g., \citealt{Shen2023, Koehler2025}). A detailed investigation of this discrepancy, however, is beyond the scope of the present work.} We explore further the co-evolution between the central BHs and their host, as well as the mass assembly history for both BHs and stellar masses in the following two subsections.

\subsection{Black hole -- stellar mass relation}
\label{sect:MBH-Ms}

\begin{figure}
    \includegraphics[width=0.5\textwidth]{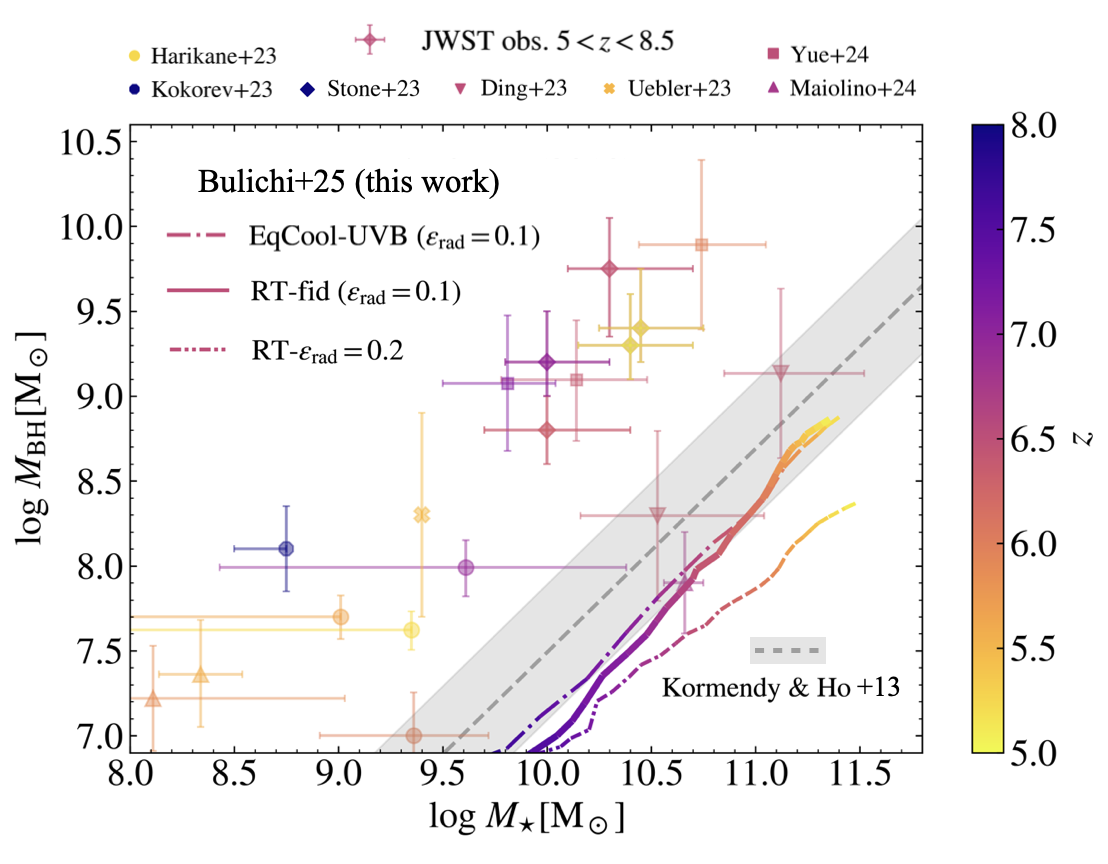}
    \caption{\textbf{The $M_{\rm{BH}}$--$M_\star$ relation for the central black hole in our simulation over $\bm{5 < z < 8}$.} Lines are color-coded by redshift and represent the fiducial RT model (solid line), the RT model with $\epsilon_{\mathrm{rad}} = 0.2$ (double-dot-dashed line), and the equilibrium cooling-UVB model (dot-dashed line). The \protect\cite{KormendyHo2013} local relation is shown as a grey dashed line with a shaded region, indicating the scatter. Observational data \tb{for $5 < z < 8.5$} from \textit{JWST}, color-coded by redshift, are also overplotted \tb{(with references shown in the plot)}. This figure illustrates that galaxies assemble their stellar mass earlier, with the central black holes consistently lying below the local $M_\mathrm{BH}$–$M_\star$ relation, and significantly less massive than the \tb{AGN} observed with \textit{JWST}. The choice of radiative efficiency leads to distinct co-evolutionary tracks: while the stellar masses remain comparable, black hole growth becomes increasingly suppressed for $\epsilon_\mathrm{rad} = 0.2$ at $z \lesssim 7.5$ (see also Fig.~\ref{fig:centralBH_z}). As seen in Fig.~\ref{fig:BHMF6}, the overall growth of both black holes and galaxies is largely insensitive to the specific radiation model. Here, we include only two representative models (fiducial and equilibrium cooling–UVB) to highlight the relatively minor effects of feedback and stochastic variability.}
    \label{fig:MBH-Ms}
\end{figure}

In Fig.~\ref{fig:MBH-Ms}, we investigate the redshift evolution of the $M_{\rm{BH}}$--$M_\star$ relation for the central BH in our simulation, for two radiative efficiencies: $\epsilon_{\rm{rad}} = 0.1$ (fiducial value) and $\epsilon_{\rm{rad}} = 0.2$ (the TNG model's fiducial value), and the fiducial RT set-up (Table~\ref{tab:prop}). We exclude the other RT models ($\epsilon_\mathrm{rad} = 0.1$) from the figure, as they all show good convergence with the RT-fid model (Sec.~\ref{sect:rad_effects}), but include the ``EqCool-UVB'' ($\epsilon_\mathrm{rad} = 0.1$), to contrast the fiducial TNG model with the predictions from TNG + on-the-fly RT. 
We overplot the recent \textit{JWST} observations from \tb{\cite{Stone2023, Ding2023, Harikane2023, Kokorev2023, Uebler2023, Maiolino2024}} and \cite{Yue2023}, for comparison. 
\tb{With only one exception in \cite{Maiolino2024},} all \tb{other} determinations are above our $M_\mathrm{BH}$--$M_\star$ predictions.
We note that the BH considered in our plot is expected to be ``tip of the iceberg'' of our BH population, since it is the most massive. \tb{Yet, it remains several orders of magnitude less massive than some of the black hole masses inferred from observations at the same stellar mass.}
This result highlights once again the difficulty in modeling \tb{massive BHs} self-consistently, while ensuring accurate galaxy properties across cosmic time (see Sec.~\ref{sect:TNG}), \tb{with current models. To address this discrepancy, prior work has proposed a range of possibilities, including super-Eddington accretion \citep{Husko2024}, revised disk models \citep{Hopkins2024,Shi2024}, and massive black hole seeds \citep{Bhowmick2024}, while also noting that observational selection biases may play a role \citep{Li2024}, as well as the impact of SED assumptions on BH mass estimates \citep{Naidu2025}, which can lead to significant overestimates.}
\tb{In this work, in order to incorporate the missing population at the high luminosity end (comparable to \citealt{Stone2023, Ding2023, Yue2023}), we will explore a modified model that mimics quasars in Sec.~\ref{sect:quasar_model}.}


As shown in Fig.~\ref{fig:centralBH_z}, the co-evolutionary trajectories of the $\epsilon_\mathrm{rad} = 0.1$ and $\epsilon_\mathrm{rad} = 0.2$ models exhibit significant differences. Whilst the stellar mass of the host galaxy remains similar in both cases, the black hole mass is consistently lower for $\epsilon_\mathrm{rad} = 0.2$ once $M_\mathrm{BH} \gtrsim 10^7\,\mathrm{M_\odot}$ ($z \lesssim 7.5)$, with the discrepancy becoming more pronounced at higher BH masses/ lower redshifts. However, as the black hole growth is self-regulated (see Fig.~\ref{fig:centralBH_z} and Sec.~\ref{sect:TNG}), the BH population is expected to lie on the \cite{KormendyHo2013} local relation at $z = 0$. As such, changing $\epsilon_\mathrm{rad}$  will only affect the $M_\mathrm{BH}-M_\star$ ratio at high-$z$. We note \tb{again} that the TNG model is by design prone to produce ``under-massive'' BHs that grow to the local relation at $z \approx 0$; and the value, and potential time evolution of radiative efficiency, as well as observational biases and uncertainties can also play a role in the discrepancy between our results and \textit{JWST} observations.

The EqCool-UVB shows some small differences in the BH/stellar mass growth track compared to the fiducial RT model. Whilst the models produce very similar AGN populations (Sec.~\ref{sect:rad_effects}), the most massive BH, and its host appear to be mildly influenced by the the two different models. Given that the accretion rate is most efficient for the highest mass BH, it is not surprising to see these differences that we attribute to the different cooling prescriptions and feedback, but also note that the effect is overall very weak. Thus, we conclude that coupling on-the-fly radiative transfer with the TNG galaxy formation model has a minimal impact on the BH and galaxy growth (see also Sec.~\ref{sect:diff_thesan}).

\subsection{Black hole accretion and star formation rate history}
\label{sect:BHAD}
\begin{figure}
    \includegraphics[width=0.5\textwidth]{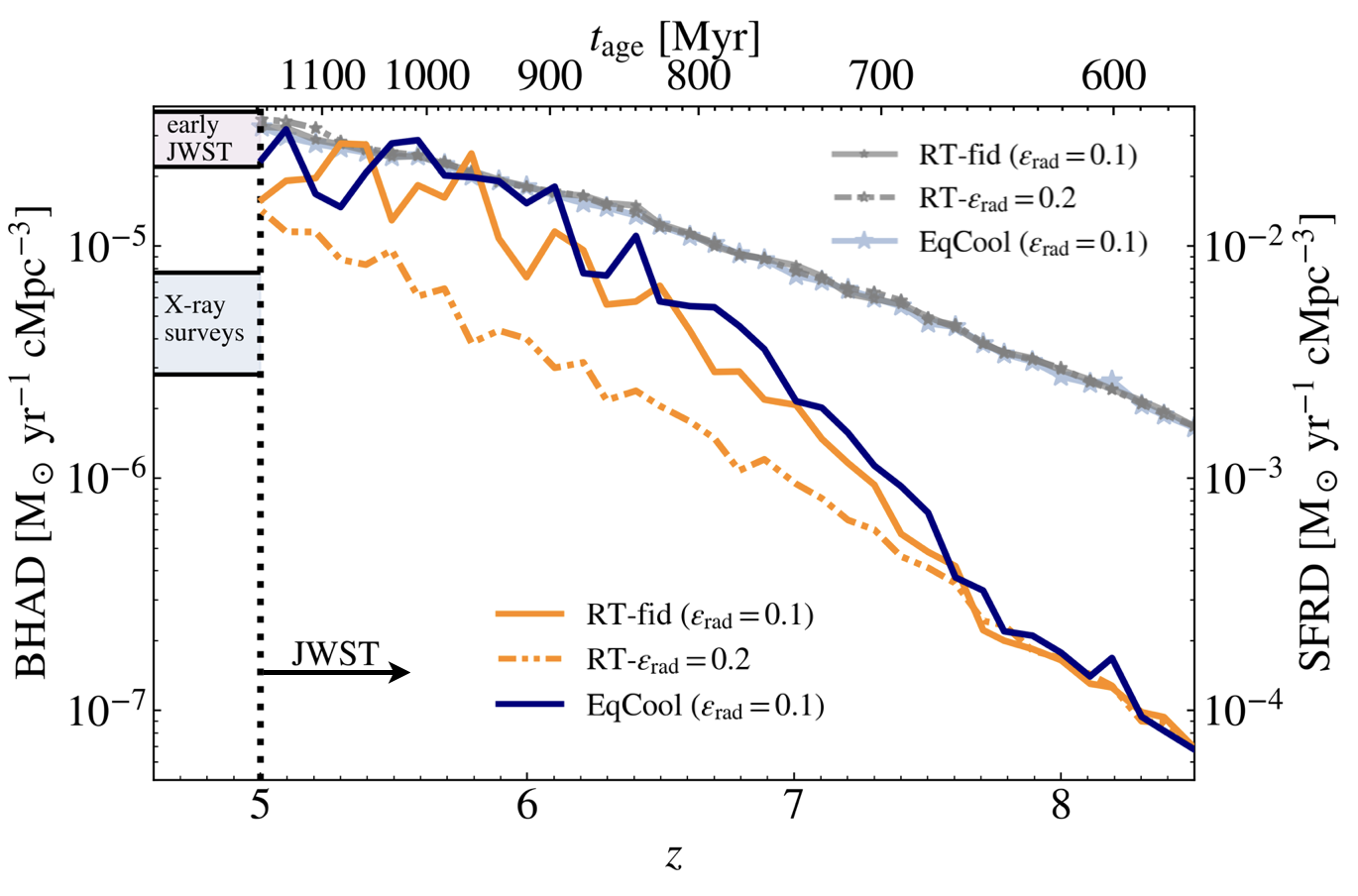}
    \caption{\textbf{Black hole accretion and star formation rate densities over cosmic time.} We compare three models: RT-$\epsilon_{\mathrm{rad}} = 0.1$ (fiducial, solid line), RT-$\epsilon_{\mathrm{rad}} = 0.2$ (double dot-dashed line), and equilibrium cooling-UVB (solid line, blue). As in Fig.~\ref{fig:BHMF6} and~\ref{fig:MBH-Ms}, galaxies assemble their stellar mass earlier and faster than black holes. A lower radiative efficiency ($\epsilon_{\mathrm{rad}} = 0.1$) leads to higher and more bursty BH accretion at $z \lesssim 7.5$ (see Sec.~\ref{sect:erad}), driven by stronger AGN feedback. Differences between RT-fid and EqCool reflect different cooling models and minor feedback and stochastic effects. We also compare the BHAD at $z = 5$ with early \textit{JWST}/MIRI data (\citealt{Yang2023}) and X-ray surveys \citep{Ueda2014,Aird2015,Vito2018,Ananna2019}, which may miss heavily obscured AGN. Future \textit{JWST} programs (e.g., MEOW) will help constrain the BHAD at $z \gtrsim 5$.}
    
   \label{fig:BHAD}
\end{figure} 

In this subsection, we highlight the mass assembly of BHs and galaxies in our box, over time $z = 8.5 \rightarrow 5$, in Fig.~\ref{fig:BHAD}. Similar to Sec.~\ref{sect:MBH-Ms}, we only show three of the models explored in the study: RT-fiducial ($\epsilon_\mathrm{rad} = 0.1$), RT-$\epsilon_\mathrm{rad} = 0.2$, and the TNG model (EqCool-UVB, $\epsilon_\mathrm{rad} = 0.1$). 

As already seen in Fig.~\ref{fig:centralBH_z} and Fig.~\ref{fig:BHMF6}, the value of the radiative efficiency does not make a strong difference on the overall accretion onto the BHs at high redshift ($z \gtrsim 7.5$), and has a negligible effect on the assembly of stellar mass. After $z \approx 7.5$, the most massive BHs accrete more efficiently when $\epsilon_\mathrm{rad} = 0.1$, and the accretion is also more bursty. 
We attribute the burstiness in accretion to the stronger AGN kinetic feedback in the $\epsilon_\mathrm{rad} =0.1$ run (see also Sec.~\ref{sect:erad}), leading to more pronounced episodic accretion, than in the RT-$\epsilon_\mathrm{rad} = 0.2$ run which displays a smooth BH mass assembly history. Similar to Sec.~\ref{sect:MBH-Ms}, the different cooling prescriptions between EqCool and RT (both $\epsilon_\mathrm{rad} = 0.1$) lead to small, stochastic differences in BH growth. Even though observational constraints at $z > 5$ have been difficult pre-\textit{JWST}, due to the presence of obscured, Compton-thick AGN, our results appear in reasonable agreement with the early \textit{JWST} results from \cite{Yang2023}, 
and significantly above X-ray surveys results \citep{Ueda2014,Aird2015,Vito2018,Ananna2019}, even for $\epsilon_\mathrm{rad} = 0.2$. At face value, our findings reinforce the conclusion that X-ray surveys may fail to detect a substantial population of obscured AGN, and upcoming \textit{JWST} surveys
are expected to provide further constraints.

The stellar mass assembly of galaxies follows a smooth track between $z = 8.5 \rightarrow 5$, independent on the value of the radiative efficiency, or on the cooling prescription, in agreement with Fig.~\ref{fig:MBH-Ms} and Fig.~\ref{fig:BHMF6}. 
It also displays a flatter trend than the BHAD, showing high levels of star formation even at high redshift. This finding shows once again that galaxies assemble first, and grow faster than BHs, whose masses remain close to the seed mass at $z \sim 9 \rightarrow 7.5$ (Sec.~\ref{sect:env}, Sec.~\ref{sect:erad}, Sec.~\ref{sect:rad_effects}).


\section{Radiation modeling} 
\label{sect:rad_mod}
In this section we focus on the gas response to radiation, under some of the different scenarios described in Table~\ref{tab:prop}. We first show a broad overview of the IGM properties, as a response to radiation. We then focus on how different sources (AGN and stars) shape the gas properties, disentangling their individual contributions. Lastly, to account for the quasar population not captured by our simulations, we introduce a quasar boosted model and compare its impact on gas properties to those in our fiducial RT model and findings from previous studies.

\subsection{IGM structure}
\label{sect:IGM}
\begin{figure*}
    \includegraphics[width=\textwidth]{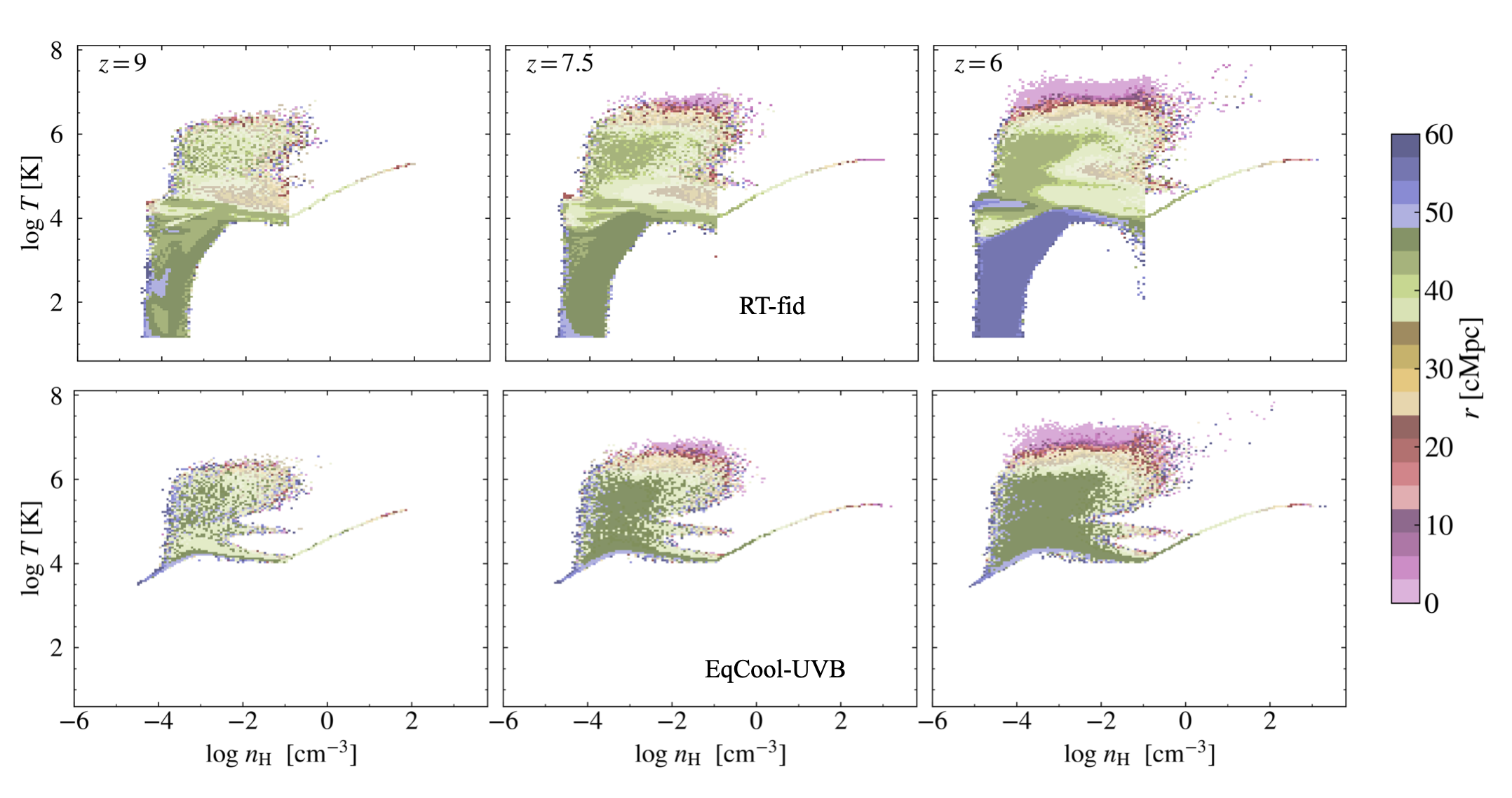}
    \caption{\textbf{Phase-space diagrams, illustrating the structure of the IGM.} We show three different redshifts: $z = 9$ (\textit{left column}), $z = 7.5$ (\textit{middle column}), and $z = 6$ (\textit{right column}), for the fiducial RT model (\textit{top row}), and the equilibrium cooling UVB (i.e., TNG model; \textit{bottom row}). Pixels are colored by the mass-weighted mean distance from the central halo (protocluster), illustrating that the hottest gas is heated by the most massive AGN and galaxies at the center of the box, as well as shock heating. This figure also clearly shows that a spatially uniform UV background eliminates all cold, low-density gas, whereas the self-consistent radiation modeling retains such gas in regions where the ionization front has yet to propagate.}
    \label{fig:phase-space}

\end{figure*} 

\begin{figure}
    \includegraphics[width=\columnwidth]{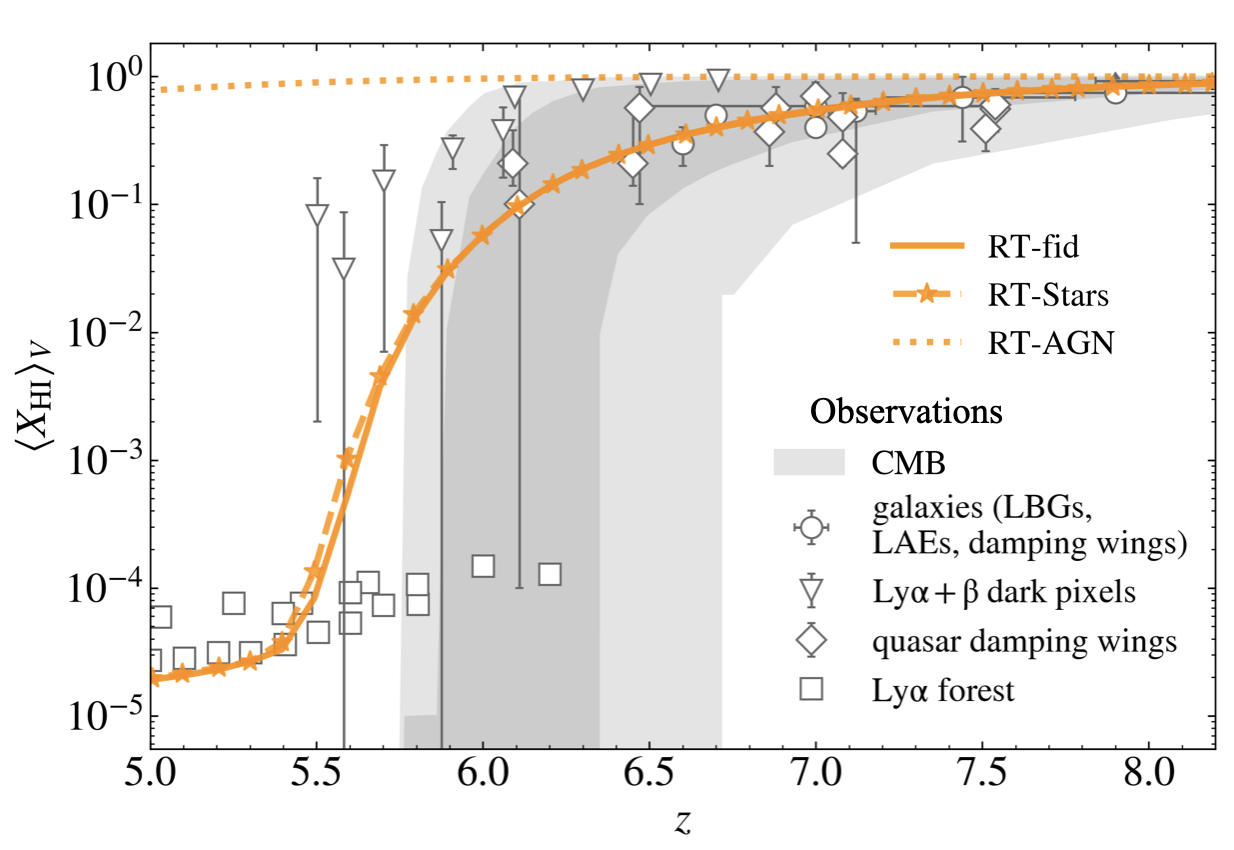}
    \caption{\textbf{Reionization history of the zoom-in region.} The solid line shows the fiducial model, the dashed-stared line shows the RT run without AGN radiation (i.e., stars only), and dotted line the RT run without stellar radiation (i.e., AGN only). It can be clearly seen that the galaxies are the main drivers of reionization, with the AGN effects being visible only at $z\approx 6$, and our simulated results are in good agreement with observations from quasar damping wings \citep{Banados2018,Davies2018,Yang2020-dw,Durovcikova2020,Wang2021,Durovcikova2024}; galaxies damping wings \citep{Mason2018, Ouchi2010, SobacchiMesinger2015, Mason2019, Ning2022, Umeda2023}; $\rm{Ly\alpha}$ forest \citep{Fan2006,Yang2020,Bosman2021}; $\mathrm{Ly\alpha+\beta}$ dark pixels \citep{McGreer2015,Jin2023}; and CMB \citep{Planck2020}.}
    \label{fig:xHIV}

\end{figure}

\begin{figure*}
    \includegraphics[width=\textwidth]{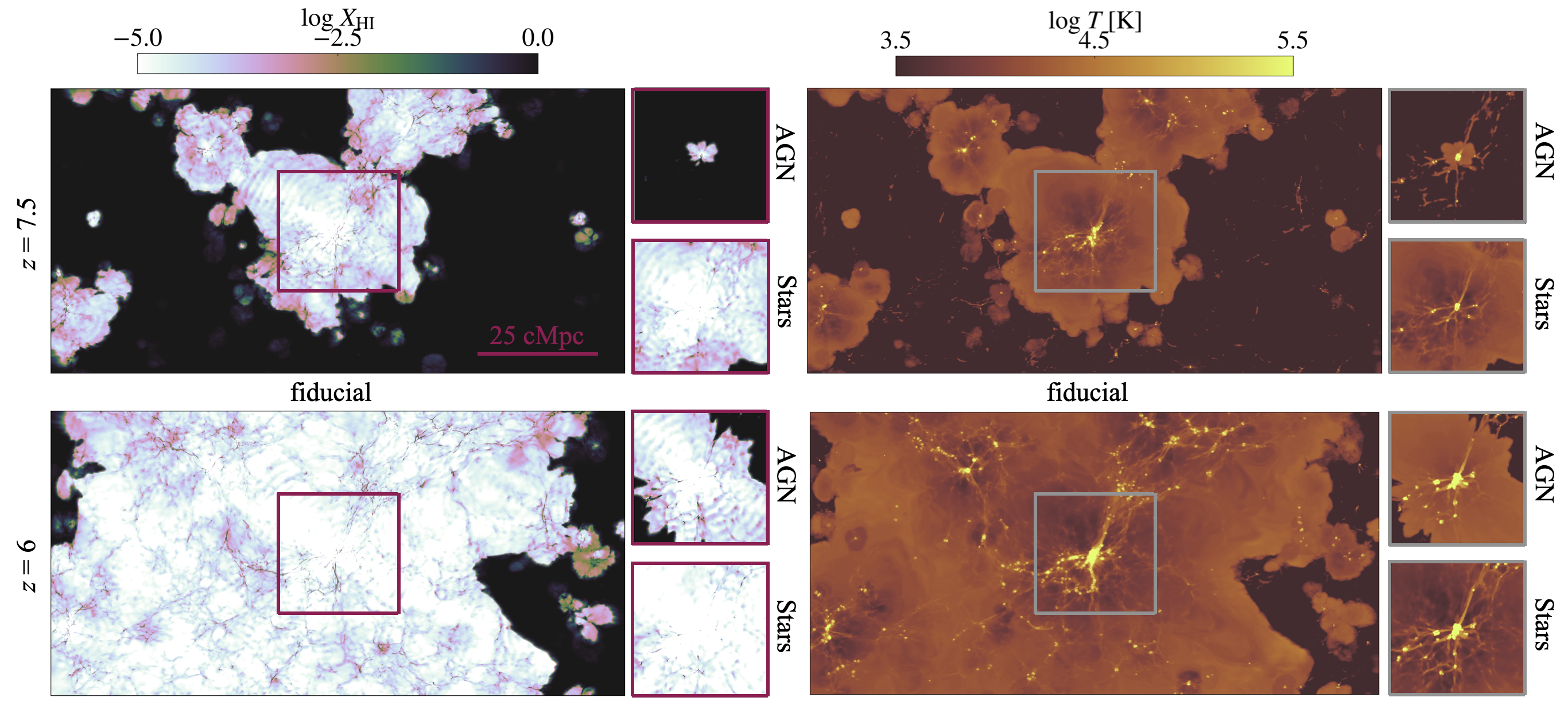}
    \caption{\textbf{Visual representation of the gas responses to the radiation.} The big panels ($120 \times 60\times 2 \, \rm{cMpc^3}$), show the fiducial model, including radiation from stars and AGN. The subpanels  ($25 \times 25\times 0.2 \, \rm{cMpc^3}$) show the runs with radiation from AGN only (\textit{top}) and stars only (\textit{bottom}). The top and bottom rows show $z = 7.5$ and $z = 6$ respectively, and the left and right column show the neutral H fraction and temperature of gas, respectively. It can be seen that the gas ionization is inhomogeneous, with ionized patches concentrated around sources (see also Fig.~\ref{fig:env_proj}) propagating ``inside-out''. Dense gas remains neutral longer due to self-shielding. Additionally, in this set-up, the stars contribute to most of the ionizing budget, even in the vicinity of the central SMBH. The AGN alone is able to form an ionized ``bubble'', whose irregular shape reflects the anisotropy in gas density. A similar trend is observed for the gas' temperature also, with stars having the strongest contribution, and the ionized gas being hotter than the gas in neutral regions.}
    \label{fig:maps}

\end{figure*} 
In order to illustrate the gas' responses to radiation, we show an overview of the global states of the gas in the density-temperature phase-space diagram (Fig.~\ref{fig:phase-space}), for two net different models: RT-fid (fiducial model, top panel), and EqCool-UVB (TNG model, bottom panel), at the same three redshifts as Fig.~\ref{fig:env_proj} and Fig.~\ref{fig:BHMF6}: $z = 9$, $z = 7.5$ and $z = 6$. We color-code the data points by the distance from the central AGN, located at the center of the most massive halo/protocluster of our box. Whilst here we only show a broad overview of the different gas responses to radiation prescriptions, we will focus on the AGN and galaxies contributions to shaping the gas properties in the next subsections.

The main difference between the two models, clearly visible in Fig.~\ref{fig:phase-space}, is that the uniform UV background heats up all the gas above $T \gtrsim 10^4 \, \mathrm{K}$, while the self-consistent radiative transfer treatment heats up the gas gradually, with the gas in the protocluster (center of the box) being heated up first by the central AGN and stars, and the outskirts remaining at $T \sim 10^2-10^3 \,\mathrm{K}$ by $z = 6$, with some gas still at the $T \approx 20\, \rm{K}$ temperature floor.

The figure also depicts other well-known features of the gas properties (e.g., \citealt{Dave2001,Vogelsberger2012}), for both RT-fid, and EqCool-UVB. First, as explained in Sec.~\ref{sect:TNG}, once the gas becomes sufficiently dense and crosses the star formation threshold ($n_\mathrm{H} > 0.11 \,\mathrm{cm}^{-3}$, we model it using the effective equation of state \citep{SpringelHernquist2003}. This equation captures the average thermal energy density of a two-phase medium composed of hot and cold gas maintained by unresolved stellar feedback in the ISM. Additionally, low-density gas heated to $T \sim 10^{4}\,\mathrm{K}$ forms a narrow ridge at higher temperatures, tracing highly photo-ionized gas in the IGM. This reflects the balance between photo-ionization heating and adiabatic cooling, due to cosmic expansion, as well as inverse Compton cooling. Gas with $T \sim 10^5 - 10^7\,\mathrm{K}$ and low density (i.e., less dense than the eEOS gas), is shock heated in virialized halos, in and around filaments. The downward slope at $T\sim 10^4 \, \mathrm{K}$ and $n_\mathrm{H} \gtrsim 10^{-2}\,\mathrm{cm^{-3}}$ corresponds to dense gas undergoing radiative cooling within galaxies. Given that the cooling time is short, the gas remains close to the equilibrium temperature: $T \sim 10^4\,\mathrm{K}$, and lower at higher densities, defined as the temperature at which photo-ionisation heating equates radiative cooling.

\subsection{Ionization of the zoom-in region}
\label{sect:ioniz}

The gas overdensity of our whole zoom region is $\delta_{\rm{g}} = \rho /\bar{\rho} -1 \approx 0.1$. As such, we expect the ionization timing of the whole region to follow roughly the Universe's reionization, as reported in \cite{Kannan2022}, though some bias can still be at play.  We show the volume weighted $X_{\rm{HI}}$ redshift evolution, measured within the full extent of our ``zoom-in'' region, in Fig.~\ref{fig:xHIV}. 
In order to disentangle the individual contributions of AGN and stars, we show three different runs: RT-fid (AGN+stars), RT-Stars (stars only; AGN turned off) and RT-AGN (AGN only; stars turned off).

It is important to note that the overall ionization of the box is highly dependent on the value of $f_\mathrm{esc,\star}$ (see Sec.~\ref{sect:RT}). Additionally, we also employ an obscuration model for the AGN, according to the \cite{Hopkins2008} parametrization (see also \citealt{Vogelsberger2013}), which introduces extra free parameters that influence the reionization history, as they restrict the AGN ionizing photon budget. 

For this study, we select $f_\mathrm{esc,\star} = 25\%$, and show the effects of different escape fractions in Appendix~\ref{app:CalibrationEsc}. AGN obscuration is modeled using the parameters from \cite{Vogelsberger2013}. For the fiducial run, the gas is still neutral at $z = 8$, and becomes ionized by $z = 5$, in good agreement with other observational constraints from quasar damping wings \citep{Banados2018,Davies2018,Yang2020-dw,Durovcikova2020,Wang2021,Durovcikova2024}; galaxies damping wings \citep{Mason2018, Ouchi2010, SobacchiMesinger2015, Mason2019, Ning2022, Umeda2023}; $\rm{Ly\alpha}$ forest \citep{Fan2006,Yang2020,Bosman2021}; $\mathrm{Ly\alpha+\beta}$ dark pixels \citep{McGreer2015,Jin2023}; and CMB \citep{Planck2020}.

Under this scheme, Fig.~\ref{fig:xHIV} shows that galaxies are the main drivers of reionization, with the AGN alone being unable to ionize the gas on large scales. This finding is in good agreement with many previous studies, e.g., \cite{Eide2020,Trebitsch2021,Yeh2023,Jiang2025,Dayal2025}. The only notable differences between RT-Stars (stars) and RT-fid (stars+AGN) are visible at $z \approx 5.7$, when the AGN become bright enough, resulting in a slightly higher ionization fraction. However, this effect washes out by $z \approx 5$, as the accretion onto the most massive BHs quenches (see Fig.~\ref{fig:centralBH_z}, Fig.~\ref{fig:BHAD}) due to the AGN's kinetic feedback mode. This results in less radiation from AGN (Fig.~\ref{fig:QLF}), unable to leave a strong imprint on the ionization state of the gas.

\subsection{Ionizing front and effects on surrounding gas}
\label{sect:effects_gas}

\begin{figure}
    \includegraphics[width=0.5\textwidth]{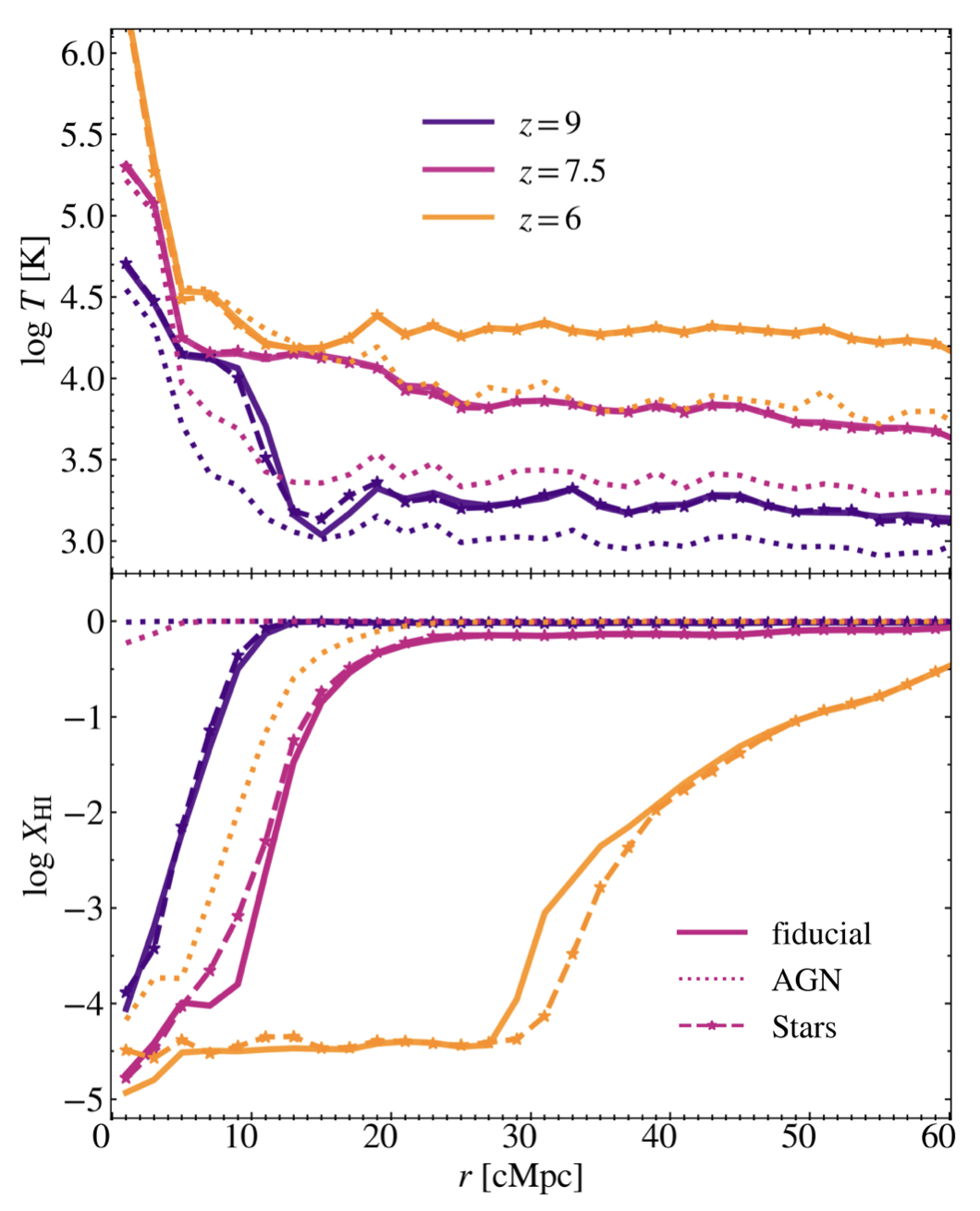}
    \caption{\textbf{Spherically averaged profiles for temperature and H neutral fraction, measured from the central AGN.} We show three different redshifts: $z =9$ (indigo), $z = 7.5$ (violet) and $z = 6$ (orange), and four different radiation models: fiducial (solid line), RT-AGN (dotted line), RT-Stars (dashed-star line). Gas gets heated up and ionized mostly from the central protocluster, with the fronts propagating inside out. Stars have the strongest contribution, with the AGN only ionizing the gas further in its vicinity at $z = 6$.}
    \label{fig:rad_prof}

\end{figure} 

\begin{figure}
    \includegraphics[width=0.5\textwidth]{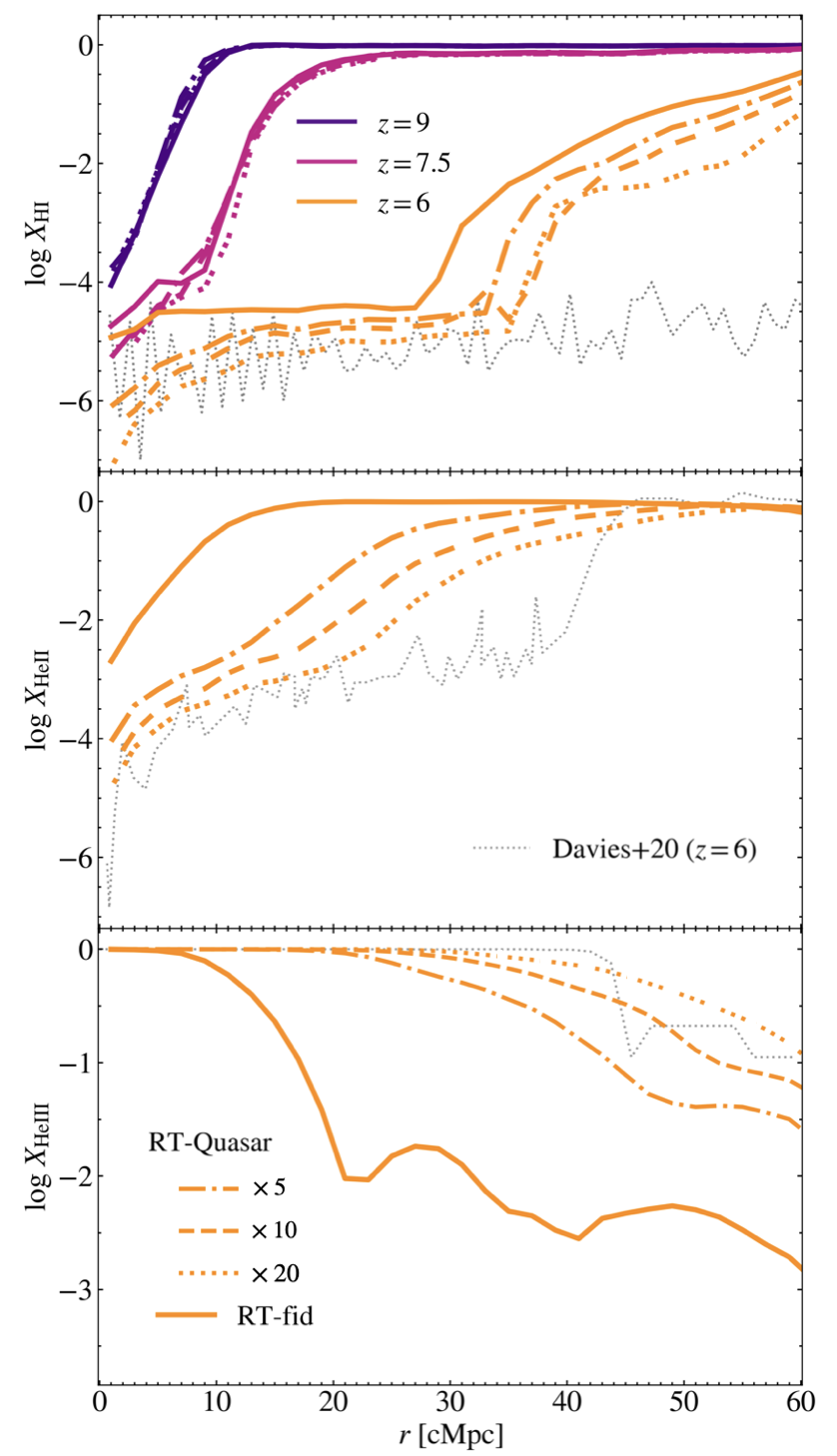}
    \caption{\textbf{Hydrogen and Helium ionization profiles, for the fiducial and RT-Quasar models.} Similar to Fig.~\ref{fig:rad_prof}, we show three redshifts for the H ionization profile: $z = 9$ (indigo), $z = 7.5$ (violet), $z = 6$ (orange), for the fiducial model (solid lines), and RT-Quasar models (dashed lines). The second and third panels depict the He ionization, at $z = 6$, and for reference, we add the profiles reported in \protect\cite{Davies2020}, which explore a quasar with $M_{1450} = -27$ at $z = 6$. This figure shows that introducing a boosted quasar model creates a clear proximity effect in the $X_\mathrm{HI}$ profile, and has a strong contribution to the He ionization levels.}
    \label{fig:rad_prof-quasar}

\end{figure} 

As seen in Sec.~\ref{sect:ioniz}, the gas within the box follows a realistic reionization history, driven primarily by galaxies. Here, we explore the topology of reionization within our box, as well as the effects of AGN and stars, on small scales. Figure~\ref{fig:maps} illustrates the distribution of neutral Hydrogen and gas temperature, within a slice of dimensions $120 \times 60 \times 2 \, \rm{cMpc^3}$, centered around the protocluster in the center of our box. We show both $z = 7.5$ (IGM mostly neutral) and $z = 6$ (ionized IGM), and zoom into a region spanning 25 cMpc on a side around the center of the box, showing the individual contributions from stars alone and AGN alone. 

Firstly, it can be seen that reionization is patchy, with the overdensities ionizing first due to the higher abundance of sources (see also Fig.~\ref{fig:env_proj}). Self shielding is also taken into account, causing the high density gas within filaments to ionize more slowly than the surrounding low density regions. Figure~\ref{fig:maps} also shows that the ionization of the gas is dominated by stars, even in the proximity of the central AGN. The central AGN alone is able to ionize the low density gas in its surroundings, but given its luminosity, its overall ionizing contribution is much lower than that of the stars (in agreement with Fig.~\ref{fig:xHIV}).  

The gas temperature shows a similar trend to $X_\mathrm{HI}$, with the gas in overdensities heated up first by the abundant sources present in these environments.  As in the case of ionization, the galaxies contribute the most to the gas heating, on both small and large scales. Lastly, as expected, the ionized regions seen in the left column of Fig.~\ref{fig:maps} also correspond to high temperatures. We explore these findings further in Fig.~\ref{fig:rad_prof}, where we show the spherically averaged profiles for $X_\mathrm{HI}$ and $T$, measured from the central AGN. 

Overall, both the gas' ionization and heating happen ``inside-out'', starting from the central region, that hosts the most massive galaxy and AGN. The AGN effects at $z = 9$ and $z = 7.5$ are negligible, but become apparent at $z =6$, for $r \lesssim 10 \, \mathrm{cMpc}$. In this regime, the central AGN is powerful enough to create an ionized bubble on its own, even in the absence of photon contributions from stars. The effect is also seen in the RT-fid vs RT-Stars profiles, with the RT-fid run producing a higher ionization fraction. Nonetheless, the trend reverses at $r \approx 30 \,\mathrm{cMpc}$, which corresponds to a low density region in the box. Namely, the stars alone ionize the gas more than stars+AGN in this region. A closer examination reveals that the RT-fid run contains slightly more low-density, cold gas ($n_\mathrm{H} \sim 10^{-6} \, \mathrm{cm}^{-3}$ and $T\sim10^2\,\mathrm{K}$). We attribute this difference to stochastic variations, as the region lies within a void where the influence of ionizing sources can fluctuate slightly between simulations. Nevertheless, this region has a negligible impact on the volume averaged $X_\mathrm{HI}$ across the entire box (see Fig.~\ref{fig:xHIV}).


\subsection{Quasar boosted model}
\label{sect:quasar_model}
As discussed in Sec.~\ref{sect:TNG} and Sec.~\ref{sect:BH_pop}, our models (Table~\ref{tab:prop}) do not reproduce the observed, rare, luminous quasars ($L_\mathrm{bol} \gtrsim 10^{47}\,\mathrm{erg~s^{-1}}$), due to box size limitations. As such, in order to capture this missing population, we introduce a quasar boosted model, in which the AGN luminosity is artificially boosted and the obscuration effects described in Sec.~\ref{sect:TNG} are removed. We explore three such artificial luminosity boosts: x5, corresponding to a maximum luminosity $L_\mathrm{bol} \approx 5 \times 10^{46}\,\mathrm{erg~s^{-1}}$, at $z =6$; x10 ($L_\mathrm{bol} \approx 10^{47}\,\mathrm{erg~s^{-1}}$ @ $z =6$) and x20 ($L_\mathrm{bol} \approx 2 \times10^{47}\,\mathrm{erg~s^{-1}}$ @ $z =6$). We keep the radiation from stars the same as in the fiducial run, RT-fid (see Table~\ref{tab:prop}). 

Figure~\ref{fig:rad_prof-quasar} shows how the quasar boosted models impact the gas ionization states, compared to our fiducial model. The top panel indicates the neutral Hydrogen fraction, as a function of distance from the central AGN, at three redshifts: $z = 9$, $z = 7.5$ and $z = 6$ (same as in Fig.~\ref{fig:rad_prof}). It can be clearly seen that the boosted quasars clearly outshine their host galaxies and produce a pronounced proximity effect, stronger for higher luminosity. Although the quasar’s impact is weaker at higher redshifts, it still produces an ``ionization surplus'' at $z = 7.5$. We will investigate this further in a subsequent paper (Bulichi et al., in prep), connecting our findings to observed proximity zones around similarly rare, luminous sources. For comparison, we also show the one-dimensional, post processing findings reported in \cite{Davies2020}, namely the radial $X_\mathrm{HI}$, $X_\mathrm{HeII}$ and $X_\mathrm{HeIII}$ profiles for a $M_{1450} = -27$ ($L_\mathrm{bol} \approx 2 \times 10^{47}\,\mathrm{erg~s^{-1}}$) quasar at $z = 6$. In the quasar's proximity ($r \lesssim 30\,\mathrm{cMpc}$), our results are in qualitative agreement with \cite{Davies2020}, with differences expected to arise due to the different gas properties (e.g., density, temperature distributions), as well as RT prescriptions (on-the-fly vs post-processing). Outside of the quasar's influence, the Hydrogen in our runs is significantly more neutral. We attribute this to the uniform UV background modeling in \cite{Davies2020}, as opposed to the self-consistent radiation modeling in our case that leads to a gradual, late reionization (see Sec.~\ref{sect:IGM} and  Sec.~\ref{sect:ioniz}).

The quasar boosted models also play an important role in the He reionization, as illustrated in the two bottom panels of Fig.~\ref{fig:rad_prof-quasar}. Given that He ionizes later, we only show $z = 6$ in the figure, and again provide a comparison with the $X_\mathrm{HeII}$ and $X_\mathrm{HeIII}$ profiles reported in \cite{Davies2020}, for the same object: a $M_{1450} = -27$ ($L_\mathrm{bol} \approx 2 \times 10^{47}\,\mathrm{erg~s^{-1}}$) quasar at $z = 6$. Whilst we find good agreement between our RT-Quasar x20 boosted model and the results of \cite{Davies2020}, we note that the He-ionized region in \cite{Davies2020} extends farther and exhibits a sharper transition between ionized and neutral. This can also be a result of the different RT prescriptions, AGN and gas properties, as well as the UVB model in \cite{Davies2020} that does not ionize the He outside of quasar's proximity zone. Additionally, less massive AGN in the outskirts of our simulations can also contribute to this finding, resulting in a smoother He ionization profile.

In relation to our RT-fiducial model, the effects of the quasar boosted models on He ionized fractions are evident. This is because most of the HeIII ionizing photons (energies $> 54.4\,\mathrm{eV}$) are produced by AGN, thus highly sensitive to the AGN luminosity. Overall, helium near the quasar is more ionized, predominantly as HeIII, and transitions gradually to a neutral state beyond the quasar’s influence. This also plays a role in the $\rm{Ly\alpha}$ transmission around the quasar, via the so-called thermal proximity effect \citep{Bolton2010,Bolton2012, Meiksin2010,Khrykin2017}, which we will also explore in a follow-up paper (Bulichi et al., in prep).

\section{Summary}
\label{sect:summary}
In this work, we made use of the well tested IllustrisTNG galaxy formation model, coupled with on-the-fly radiative transfer via {\small AREPO-RT}, in order to study the high\tb{-}redshift AGN population in a protocluster selected from the MillenniumTNG simulation box. We focused on the co-evolution between AGN and their host galaxies, as well as the effects of radiation on the gas properties. We summarize our main findings below:

\begin{itemize}
    \item Under the IllustrisTNG galaxy formation model, black holes grow in the densest environments and remain below the local $M_\mathrm{BH}$--$M_\star$ relation at high redshift, indicating that galaxies build up their stellar mass earlier and more rapidly than their central black holes (Figs.~\ref{fig:env_proj}, \ref{fig:BHMF6}, and \ref{fig:MBH-Ms}). 
    \item In order to enhance black hole growth, we modified the radiative efficiency from the fiducial TNG value $\epsilon_\mathrm{rad} = 0.2$ to $\epsilon_\mathrm{rad} = 0.1$. This change is effective only at the high-mass end $M_\mathrm{BH} \gtrsim 10^7\,\mathrm{M_\odot}$ where accretion is efficient under the Bondi--Hoyle prescription ($\dot{M} \propto {M_\mathrm{BH}^2}$) at the resolution employed in this study. This results in a $\sim 0.5 \, \mathrm{dex}$ more massive central BH by the end of the simulation (Fig.~\ref{fig:centralBH_z}), while the growth of the host galaxy is mostly unaffected (Fig.~\ref{fig:centralBH_z}, Fig.~\ref{fig:MBH-Ms}). 
    \item For $\epsilon_\mathrm{rad} = 0.1$ accretion onto the black holes is more bursty than for $\epsilon_\mathrm{rad} = 0.2$, due to the stronger feedback, and it also quenches at $z \approx 5.7$ for the most massive, central black hole, due to the activation of AGN kinetic feedback (Fig.~\ref{fig:BHAD}). Changing the radiative efficiency also enhances the AGN luminosity in the regime where accretion is efficient ($M_\mathrm{BH} \gtrsim 10^7\,\mathrm{M_\odot}$), and before the onset of the AGN kinetic feedback mode (Fig.~\ref{fig:QLF}), resulting in good agreement with observations of ``little red dots'' \citep{Matthee2023}.
    \item We explored several different radiation models (Table~\ref{tab:prop}), showing good convergence in galaxy and black hole properties, with small differences due to feedback and stochastic effects (Fig.~\ref{fig:BHMF6}, Fig.~\ref{fig:MBH-Ms}, Fig.~\ref{fig:BHAD}).
    \item Overall, our simulated AGN are significantly less massive than \textit{JWST} observed \textit{AGN} (Fig.~\ref{fig:QLF}, Fig.~\ref{fig:MBH-Ms}), showing the difficulty in modeling these rare objects, even in very large-volume cosmological simulations. At a population level, we show the mass assembly history of both black holes and galaxies in Fig.~\ref{fig:BHAD}, at the high\tb{-}redshift regime $z > 5$, expected to be observed by \textit{JWST}. 
    
    \item We model the gas responses to the radiation on-the-fly, summarizing the differences between this approach and an uniform UV background in Fig.~\ref{fig:phase-space}: while an uniform UV background eliminates all cold, low-density gas, the self-consistent radiation modeling retains such gas in regions where the ionization front has yet to propagate. 
    \item Given the low AGN luminosities, the ionizing output comes primarily from stars, on both small and large scales, contributing the most to the reionization history (Fig.~\ref{fig:xHIV}), and the formation and sizes of ionized bubbles (Fig.~\ref{fig:maps}, Fig.~\ref{fig:rad_prof}).
    \item Lastly, since our model does not reproduce rare, luminous quasars, we introduce three quasar boosted models, where we artificially enhance the central AGN luminosity to match that of observed quasars, and eliminate the obscuration attenuation. Such objects outshine their host galaxies (Fig.~\ref{fig:rad_prof-quasar}), create a strong proximity effect, and have the strongest contribution to the He ionization, at late times ($z \lesssim 6$).
    
\end{itemize}

Overall, this work underscores the challenges of self-consistently reproducing the full range of observed black hole masses and luminosities, while also matching their redshift evolution down to the local Universe. After enhancing BH growth by altering the radiative efficiency, and introducing a new quasar boosted model, we successfully captured a wide range of AGN properties, and modeled the corresponding gas responses on-the-fly, finding realistic results on both small and large scales. The on-the-fly radiative transfer prescriptions mark a significant step forward, as they allow radiation to interact with and shape the gas dynamics and thermochemistry throughout the simulation, capturing feedback effects in real time--an advancement we will build upon in future work.


\section*{Acknowledgements}
\tb{We thank the anonymous referee for their constructive report, which has helped to strengthen this paper. We also thank Dominika \v{D}urov{\v{c}}{\'\i}kov{\'a}, Meredith Neyer, and Ruediger Pakmor for valuable discussions, as well as Volker Springel for insightful input and for granting us access to {\small AREPO}.}
An award of computer time was provided by the INCITE program. This research also used resources of the Oak Ridge Leadership Computing Facility, which is a DOE Office of Science User Facility supported under Contract DE-AC05-00OR22725. Support for OZ was provided by Harvard University through the Institute for Theory and Computation Fellowship. RK acknowledges support of the Natural Sciences and Engineering Research Council of Canada (NSERC) through a Discovery Grant and a Discovery Launch Supplement (funding reference numbers RGPIN-2024-06222 and DGECR-2024-00144) and York University's Global Research Excellence Initiative. XS acknowledges the support from the NASA theory grant JWST-AR-04814.

\section*{Data Availability}
The data underlying this paper will be shared upon reasonable request to the corresponding author.



\bibliographystyle{mnras}
\bibliography{references} 

\begin{thebibliography}{}
\makeatletter
\relax
\def\mn@urlcharsother{\let\do\@makeother \do\$\do\&\do\#\do\^\do\_\do\%\do\~}
\def\mn@doi{\begingroup\mn@urlcharsother \@ifnextchar [ {\mn@doi@} {\mn@doi@[]}}
\def\mn@doi@[#1]#2{\def\@tempa{#1}\ifx\@tempa\@empty \href {http://dx.doi.org/#2} {doi:#2}\else \href {http://dx.doi.org/#2} {#1}\fi \endgroup}
\def\mn@eprint#1#2{\mn@eprint@#1:#2::\@nil}
\def\mn@eprint@arXiv#1{\href {http://arxiv.org/abs/#1} {{\tt arXiv:#1}}}
\def\mn@eprint@dblp#1{\href {http://dblp.uni-trier.de/rec/bibtex/#1.xml} {dblp:#1}}
\def\mn@eprint@#1:#2:#3:#4\@nil{\def\@tempa {#1}\def\@tempb {#2}\def\@tempc {#3}\ifx \@tempc \@empty \let \@tempc \@tempb \let \@tempb \@tempa \fi \ifx \@tempb \@empty \def\@tempb {arXiv}\fi \@ifundefined {mn@eprint@\@tempb}{\@tempb:\@tempc}{\expandafter \expandafter \csname mn@eprint@\@tempb\endcsname \expandafter{\@tempc}}}

\bibitem[\protect\citeauthoryear{{Aarseth}}{{Aarseth}}{2003}]{Aarseth2003}
{Aarseth} S.~J.,  2003, {Gravitational N-Body Simulations}.
{Cambridge University Press}

\bibitem[\protect\citeauthoryear{{Aird}, {Coil}, {Georgakakis}, {Nandra}, {Barro}  \& {P{\'e}rez-Gonz{\'a}lez}}{{Aird} et~al.}{2015}]{Aird2015}
{Aird} J.,  {Coil} A.~L.,  {Georgakakis} A.,  {Nandra} K.,  {Barro} G.,   {P{\'e}rez-Gonz{\'a}lez} P.~G.,  2015, \mn@doi [\mnras] {10.1093/mnras/stv1062}, \href {https://ui.adsabs.harvard.edu/abs/2015MNRAS.451.1892A} {451, 1892}

\bibitem[\protect\citeauthoryear{{Ananna} et~al.,}{{Ananna} et~al.}{2019}]{Ananna2019}
{Ananna} T.~T.,  et~al., 2019, \mn@doi [\apj] {10.3847/1538-4357/aafb77}, \href {https://ui.adsabs.harvard.edu/abs/2019ApJ...871..240A} {871, 240}

\bibitem[\protect\citeauthoryear{{Asthana}, {Haehnelt}, {Kulkarni}, {Aubert}, {Bolton}  \& {Keating}}{{Asthana} et~al.}{2024}]{Asthana2024}
{Asthana} S.,  {Haehnelt} M.~G.,  {Kulkarni} G.,  {Aubert} D.,  {Bolton} J.~S.,   {Keating} L.~C.,  2024, \mn@doi [\mnras] {10.1093/mnras/stae1945}, \href {https://ui.adsabs.harvard.edu/abs/2024MNRAS.533.2843A} {533, 2843}

\bibitem[\protect\citeauthoryear{{Ba{\~n}ados} et~al.,}{{Ba{\~n}ados} et~al.}{2018}]{Banados2018}
{Ba{\~n}ados} E.,  et~al., 2018, \mn@doi [\nat] {10.1038/nature25180}, \href {https://ui.adsabs.harvard.edu/abs/2018Natur.553..473B} {553, 473}

\bibitem[\protect\citeauthoryear{{Ba{\~n}ados} et~al.,}{{Ba{\~n}ados} et~al.}{2023}]{Banados2023}
{Ba{\~n}ados} E.,  et~al., 2023, \mn@doi [\apjs] {10.3847/1538-4365/acb3c7}, \href {https://ui.adsabs.harvard.edu/abs/2023ApJS..265...29B} {265, 29}

\bibitem[\protect\citeauthoryear{{Bagla}}{{Bagla}}{2002}]{Bagla2022}
{Bagla} J.~S.,  2002, \mn@doi [Journal of Astrophysics and Astronomy] {10.1007/BF02702282}, \href {https://ui.adsabs.harvard.edu/abs/2002JApA...23..185B} {23, 185}

\bibitem[\protect\citeauthoryear{{Barnes} \& {Hut}}{{Barnes} \& {Hut}}{1986}]{Barnes1986}
{Barnes} J.,  {Hut} P.,  1986, \mn@doi [\nat] {10.1038/324446a0}, \href {https://ui.adsabs.harvard.edu/abs/1986Natur.324..446B} {324, 446}

\bibitem[\protect\citeauthoryear{{Beckmann}, {Slyz}  \& {Devriendt}}{{Beckmann} et~al.}{2018}]{Beckmann2018}
{Beckmann} R.~S.,  {Slyz} A.,   {Devriendt} J.,  2018, \mn@doi [\mnras] {10.1093/mnras/sty931}, \href {https://ui.adsabs.harvard.edu/abs/2018MNRAS.478..995B} {478, 995}

\bibitem[\protect\citeauthoryear{{Bennett}, {Sijacki}, {Costa}, {Laporte}  \& {Witten}}{{Bennett} et~al.}{2024}]{Bennett2024}
{Bennett} J.~S.,  {Sijacki} D.,  {Costa} T.,  {Laporte} N.,   {Witten} C.,  2024, \mn@doi [\mnras] {10.1093/mnras/stad3179}, \href {https://ui.adsabs.harvard.edu/abs/2024MNRAS.527.1033B} {527, 1033}

\bibitem[\protect\citeauthoryear{{Bhowmick}, {Blecha}  \& {Thomas}}{{Bhowmick} et~al.}{2020}]{Bhowmick2020}
{Bhowmick} A.~K.,  {Blecha} L.,   {Thomas} J.,  2020, \mn@doi [\apj] {10.3847/1538-4357/abc1e6}, \href {https://ui.adsabs.harvard.edu/abs/2020ApJ...904..150B} {904, 150}

\bibitem[\protect\citeauthoryear{{Bhowmick} et~al.,}{{Bhowmick} et~al.}{2024}]{Bhowmick2024}
{Bhowmick} A.~K.,  et~al., 2024, \mn@doi [\mnras] {10.1093/mnras/stae1819}, \href {https://ui.adsabs.harvard.edu/abs/2024MNRAS.533.1907B} {533, 1907}

\bibitem[\protect\citeauthoryear{{Bolton}, {Becker}, {Wyithe}, {Haehnelt}  \& {Sargent}}{{Bolton} et~al.}{2010}]{Bolton2010}
{Bolton} J.~S.,  {Becker} G.~D.,  {Wyithe} J. S.~B.,  {Haehnelt} M.~G.,   {Sargent} W. L.~W.,  2010, \mn@doi [\mnras] {10.1111/j.1365-2966.2010.16701.x}, \href {https://ui.adsabs.harvard.edu/abs/2010MNRAS.406..612B} {406, 612}

\bibitem[\protect\citeauthoryear{{Bolton}, {Becker}, {Raskutti}, {Wyithe}, {Haehnelt}  \& {Sargent}}{{Bolton} et~al.}{2012}]{Bolton2012}
{Bolton} J.~S.,  {Becker} G.~D.,  {Raskutti} S.,  {Wyithe} J. S.~B.,  {Haehnelt} M.~G.,   {Sargent} W. L.~W.,  2012, \mn@doi [\mnras] {10.1111/j.1365-2966.2011.19929.x}, \href {https://ui.adsabs.harvard.edu/abs/2012MNRAS.419.2880B} {419, 2880}

\bibitem[\protect\citeauthoryear{{Bondi}}{{Bondi}}{1952}]{Bondi1952}
{Bondi} H.,  1952, \mn@doi [\mnras] {10.1093/mnras/112.2.195}, \href {https://ui.adsabs.harvard.edu/abs/1952MNRAS.112..195B} {112, 195}

\bibitem[\protect\citeauthoryear{{Bondi} \& {Hoyle}}{{Bondi} \& {Hoyle}}{1944}]{Bondi1944}
{Bondi} H.,  {Hoyle} F.,  1944, \mn@doi [\mnras] {10.1093/mnras/104.5.273}, \href {https://ui.adsabs.harvard.edu/abs/1944MNRAS.104..273B} {104, 273}

\bibitem[\protect\citeauthoryear{{Bosman}, {{\v{D}}urov{\v{c}}{\'\i}kov{\'a}}, {Davies}  \& {Eilers}}{{Bosman} et~al.}{2021}]{Bosman2021}
{Bosman} S. E.~I.,  {{\v{D}}urov{\v{c}}{\'\i}kov{\'a}} D.,  {Davies} F.~B.,   {Eilers} A.-C.,  2021, \mn@doi [\mnras] {10.1093/mnras/stab572}, \href {https://ui.adsabs.harvard.edu/abs/2021MNRAS.503.2077B} {503, 2077}

\bibitem[\protect\citeauthoryear{{Bower}, {Schaye}, {Frenk}, {Theuns}, {Schaller}, {Crain}  \& {McAlpine}}{{Bower} et~al.}{2017}]{Bower2017}
{Bower} R.~G.,  {Schaye} J.,  {Frenk} C.~S.,  {Theuns} T.,  {Schaller} M.,  {Crain} R.~A.,   {McAlpine} S.,  2017, \mn@doi [\mnras] {10.1093/mnras/stw2735}, \href {https://ui.adsabs.harvard.edu/abs/2017MNRAS.465...32B} {465, 32}

\bibitem[\protect\citeauthoryear{{Burger} et~al.,}{{Burger} et~al.}{2025}]{Burger2025}
{Burger} J.~D.,  et~al., 2025, \mn@doi [arXiv e-prints] {10.48550/arXiv.2502.13244}, \href {https://ui.adsabs.harvard.edu/abs/2025arXiv250213244B} {p. arXiv:2502.13244}

\bibitem[\protect\citeauthoryear{{Cen}}{{Cen}}{1992}]{Cen1992}
{Cen} R.,  1992, \mn@doi [\apjs] {10.1086/191630}, \href {https://ui.adsabs.harvard.edu/abs/1992ApJS...78..341C} {78, 341}

\bibitem[\protect\citeauthoryear{{Chabrier}}{{Chabrier}}{2003}]{Chabrier2003}
{Chabrier} G.,  2003, \mn@doi [\pasp] {10.1086/376392}, \href {https://ui.adsabs.harvard.edu/abs/2003PASP..115..763C} {115, 763}

\bibitem[\protect\citeauthoryear{{Chen} \& {Gnedin}}{{Chen} \& {Gnedin}}{2021}]{ChenGnedin2021}
{Chen} H.,  {Gnedin} N.~Y.,  2021, \mn@doi [\apj] {10.3847/1538-4357/abe7e7}, \href {https://ui.adsabs.harvard.edu/abs/2021ApJ...911...60C} {911, 60}

\bibitem[\protect\citeauthoryear{{Chittenden}, {Glazebrook}, {Nanayakkara}, {Kawinwanichakij}, {Lagos}, {Kimmig}  \& {Remus}}{{Chittenden} et~al.}{2025}]{Chittenden2025}
{Chittenden} H.~G.,  {Glazebrook} K.,  {Nanayakkara} T.,  {Kawinwanichakij} L.,  {Lagos} C.,  {Kimmig} L.,   {Remus} R.-S.,  2025, \mn@doi [arXiv e-prints] {10.48550/arXiv.2504.19696}, \href {https://ui.adsabs.harvard.edu/abs/2025arXiv250419696C} {p. arXiv:2504.19696}

\bibitem[\protect\citeauthoryear{{Cicone} et~al.,}{{Cicone} et~al.}{2014}]{Cicone2014}
{Cicone} C.,  et~al., 2014, \mn@doi [\aap] {10.1051/0004-6361/201322464}, \href {https://ui.adsabs.harvard.edu/abs/2014A&A...562A..21C} {562, A21}

\bibitem[\protect\citeauthoryear{{Costa}, {Sijacki}, {Trenti}  \& {Haehnelt}}{{Costa} et~al.}{2014}]{Costa2014}
{Costa} T.,  {Sijacki} D.,  {Trenti} M.,   {Haehnelt} M.~G.,  2014, \mn@doi [\mnras] {10.1093/mnras/stu101}, \href {https://ui.adsabs.harvard.edu/abs/2014MNRAS.439.2146C} {439, 2146}

\bibitem[\protect\citeauthoryear{{Dav{\'e}} et~al.,}{{Dav{\'e}} et~al.}{2001}]{Dave2001}
{Dav{\'e}} R.,  et~al., 2001, \mn@doi [\apj] {10.1086/320548}, \href {https://ui.adsabs.harvard.edu/abs/2001ApJ...552..473D} {552, 473}

\bibitem[\protect\citeauthoryear{{Davies} et~al.,}{{Davies} et~al.}{2018}]{Davies2018}
{Davies} F.~B.,  et~al., 2018, \mn@doi [\apj] {10.3847/1538-4357/aad6dc}, \href {https://ui.adsabs.harvard.edu/abs/2018ApJ...864..142D} {864, 142}

\bibitem[\protect\citeauthoryear{{Davies}, {Hennawi}  \& {Eilers}}{{Davies} et~al.}{2020}]{Davies2020}
{Davies} F.~B.,  {Hennawi} J.~F.,   {Eilers} A.-C.,  2020, \mn@doi [\mnras] {10.1093/mnras/stz3303}, \href {https://ui.adsabs.harvard.edu/abs/2020MNRAS.493.1330D} {493, 1330}

\bibitem[\protect\citeauthoryear{{Davis}, {Efstathiou}, {Frenk}  \& {White}}{{Davis} et~al.}{1985}]{Davis1985}
{Davis} M.,  {Efstathiou} G.,  {Frenk} C.~S.,   {White} S.~D.~M.,  1985, \mn@doi [\apj] {10.1086/163168}, \href {https://ui.adsabs.harvard.edu/abs/1985ApJ...292..371D} {292, 371}

\bibitem[\protect\citeauthoryear{{Dayal} et~al.,}{{Dayal} et~al.}{2025}]{Dayal2025}
{Dayal} P.,  et~al., 2025, \mn@doi [\aap] {10.1051/0004-6361/202449331}, \href {https://ui.adsabs.harvard.edu/abs/2025A&A...697A.211D} {697, A211}

\bibitem[\protect\citeauthoryear{{Ding} et~al.,}{{Ding} et~al.}{2023}]{Ding2023}
{Ding} X.,  et~al., 2023, \mn@doi [\nat] {10.1038/s41586-023-06345-5}, \href {https://ui.adsabs.harvard.edu/abs/2023Natur.621...51D} {621, 51}

\bibitem[\protect\citeauthoryear{{Dubroca} \& {Feugeas}}{{Dubroca} \& {Feugeas}}{1999}]{DubrocaFeugeas1999}
{Dubroca} B.,  {Feugeas} J.,  1999, \mn@doi [Academie des Sciences Paris Comptes Rendus Serie Sciences Mathematiques] {10.1016/S0764-4442(00)87499-6}, \href {https://ui.adsabs.harvard.edu/abs/1999CRASM.329..915D} {329, 915}

\bibitem[\protect\citeauthoryear{{Eide}, {Ciardi}, {Graziani}, {Busch}, {Feng}  \& {Di Matteo}}{{Eide} et~al.}{2020}]{Eide2020}
{Eide} M.~B.,  {Ciardi} B.,  {Graziani} L.,  {Busch} P.,  {Feng} Y.,   {Di Matteo} T.,  2020, \mn@doi [\mnras] {10.1093/mnras/staa2774}, \href {https://ui.adsabs.harvard.edu/abs/2020MNRAS.498.6083E} {498, 6083}

\bibitem[\protect\citeauthoryear{{Eilers} et~al.,}{{Eilers} et~al.}{2024}]{Eilers2024}
{Eilers} A.-C.,  et~al., 2024, \mn@doi [\apj] {10.3847/1538-4357/ad778b}, \href {https://ui.adsabs.harvard.edu/abs/2024ApJ...974..275E} {974, 275}

\bibitem[\protect\citeauthoryear{{Eldridge}, {Stanway}, {Xiao}, {McClelland}, {Taylor}, {Ng}, {Greis}  \& {Bray}}{{Eldridge} et~al.}{2017}]{Eldridge2017}
{Eldridge} J.~J.,  {Stanway} E.~R.,  {Xiao} L.,  {McClelland} L.~A.~S.,  {Taylor} G.,  {Ng} M.,  {Greis} S.~M.~L.,   {Bray} J.~C.,  2017, \mn@doi [\pasa] {10.1017/pasa.2017.51}, \href {https://ui.adsabs.harvard.edu/abs/2017PASA...34...58E} {34, e058}

\bibitem[\protect\citeauthoryear{{Fabian}}{{Fabian}}{2012}]{Fabian2012}
{Fabian} A.~C.,  2012, \mn@doi [\araa] {10.1146/annurev-astro-081811-125521}, \href {https://ui.adsabs.harvard.edu/abs/2012ARA&A..50..455F} {50, 455}

\bibitem[\protect\citeauthoryear{{Fan} et~al.,}{{Fan} et~al.}{2006}]{Fan2006}
{Fan} X.,  et~al., 2006, \mn@doi [\aj] {10.1086/504836}, \href {https://ui.adsabs.harvard.edu/abs/2006AJ....132..117F} {132, 117}

\bibitem[\protect\citeauthoryear{{Fan}, {Ba{\~n}ados}  \& {Simcoe}}{{Fan} et~al.}{2023}]{FanBanadosSimcoe2023}
{Fan} X.,  {Ba{\~n}ados} E.,   {Simcoe} R.~A.,  2023, \mn@doi [\araa] {10.1146/annurev-astro-052920-102455}, \href {https://ui.adsabs.harvard.edu/abs/2023ARA&A..61..373F} {61, 373}

\bibitem[\protect\citeauthoryear{{Farcy} et~al.,}{{Farcy} et~al.}{2025}]{Farcy2025}
{Farcy} M.,  et~al., 2025, \mn@doi [\mnras] {10.1093/mnras/staf1464}, \href {https://ui.adsabs.harvard.edu/abs/2025MNRAS.543..967F} {543, 967}

\bibitem[\protect\citeauthoryear{{Faucher-Gigu{\`e}re}, {Lidz}, {Zaldarriaga}  \& {Hernquist}}{{Faucher-Gigu{\`e}re} et~al.}{2009}]{Faucher2009}
{Faucher-Gigu{\`e}re} C.-A.,  {Lidz} A.,  {Zaldarriaga} M.,   {Hernquist} L.,  2009, \mn@doi [\apj] {10.1088/0004-637X/703/2/1416}, \href {https://ui.adsabs.harvard.edu/abs/2009ApJ...703.1416F} {703, 1416}

\bibitem[\protect\citeauthoryear{{Furtak} et~al.,}{{Furtak} et~al.}{2023}]{Furtak2023}
{Furtak} L.~J.,  et~al., 2023, \mn@doi [\apj] {10.3847/1538-4357/acdc9d}, \href {https://ui.adsabs.harvard.edu/abs/2023ApJ...952..142F} {952, 142}

\bibitem[\protect\citeauthoryear{{Garaldi}, {Kannan}, {Smith}, {Springel}, {Pakmor}, {Vogelsberger}  \& {Hernquist}}{{Garaldi} et~al.}{2022}]{Garaldi2022}
{Garaldi} E.,  {Kannan} R.,  {Smith} A.,  {Springel} V.,  {Pakmor} R.,  {Vogelsberger} M.,   {Hernquist} L.,  2022, \mn@doi [\mnras] {10.1093/mnras/stac257}, \href {https://ui.adsabs.harvard.edu/abs/2022MNRAS.512.4909G} {512, 4909}

\bibitem[\protect\citeauthoryear{{Garaldi} et~al.,}{{Garaldi} et~al.}{2024}]{Garaldi2023}
{Garaldi} E.,  et~al., 2024, \mn@doi [\mnras] {10.1093/mnras/stae839}, \href {https://ui.adsabs.harvard.edu/abs/2024MNRAS.530.3765G} {530, 3765}

\bibitem[\protect\citeauthoryear{{Garc{\'\i}a-Vergara} et~al.,}{{Garc{\'\i}a-Vergara} et~al.}{2022}]{GarciaVergara2022}
{Garc{\'\i}a-Vergara} C.,  et~al., 2022, \mn@doi [\apj] {10.3847/1538-4357/ac469d}, \href {https://ui.adsabs.harvard.edu/abs/2022ApJ...927...65G} {927, 65}

\bibitem[\protect\citeauthoryear{{Gaspari}, {Ruszkowski}  \& {Oh}}{{Gaspari} et~al.}{2013}]{Gaspari2013}
{Gaspari} M.,  {Ruszkowski} M.,   {Oh} S.~P.,  2013, \mn@doi [\mnras] {10.1093/mnras/stt692}, \href {https://ui.adsabs.harvard.edu/abs/2013MNRAS.432.3401G} {432, 3401}

\bibitem[\protect\citeauthoryear{{Geris} et~al.,}{{Geris} et~al.}{2025}]{Geris2025}
{Geris} S.,  et~al., 2025, \mn@doi [arXiv e-prints] {10.48550/arXiv.2506.22147}, \href {https://ui.adsabs.harvard.edu/abs/2025arXiv250622147G} {p. arXiv:2506.22147}

\bibitem[\protect\citeauthoryear{{Habouzit}}{{Habouzit}}{2025}]{Habouzit2024}
{Habouzit} M.,  2025, \mn@doi [\mnras] {10.1093/mnras/staf167}, \href {https://ui.adsabs.harvard.edu/abs/2025MNRAS.537.2323H} {537, 2323}

\bibitem[\protect\citeauthoryear{{Habouzit}, {Pisani}, {Goulding}, {Dubois}, {Somerville}  \& {Greene}}{{Habouzit} et~al.}{2020}]{Habouzit2020}
{Habouzit} M.,  {Pisani} A.,  {Goulding} A.,  {Dubois} Y.,  {Somerville} R.~S.,   {Greene} J.~E.,  2020, \mn@doi [\mnras] {10.1093/mnras/staa219}, \href {https://ui.adsabs.harvard.edu/abs/2020MNRAS.493..899H} {493, 899}

\bibitem[\protect\citeauthoryear{{Habouzit} et~al.,}{{Habouzit} et~al.}{2022}]{Habouzit2022}
{Habouzit} M.,  et~al., 2022, \mn@doi [\mnras] {10.1093/mnras/stab3147}, \href {https://ui.adsabs.harvard.edu/abs/2022MNRAS.509.3015H} {509, 3015}

\bibitem[\protect\citeauthoryear{{Harikane} et~al.,}{{Harikane} et~al.}{2023}]{Harikane2023}
{Harikane} Y.,  et~al., 2023, \mn@doi [\apj] {10.3847/1538-4357/ad029e}, \href {https://ui.adsabs.harvard.edu/abs/2023ApJ...959...39H} {959, 39}

\bibitem[\protect\citeauthoryear{{Harvey} et~al.,}{{Harvey} et~al.}{2025}]{Harvey2024}
{Harvey} T.,  et~al., 2025, \mn@doi [\apj] {10.3847/1538-4357/ad8c29}, \href {https://ui.adsabs.harvard.edu/abs/2025ApJ...978...89H} {978, 89}

\bibitem[\protect\citeauthoryear{{Hennawi} et~al.,}{{Hennawi} et~al.}{2006}]{Hennawi2006}
{Hennawi} J.~F.,  et~al., 2006, \mn@doi [\aj] {10.1086/498235}, \href {https://ui.adsabs.harvard.edu/abs/2006AJ....131....1H} {131, 1}

\bibitem[\protect\citeauthoryear{{Hopkins} \& {Quataert}}{{Hopkins} \& {Quataert}}{2011}]{Hopkins2011}
{Hopkins} P.~F.,  {Quataert} E.,  2011, \mn@doi [\mnras] {10.1111/j.1365-2966.2011.18542.x}, \href {https://ui.adsabs.harvard.edu/abs/2011MNRAS.415.1027H} {415, 1027}

\bibitem[\protect\citeauthoryear{{Hopkins}, {Hernquist}, {Cox}  \& {Kere{\v{s}}}}{{Hopkins} et~al.}{2008}]{Hopkins2008}
{Hopkins} P.~F.,  {Hernquist} L.,  {Cox} T.~J.,   {Kere{\v{s}}} D.,  2008, \mn@doi [\apjs] {10.1086/524362}, \href {https://ui.adsabs.harvard.edu/abs/2008ApJS..175..356H} {175, 356}

\bibitem[\protect\citeauthoryear{{Hopkins} et~al.,}{{Hopkins} et~al.}{2024}]{Hopkins2024}
{Hopkins} P.~F.,  et~al., 2024, \mn@doi [The Open Journal of Astrophysics] {10.21105/astro.2310.04507}, \href {https://ui.adsabs.harvard.edu/abs/2024OJAp....7E..20H} {7, 20}

\bibitem[\protect\citeauthoryear{{Hu{\v{s}}ko}, {Lacey}, {Roper}, {Schaye}, {Briggs}  \& {Schaller}}{{Hu{\v{s}}ko} et~al.}{2025}]{Husko2024}
{Hu{\v{s}}ko} F.,  {Lacey} C.~G.,  {Roper} W.~J.,  {Schaye} J.,  {Briggs} J.~M.,   {Schaller} M.,  2025, \mn@doi [\mnras] {10.1093/mnras/staf146}, \href {https://ui.adsabs.harvard.edu/abs/2025MNRAS.537.2559H} {537, 2559}

\bibitem[\protect\citeauthoryear{{Inayoshi}, {Visbal}  \& {Haiman}}{{Inayoshi} et~al.}{2020}]{Inayoshi2020}
{Inayoshi} K.,  {Visbal} E.,   {Haiman} Z.,  2020, \mn@doi [\araa] {10.1146/annurev-astro-120419-014455}, \href {https://ui.adsabs.harvard.edu/abs/2020ARA&A..58...27I} {58, 27}

\bibitem[\protect\citeauthoryear{{Jamieson} et~al.,}{{Jamieson} et~al.}{2025}]{Jamieson2024}
{Jamieson} N.,  et~al., 2025, \mn@doi [\mnras] {10.1093/mnras/staf996}, \href {https://ui.adsabs.harvard.edu/abs/2025MNRAS.541.1088J} {541, 1088}

\bibitem[\protect\citeauthoryear{{Jiang} et~al.,}{{Jiang} et~al.}{2016}]{Jiang2016}
{Jiang} L.,  et~al., 2016, \mn@doi [\apj] {10.3847/1538-4357/833/2/222}, \href {https://ui.adsabs.harvard.edu/abs/2016ApJ...833..222J} {833, 222}

\bibitem[\protect\citeauthoryear{{Jiang}, {Jiang}, {Sun}, {Liu}  \& {Fu}}{{Jiang} et~al.}{2025}]{Jiang2025}
{Jiang} D.,  {Jiang} L.,  {Sun} S.,  {Liu} W.,   {Fu} S.,  2025, \mn@doi [arXiv e-prints] {10.48550/arXiv.2502.03683}, \href {https://ui.adsabs.harvard.edu/abs/2025arXiv250203683J} {p. arXiv:2502.03683}

\bibitem[\protect\citeauthoryear{{Jin} et~al.,}{{Jin} et~al.}{2023}]{Jin2023}
{Jin} X.,  et~al., 2023, \mn@doi [\apj] {10.3847/1538-4357/aca678}, \href {https://ui.adsabs.harvard.edu/abs/2023ApJ...942...59J} {942, 59}

\bibitem[\protect\citeauthoryear{{Juod{\v{z}}balis} et~al.,}{{Juod{\v{z}}balis} et~al.}{2025}]{Juodvzbalis2025}
{Juod{\v{z}}balis} I.,  et~al., 2025, \mn@doi [arXiv e-prints] {10.48550/arXiv.2504.03551}, \href {https://ui.adsabs.harvard.edu/abs/2025arXiv250403551J} {p. arXiv:2504.03551}

\bibitem[\protect\citeauthoryear{{Kannan}, {Vogelsberger}, {Marinacci}, {McKinnon}, {Pakmor}  \& {Springel}}{{Kannan} et~al.}{2019}]{Kannan2019}
{Kannan} R.,  {Vogelsberger} M.,  {Marinacci} F.,  {McKinnon} R.,  {Pakmor} R.,   {Springel} V.,  2019, \mn@doi [\mnras] {10.1093/mnras/stz287}, \href {https://ui.adsabs.harvard.edu/abs/2019MNRAS.485..117K} {485, 117}

\bibitem[\protect\citeauthoryear{{Kannan}, {Garaldi}, {Smith}, {Pakmor}, {Springel}, {Vogelsberger}  \& {Hernquist}}{{Kannan} et~al.}{2022}]{Kannan2022}
{Kannan} R.,  {Garaldi} E.,  {Smith} A.,  {Pakmor} R.,  {Springel} V.,  {Vogelsberger} M.,   {Hernquist} L.,  2022, \mn@doi [\mnras] {10.1093/mnras/stab3710}, \href {https://ui.adsabs.harvard.edu/abs/2022MNRAS.511.4005K} {511, 4005}

\bibitem[\protect\citeauthoryear{{Kannan} et~al.,}{{Kannan} et~al.}{2025}]{Kannan2025}
{Kannan} R.,  et~al., 2025, \mn@doi [arXiv e-prints] {10.48550/arXiv.2502.20437}, \href {https://ui.adsabs.harvard.edu/abs/2025arXiv250220437K} {p. arXiv:2502.20437}

\bibitem[\protect\citeauthoryear{{Katz}, {Weinberg}  \& {Hernquist}}{{Katz} et~al.}{1996}]{Katz1996}
{Katz} N.,  {Weinberg} D.~H.,   {Hernquist} L.,  1996, \mn@doi [\apjs] {10.1086/192305}, \href {https://ui.adsabs.harvard.edu/abs/1996ApJS..105...19K} {105, 19}

\bibitem[\protect\citeauthoryear{{Kho}, {Bhowmick}, {Torrey}, {Garcia}, {Ahvazi}, {Blecha}  \& {Vogelsberger}}{{Kho} et~al.}{2025}]{Kho2025}
{Kho} J.,  {Bhowmick} A.~K.,  {Torrey} P.,  {Garcia} A.~M.,  {Ahvazi} N.,  {Blecha} L.,   {Vogelsberger} M.,  2025, \mn@doi [arXiv e-prints] {10.48550/arXiv.2506.17476}, \href {https://ui.adsabs.harvard.edu/abs/2025arXiv250617476K} {p. arXiv:2506.17476}

\bibitem[\protect\citeauthoryear{{Khrykin}, {Hennawi}  \& {McQuinn}}{{Khrykin} et~al.}{2017}]{Khrykin2017}
{Khrykin} I.~S.,  {Hennawi} J.~F.,   {McQuinn} M.,  2017, \mn@doi [\apj] {10.3847/1538-4357/aa6621}, \href {https://ui.adsabs.harvard.edu/abs/2017ApJ...838...96K} {838, 96}

\bibitem[\protect\citeauthoryear{{King} \& {Pounds}}{{King} \& {Pounds}}{2015}]{KingPounds2015}
{King} A.,  {Pounds} K.,  2015, \mn@doi [\araa] {10.1146/annurev-astro-082214-122316}, \href {https://ui.adsabs.harvard.edu/abs/2015ARA&A..53..115K} {53, 115}

\bibitem[\protect\citeauthoryear{{Kocevski} et~al.,}{{Kocevski} et~al.}{2023}]{Kocevski2023}
{Kocevski} D.~D.,  et~al., 2023, \mn@doi [\apjl] {10.3847/2041-8213/acad00}, \href {https://ui.adsabs.harvard.edu/abs/2023ApJ...946L..14K} {946, L14}

\bibitem[\protect\citeauthoryear{{Koehler}, {Jiao}  \& {Kannan}}{{Koehler} et~al.}{2024}]{Koehler2025}
{Koehler} S.~M.,  {Jiao} H.,   {Kannan} R.,  2024, \mn@doi [arXiv e-prints] {10.48550/arXiv.2412.00182}, \href {https://ui.adsabs.harvard.edu/abs/2024arXiv241200182K} {p. arXiv:2412.00182}

\bibitem[\protect\citeauthoryear{{Kokorev} et~al.,}{{Kokorev} et~al.}{2023}]{Kokorev2023}
{Kokorev} V.,  et~al., 2023, \mn@doi [\apjl] {10.3847/2041-8213/ad037a}, \href {https://ui.adsabs.harvard.edu/abs/2023ApJ...957L...7K} {957, L7}

\bibitem[\protect\citeauthoryear{{Kormendy} \& {Ho}}{{Kormendy} \& {Ho}}{2013}]{KormendyHo2013}
{Kormendy} J.,  {Ho} L.~C.,  2013, \mn@doi [\araa] {10.1146/annurev-astro-082708-101811}, \href {https://ui.adsabs.harvard.edu/abs/2013ARA&A..51..511K} {51, 511}

\bibitem[\protect\citeauthoryear{{Latif}, {Whalen}, {Khochfar}, {Herrington}  \& {Woods}}{{Latif} et~al.}{2022}]{Latif2022}
{Latif} M.~A.,  {Whalen} D.~J.,  {Khochfar} S.,  {Herrington} N.~P.,   {Woods} T.~E.,  2022, \mn@doi [\nat] {10.1038/s41586-022-04813-y}, \href {https://ui.adsabs.harvard.edu/abs/2022Natur.607...48L} {607, 48}

\bibitem[\protect\citeauthoryear{{Levermore}}{{Levermore}}{1984}]{Levermore1984}
{Levermore} C.~D.,  1984, \mn@doi [\jqsrt] {10.1016/0022-4073(84)90112-2}, \href {https://ui.adsabs.harvard.edu/abs/1984JQSRT..31..149L} {31, 149}

\bibitem[\protect\citeauthoryear{{Li} et~al.,}{{Li} et~al.}{2025}]{Li2024}
{Li} J.,  et~al., 2025, \mn@doi [\apj] {10.3847/1538-4357/ada603}, \href {https://ui.adsabs.harvard.edu/abs/2025ApJ...981...19L} {981, 19}

\bibitem[\protect\citeauthoryear{{Lusso}, {Worseck}, {Hennawi}, {Prochaska}, {Vignali}, {Stern}  \& {O'Meara}}{{Lusso} et~al.}{2015}]{Lusso2015}
{Lusso} E.,  {Worseck} G.,  {Hennawi} J.~F.,  {Prochaska} J.~X.,  {Vignali} C.,  {Stern} J.,   {O'Meara} J.~M.,  2015, \mn@doi [\mnras] {10.1093/mnras/stv516}, \href {https://ui.adsabs.harvard.edu/abs/2015MNRAS.449.4204L} {449, 4204}

\bibitem[\protect\citeauthoryear{{Maiolino} et~al.,}{{Maiolino} et~al.}{2024a}]{Maiolino2023}
{Maiolino} R.,  et~al., 2024a, \mn@doi [\nat] {10.1038/s41586-024-07052-5}, \href {https://ui.adsabs.harvard.edu/abs/2024Natur.627...59M} {627, 59}

\bibitem[\protect\citeauthoryear{{Maiolino} et~al.,}{{Maiolino} et~al.}{2024b}]{Maiolino2024}
{Maiolino} R.,  et~al., 2024b, \mn@doi [\aap] {10.1051/0004-6361/202347640}, \href {https://ui.adsabs.harvard.edu/abs/2024A&A...691A.145M} {691, A145}

\bibitem[\protect\citeauthoryear{{Marinacci} et~al.,}{{Marinacci} et~al.}{2018}]{Marinacci2018}
{Marinacci} F.,  et~al., 2018, \mn@doi [\mnras] {10.1093/mnras/sty2206}, \href {https://ui.adsabs.harvard.edu/abs/2018MNRAS.480.5113M} {480, 5113}

\bibitem[\protect\citeauthoryear{{Mason}, {Treu}, {Dijkstra}, {Mesinger}, {Trenti}, {Pentericci}, {de Barros}  \& {Vanzella}}{{Mason} et~al.}{2018}]{Mason2018}
{Mason} C.~A.,  {Treu} T.,  {Dijkstra} M.,  {Mesinger} A.,  {Trenti} M.,  {Pentericci} L.,  {de Barros} S.,   {Vanzella} E.,  2018, \mn@doi [\apj] {10.3847/1538-4357/aab0a7}, \href {https://ui.adsabs.harvard.edu/abs/2018ApJ...856....2M} {856, 2}

\bibitem[\protect\citeauthoryear{{Mason} et~al.,}{{Mason} et~al.}{2019}]{Mason2019}
{Mason} C.~A.,  et~al., 2019, \mn@doi [\mnras] {10.1093/mnras/stz632}, \href {https://ui.adsabs.harvard.edu/abs/2019MNRAS.485.3947M} {485, 3947}

\bibitem[\protect\citeauthoryear{{Matsuoka} et~al.,}{{Matsuoka} et~al.}{2018}]{Matsuoka2018}
{Matsuoka} Y.,  et~al., 2018, \mn@doi [\apj] {10.3847/1538-4357/aaee7a}, \href {https://ui.adsabs.harvard.edu/abs/2018ApJ...869..150M} {869, 150}

\bibitem[\protect\citeauthoryear{{Matthee} et~al.,}{{Matthee} et~al.}{2024}]{Matthee2023}
{Matthee} J.,  et~al., 2024, \mn@doi [\apj] {10.3847/1538-4357/ad2345}, \href {https://ui.adsabs.harvard.edu/abs/2024ApJ...963..129M} {963, 129}

\bibitem[\protect\citeauthoryear{{McClymont} et~al.,}{{McClymont} et~al.}{2025a}]{McClymont2025a}
{McClymont} W.,  et~al., 2025a, \mn@doi [arXiv e-prints] {10.48550/arXiv.2503.00106}, \href {https://ui.adsabs.harvard.edu/abs/2025arXiv250300106M} {p. arXiv:2503.00106}

\bibitem[\protect\citeauthoryear{{McClymont} et~al.,}{{McClymont} et~al.}{2025b}]{McClymont2025b}
{McClymont} W.,  et~al., 2025b, \mn@doi [arXiv e-prints] {10.48550/arXiv.2503.04894}, \href {https://ui.adsabs.harvard.edu/abs/2025arXiv250304894M} {p. arXiv:2503.04894}

\bibitem[\protect\citeauthoryear{{McGreer}, {Mesinger}  \& {D'Odorico}}{{McGreer} et~al.}{2015}]{McGreer2015}
{McGreer} I.~D.,  {Mesinger} A.,   {D'Odorico} V.,  2015, \mn@doi [\mnras] {10.1093/mnras/stu2449}, \href {https://ui.adsabs.harvard.edu/abs/2015MNRAS.447..499M} {447, 499}

\bibitem[\protect\citeauthoryear{{McGreer}, {Eftekharzadeh}, {Myers}  \& {Fan}}{{McGreer} et~al.}{2016}]{McGreer2016}
{McGreer} I.~D.,  {Eftekharzadeh} S.,  {Myers} A.~D.,   {Fan} X.,  2016, \mn@doi [\aj] {10.3847/0004-6256/151/3/61}, \href {https://ui.adsabs.harvard.edu/abs/2016AJ....151...61M} {151, 61}

\bibitem[\protect\citeauthoryear{{Meiksin}, {Tittley}  \& {Brown}}{{Meiksin} et~al.}{2010}]{Meiksin2010}
{Meiksin} A.,  {Tittley} E.~R.,   {Brown} C.~K.,  2010, \mn@doi [\mnras] {10.1111/j.1365-2966.2009.15667.x}, \href {https://ui.adsabs.harvard.edu/abs/2010MNRAS.401...77M} {401, 77}

\bibitem[\protect\citeauthoryear{{Mortlock} et~al.,}{{Mortlock} et~al.}{2011}]{Mortlock2011}
{Mortlock} D.~J.,  et~al., 2011, \mn@doi [\nat] {10.1038/nature10159}, \href {https://ui.adsabs.harvard.edu/abs/2011Natur.474..616M} {474, 616}

\bibitem[\protect\citeauthoryear{{Naidu} et~al.,}{{Naidu} et~al.}{2025}]{Naidu2025}
{Naidu} R.~P.,  et~al., 2025, \mn@doi [arXiv e-prints] {10.48550/arXiv.2503.16596}, \href {https://ui.adsabs.harvard.edu/abs/2025arXiv250316596N} {p. arXiv:2503.16596}

\bibitem[\protect\citeauthoryear{{Natarajan}, {Pacucci}, {Ricarte}, {Bogd{\'a}n}, {Goulding}  \& {Cappelluti}}{{Natarajan} et~al.}{2024}]{Natarajan2024}
{Natarajan} P.,  {Pacucci} F.,  {Ricarte} A.,  {Bogd{\'a}n} {\'A}.,  {Goulding} A.~D.,   {Cappelluti} N.,  2024, \mn@doi [\apjl] {10.3847/2041-8213/ad0e76}, \href {https://ui.adsabs.harvard.edu/abs/2024ApJ...960L...1N} {960, L1}

\bibitem[\protect\citeauthoryear{{Navarro-Carrera}, {Rinaldi}, {Caputi}, {Iani}, {Kokorev}  \& {van Mierlo}}{{Navarro-Carrera} et~al.}{2024}]{Navarro-Carrera2024}
{Navarro-Carrera} R.,  {Rinaldi} P.,  {Caputi} K.~I.,  {Iani} E.,  {Kokorev} V.,   {van Mierlo} S.~E.,  2024, \mn@doi [\apj] {10.3847/1538-4357/ad0df6}, \href {https://ui.adsabs.harvard.edu/abs/2024ApJ...961..207N} {961, 207}

\bibitem[\protect\citeauthoryear{{Nelson} et~al.,}{{Nelson} et~al.}{2018}]{Nelson2018}
{Nelson} D.,  et~al., 2018, \mn@doi [\mnras] {10.1093/mnras/stx3040}, \href {https://ui.adsabs.harvard.edu/abs/2018MNRAS.475..624N} {475, 624}

\bibitem[\protect\citeauthoryear{{Nelson} et~al.,}{{Nelson} et~al.}{2019}]{Nelson2019TNG50}
{Nelson} D.,  et~al., 2019, \mn@doi [\mnras] {10.1093/mnras/stz2306}, \href {https://ui.adsabs.harvard.edu/abs/2019MNRAS.490.3234N} {490, 3234}

\bibitem[\protect\citeauthoryear{{Neyer} et~al.,}{{Neyer} et~al.}{2024}]{Neyer2024}
{Neyer} M.,  et~al., 2024, \mn@doi [\mnras] {10.1093/mnras/stae1325}, \href {https://ui.adsabs.harvard.edu/abs/2024MNRAS.531.2943N} {531, 2943}

\bibitem[\protect\citeauthoryear{{Ning}, {Jiang}, {Zheng}  \& {Wu}}{{Ning} et~al.}{2022}]{Ning2022}
{Ning} Y.,  {Jiang} L.,  {Zheng} Z.-Y.,   {Wu} J.,  2022, \mn@doi [\apj] {10.3847/1538-4357/ac4268}, \href {https://ui.adsabs.harvard.edu/abs/2022ApJ...926..230N} {926, 230}

\bibitem[\protect\citeauthoryear{{Onoue} et~al.,}{{Onoue} et~al.}{2023}]{Onoue2023}
{Onoue} M.,  et~al., 2023, \mn@doi [\apjl] {10.3847/2041-8213/aca9d3}, \href {https://ui.adsabs.harvard.edu/abs/2023ApJ...942L..17O} {942, L17}

\bibitem[\protect\citeauthoryear{{Ouchi} et~al.,}{{Ouchi} et~al.}{2010}]{Ouchi2010}
{Ouchi} M.,  et~al., 2010, \mn@doi [\apj] {10.1088/0004-637X/723/1/869}, \href {https://ui.adsabs.harvard.edu/abs/2010ApJ...723..869O} {723, 869}

\bibitem[\protect\citeauthoryear{{Overzier}}{{Overzier}}{2016}]{Overzier2016}
{Overzier} R.~A.,  2016, \mn@doi [\aapr] {10.1007/s00159-016-0100-3}, \href {https://ui.adsabs.harvard.edu/abs/2016A&ARv..24...14O} {24, 14}

\bibitem[\protect\citeauthoryear{{Pacucci}, {Nguyen}, {Carniani}, {Maiolino}  \& {Fan}}{{Pacucci} et~al.}{2023}]{Pacucci2023}
{Pacucci} F.,  {Nguyen} B.,  {Carniani} S.,  {Maiolino} R.,   {Fan} X.,  2023, \mn@doi [\apjl] {10.3847/2041-8213/ad0158}, \href {https://ui.adsabs.harvard.edu/abs/2023ApJ...957L...3P} {957, L3}

\bibitem[\protect\citeauthoryear{{Pakmor}, {Springel}, {Bauer}, {Mocz}, {Munoz}, {Ohlmann}, {Schaal}  \& {Zhu}}{{Pakmor} et~al.}{2016}]{Pakmor2016}
{Pakmor} R.,  {Springel} V.,  {Bauer} A.,  {Mocz} P.,  {Munoz} D.~J.,  {Ohlmann} S.~T.,  {Schaal} K.,   {Zhu} C.,  2016, \mn@doi [\mnras] {10.1093/mnras/stv2380}, \href {https://ui.adsabs.harvard.edu/abs/2016MNRAS.455.1134P} {455, 1134}

\bibitem[\protect\citeauthoryear{{Pakmor} et~al.,}{{Pakmor} et~al.}{2023}]{pakmor2023millenniumtng}
{Pakmor} R.,  et~al., 2023, \mn@doi [\mnras] {10.1093/mnras/stac3620}, \href {https://ui.adsabs.harvard.edu/abs/2023MNRAS.524.2539P} {524, 2539}

\bibitem[\protect\citeauthoryear{{Pakmor} et~al.,}{{Pakmor} et~al.}{2024}]{Pakmor2024}
{Pakmor} R.,  et~al., 2024, \mn@doi [\mnras] {10.1093/mnras/stae112}, \href {https://ui.adsabs.harvard.edu/abs/2024MNRAS.528.2308P} {528, 2308}

\bibitem[\protect\citeauthoryear{{Pillepich} et~al.,}{{Pillepich} et~al.}{2018}]{Pillepich2018a}
{Pillepich} A.,  et~al., 2018, \mn@doi [\mnras] {10.1093/mnras/stx2656}, \href {https://ui.adsabs.harvard.edu/abs/2018MNRAS.473.4077P} {473, 4077}

\bibitem[\protect\citeauthoryear{{Piotrowska}, {Bluck}, {Maiolino}  \& {Peng}}{{Piotrowska} et~al.}{2022}]{Piotrowska2022}
{Piotrowska} J.~M.,  {Bluck} A. F.~L.,  {Maiolino} R.,   {Peng} Y.,  2022, \mn@doi [\mnras] {10.1093/mnras/stab3673}, \href {https://ui.adsabs.harvard.edu/abs/2022MNRAS.512.1052P} {512, 1052}

\bibitem[\protect\citeauthoryear{{Planck Collaboration} et~al.,}{{Planck Collaboration} et~al.}{2016}]{Planck2016}
{Planck Collaboration} et~al., 2016, \mn@doi [\aap] {10.1051/0004-6361/201525830}, \href {https://ui.adsabs.harvard.edu/abs/2016A&A...594A..13P} {594, A13}

\bibitem[\protect\citeauthoryear{{Planck Collaboration} et~al.,}{{Planck Collaboration} et~al.}{2020}]{Planck2020}
{Planck Collaboration} et~al., 2020, \mn@doi [\aap] {10.1051/0004-6361/201833910}, \href {https://ui.adsabs.harvard.edu/abs/2020A&A...641A...6P} {641, A6}

\bibitem[\protect\citeauthoryear{{Rahmati}, {Pawlik}, {Rai{\v{c}}evi{\'c}}  \& {Schaye}}{{Rahmati} et~al.}{2013}]{Rahmati2013}
{Rahmati} A.,  {Pawlik} A.~H.,  {Rai{\v{c}}evi{\'c}} M.,   {Schaye} J.,  2013, \mn@doi [\mnras] {10.1093/mnras/stt066}, \href {https://ui.adsabs.harvard.edu/abs/2013MNRAS.430.2427R} {430, 2427}

\bibitem[\protect\citeauthoryear{{Scharr{\'e}}, {Sorini}  \& {Dav{\'e}}}{{Scharr{\'e}} et~al.}{2024}]{Scharre2024}
{Scharr{\'e}} L.,  {Sorini} D.,   {Dav{\'e}} R.,  2024, \mn@doi [\mnras] {10.1093/mnras/stae2098}, \href {https://ui.adsabs.harvard.edu/abs/2024MNRAS.534..361S} {534, 361}

\bibitem[\protect\citeauthoryear{{Scoggins}, {Haiman}  \& {Wise}}{{Scoggins} et~al.}{2023}]{Scoggins2023}
{Scoggins} M.~T.,  {Haiman} Z.,   {Wise} J.~H.,  2023, \mn@doi [\mnras] {10.1093/mnras/stac3715}, \href {https://ui.adsabs.harvard.edu/abs/2023MNRAS.519.2155S} {519, 2155}

\bibitem[\protect\citeauthoryear{{Shen} et~al.,}{{Shen} et~al.}{2007}]{Shen2007}
{Shen} Y.,  et~al., 2007, \mn@doi [\aj] {10.1086/513517}, \href {https://ui.adsabs.harvard.edu/abs/2007AJ....133.2222S} {133, 2222}

\bibitem[\protect\citeauthoryear{{Shen}, {Hopkins}, {Faucher-Gigu{\`e}re}, {Alexander}, {Richards}, {Ross}  \& {Hickox}}{{Shen} et~al.}{2020}]{Shen2020}
{Shen} X.,  {Hopkins} P.~F.,  {Faucher-Gigu{\`e}re} C.-A.,  {Alexander} D.~M.,  {Richards} G.~T.,  {Ross} N.~P.,   {Hickox} R.~C.,  2020, \mn@doi [\mnras] {10.1093/mnras/staa1381}, \href {https://ui.adsabs.harvard.edu/abs/2020MNRAS.495.3252S} {495, 3252}

\bibitem[\protect\citeauthoryear{{Shen}, {Vogelsberger}, {Boylan-Kolchin}, {Tacchella}  \& {Kannan}}{{Shen} et~al.}{2023}]{Shen2023}
{Shen} X.,  {Vogelsberger} M.,  {Boylan-Kolchin} M.,  {Tacchella} S.,   {Kannan} R.,  2023, \mn@doi [\mnras] {10.1093/mnras/stad2508}, \href {https://ui.adsabs.harvard.edu/abs/2023MNRAS.525.3254S} {525, 3254}

\bibitem[\protect\citeauthoryear{{Shen} et~al.,}{{Shen} et~al.}{2024}]{Shen2024}
{Shen} X.,  et~al., 2024, \mn@doi [\mnras] {10.1093/mnras/stae2156}, \href {https://ui.adsabs.harvard.edu/abs/2024MNRAS.534.1433S} {534, 1433}

\bibitem[\protect\citeauthoryear{{Shen} et~al.,}{{Shen} et~al.}{2025}]{Shen2025}
{Shen} X.,  et~al., 2025, \mn@doi [arXiv e-prints] {10.48550/arXiv.2503.01949}, \href {https://ui.adsabs.harvard.edu/abs/2025arXiv250301949S} {p. arXiv:2503.01949}

\bibitem[\protect\citeauthoryear{{Shi}, {Kremer}  \& {Hopkins}}{{Shi} et~al.}{2024}]{Shi2024}
{Shi} Y.,  {Kremer} K.,   {Hopkins} P.~F.,  2024, \mn@doi [\apjl] {10.3847/2041-8213/ad5a95}, \href {https://ui.adsabs.harvard.edu/abs/2024ApJ...969L..31S} {969, L31}

\bibitem[\protect\citeauthoryear{{Smith} \& {Bromm}}{{Smith} \& {Bromm}}{2019}]{SmithSMBH2019}
{Smith} A.,  {Bromm} V.,  2019, \mn@doi [Contemporary Physics] {10.1080/00107514.2019.1615715}, \href {https://ui.adsabs.harvard.edu/abs/2019ConPh..60..111S} {60, 111}

\bibitem[\protect\citeauthoryear{{Smith}, {Sigurdsson}  \& {Abel}}{{Smith} et~al.}{2008}]{Smith2008}
{Smith} B.,  {Sigurdsson} S.,   {Abel} T.,  2008, \mn@doi [\mnras] {10.1111/j.1365-2966.2008.12922.x}, \href {https://ui.adsabs.harvard.edu/abs/2008MNRAS.385.1443S} {385, 1443}

\bibitem[\protect\citeauthoryear{{Smith}, {Kannan}, {Garaldi}, {Vogelsberger}, {Pakmor}, {Springel}  \& {Hernquist}}{{Smith} et~al.}{2022}]{Smith2022}
{Smith} A.,  {Kannan} R.,  {Garaldi} E.,  {Vogelsberger} M.,  {Pakmor} R.,  {Springel} V.,   {Hernquist} L.,  2022, \mn@doi [\mnras] {10.1093/mnras/stac713}, \href {https://ui.adsabs.harvard.edu/abs/2022MNRAS.512.3243S} {512, 3243}

\bibitem[\protect\citeauthoryear{{Sobacchi} \& {Mesinger}}{{Sobacchi} \& {Mesinger}}{2015}]{SobacchiMesinger2015}
{Sobacchi} E.,  {Mesinger} A.,  2015, \mn@doi [\mnras] {10.1093/mnras/stv1751}, \href {https://ui.adsabs.harvard.edu/abs/2015MNRAS.453.1843S} {453, 1843}

\bibitem[\protect\citeauthoryear{{Soltan}}{{Soltan}}{1982}]{Soltan1982}
{Soltan} A.,  1982, \mn@doi [\mnras] {10.1093/mnras/200.1.115}, \href {https://ui.adsabs.harvard.edu/abs/1982MNRAS.200..115S} {200, 115}

\bibitem[\protect\citeauthoryear{{Springel}}{{Springel}}{2010}]{Springel2010}
{Springel} V.,  2010, \mn@doi [\mnras] {10.1111/j.1365-2966.2009.15715.x}, \href {https://ui.adsabs.harvard.edu/abs/2010MNRAS.401..791S} {401, 791}

\bibitem[\protect\citeauthoryear{{Springel} \& {Hernquist}}{{Springel} \& {Hernquist}}{2003}]{SpringelHernquist2003}
{Springel} V.,  {Hernquist} L.,  2003, \mn@doi [\mnras] {10.1046/j.1365-8711.2003.06206.x}, \href {https://ui.adsabs.harvard.edu/abs/2003MNRAS.339..289S} {339, 289}

\bibitem[\protect\citeauthoryear{{Springel}, {White}, {Tormen}  \& {Kauffmann}}{{Springel} et~al.}{2001}]{springel2001populating}
{Springel} V.,  {White} S. D.~M.,  {Tormen} G.,   {Kauffmann} G.,  2001, \mn@doi [\mnras] {10.1046/j.1365-8711.2001.04912.x}, \href {https://ui.adsabs.harvard.edu/abs/2001MNRAS.328..726S} {328, 726}

\bibitem[\protect\citeauthoryear{{Springel}, {Di Matteo}  \& {Hernquist}}{{Springel} et~al.}{2005a}]{Springel2005b}
{Springel} V.,  {Di Matteo} T.,   {Hernquist} L.,  2005a, \mn@doi [\mnras] {10.1111/j.1365-2966.2005.09238.x}, \href {https://ui.adsabs.harvard.edu/abs/2005MNRAS.361..776S} {361, 776}

\bibitem[\protect\citeauthoryear{{Springel} et~al.,}{{Springel} et~al.}{2005b}]{Springel2005}
{Springel} V.,  et~al., 2005b, \mn@doi [\nat] {10.1038/nature03597}, \href {https://ui.adsabs.harvard.edu/abs/2005Natur.435..629S} {435, 629}

\bibitem[\protect\citeauthoryear{{Springel} et~al.,}{{Springel} et~al.}{2018}]{Springel2018}
{Springel} V.,  et~al., 2018, \mn@doi [\mnras] {10.1093/mnras/stx3304}, \href {https://ui.adsabs.harvard.edu/abs/2018MNRAS.475..676S} {475, 676}

\bibitem[\protect\citeauthoryear{{Springel}, {Pakmor}, {Zier}  \& {Reinecke}}{{Springel} et~al.}{2021}]{Gadget4}
{Springel} V.,  {Pakmor} R.,  {Zier} O.,   {Reinecke} M.,  2021, \mn@doi [\mnras] {10.1093/mnras/stab1855}, \href {https://ui.adsabs.harvard.edu/abs/2021MNRAS.506.2871S} {506, 2871}

\bibitem[\protect\citeauthoryear{{Stone}, {Lyu}, {Rieke}  \& {Alberts}}{{Stone} et~al.}{2023}]{Stone2023}
{Stone} M.~A.,  {Lyu} J.,  {Rieke} G.~H.,   {Alberts} S.,  2023, \mn@doi [\apj] {10.3847/1538-4357/acebe0}, \href {https://ui.adsabs.harvard.edu/abs/2023ApJ...953..180S} {953, 180}

\bibitem[\protect\citeauthoryear{{Taylor} et~al.,}{{Taylor} et~al.}{2025}]{Taylor2024}
{Taylor} A.~J.,  et~al., 2025, \mn@doi [\apj] {10.3847/1538-4357/add15b}, \href {https://ui.adsabs.harvard.edu/abs/2025ApJ...986..165T} {986, 165}

\bibitem[\protect\citeauthoryear{{Terrazas} et~al.,}{{Terrazas} et~al.}{2020}]{Terrazas2020}
{Terrazas} B.~A.,  et~al., 2020, \mn@doi [\mnras] {10.1093/mnras/staa374}, \href {https://ui.adsabs.harvard.edu/abs/2020MNRAS.493.1888T} {493, 1888}

\bibitem[\protect\citeauthoryear{{Trebitsch} et~al.,}{{Trebitsch} et~al.}{2021}]{Trebitsch2021}
{Trebitsch} M.,  et~al., 2021, \mn@doi [\aap] {10.1051/0004-6361/202037698}, \href {https://ui.adsabs.harvard.edu/abs/2021A&A...653A.154T} {653, A154}

\bibitem[\protect\citeauthoryear{{{\"U}bler} et~al.,}{{{\"U}bler} et~al.}{2023}]{Uebler2023}
{{\"U}bler} H.,  et~al., 2023, \mn@doi [\aap] {10.1051/0004-6361/202346137}, \href {https://ui.adsabs.harvard.edu/abs/2023A&A...677A.145U} {677, A145}

\bibitem[\protect\citeauthoryear{{Ueda}, {Akiyama}, {Hasinger}, {Miyaji}  \& {Watson}}{{Ueda} et~al.}{2014}]{Ueda2014}
{Ueda} Y.,  {Akiyama} M.,  {Hasinger} G.,  {Miyaji} T.,   {Watson} M.~G.,  2014, \mn@doi [\apj] {10.1088/0004-637X/786/2/104}, \href {https://ui.adsabs.harvard.edu/abs/2014ApJ...786..104U} {786, 104}

\bibitem[\protect\citeauthoryear{{Umeda}, {Ouchi}, {Nakajima}, {Harikane}, {Ono}, {Xu}, {Isobe}  \& {Zhang}}{{Umeda} et~al.}{2024}]{Umeda2023}
{Umeda} H.,  {Ouchi} M.,  {Nakajima} K.,  {Harikane} Y.,  {Ono} Y.,  {Xu} Y.,  {Isobe} Y.,   {Zhang} Y.,  2024, \mn@doi [\apj] {10.3847/1538-4357/ad554e}, \href {https://ui.adsabs.harvard.edu/abs/2024ApJ...971..124U} {971, 124}

\bibitem[\protect\citeauthoryear{{Vito} et~al.,}{{Vito} et~al.}{2018}]{Vito2018}
{Vito} F.,  et~al., 2018, \mn@doi [\mnras] {10.1093/mnras/stx2486}, \href {https://ui.adsabs.harvard.edu/abs/2018MNRAS.473.2378V} {473, 2378}

\bibitem[\protect\citeauthoryear{{Vogelsberger}, {Sijacki}, {Kere{\v{s}}}, {Springel}  \& {Hernquist}}{{Vogelsberger} et~al.}{2012}]{Vogelsberger2012}
{Vogelsberger} M.,  {Sijacki} D.,  {Kere{\v{s}}} D.,  {Springel} V.,   {Hernquist} L.,  2012, \mn@doi [\mnras] {10.1111/j.1365-2966.2012.21590.x}, \href {https://ui.adsabs.harvard.edu/abs/2012MNRAS.425.3024V} {425, 3024}

\bibitem[\protect\citeauthoryear{{Vogelsberger}, {Genel}, {Sijacki}, {Torrey}, {Springel}  \& {Hernquist}}{{Vogelsberger} et~al.}{2013}]{Vogelsberger2013}
{Vogelsberger} M.,  {Genel} S.,  {Sijacki} D.,  {Torrey} P.,  {Springel} V.,   {Hernquist} L.,  2013, \mn@doi [\mnras] {10.1093/mnras/stt1789}, \href {https://ui.adsabs.harvard.edu/abs/2013MNRAS.436.3031V} {436, 3031}

\bibitem[\protect\citeauthoryear{{Vogelsberger}, {Marinacci}, {Torrey}  \& {Puchwein}}{{Vogelsberger} et~al.}{2020}]{Vogelsberger2020}
{Vogelsberger} M.,  {Marinacci} F.,  {Torrey} P.,   {Puchwein} E.,  2020, \mn@doi [Nature Reviews Physics] {10.1038/s42254-019-0127-2}, \href {https://ui.adsabs.harvard.edu/abs/2020NatRP...2...42V} {2, 42}

\bibitem[\protect\citeauthoryear{{Wang} et~al.,}{{Wang} et~al.}{2019}]{Wang2019}
{Wang} F.,  et~al., 2019, \mn@doi [\apj] {10.3847/1538-4357/ab2be5}, \href {https://ui.adsabs.harvard.edu/abs/2019ApJ...884...30W} {884, 30}

\bibitem[\protect\citeauthoryear{{Wang} et~al.,}{{Wang} et~al.}{2021}]{Wang2021}
{Wang} F.,  et~al., 2021, \mn@doi [\apjl] {10.3847/2041-8213/abd8c6}, \href {https://ui.adsabs.harvard.edu/abs/2021ApJ...907L...1W} {907, L1}

\bibitem[\protect\citeauthoryear{{Wang} et~al.,}{{Wang} et~al.}{2024}]{Wang2024}
{Wang} F.,  et~al., 2024, \mn@doi [\apjl] {10.3847/2041-8213/ad20ef}, \href {https://ui.adsabs.harvard.edu/abs/2024ApJ...962L..11W} {962, L11}

\bibitem[\protect\citeauthoryear{{Wang} et~al.,}{{Wang} et~al.}{2025}]{Wang2025}
{Wang} Z.,  et~al., 2025, \mn@doi [arXiv e-prints] {10.48550/arXiv.2505.05554}, \href {https://ui.adsabs.harvard.edu/abs/2025arXiv250505554W} {p. arXiv:2505.05554}

\bibitem[\protect\citeauthoryear{{Weibel} et~al.,}{{Weibel} et~al.}{2024}]{Weibel2024}
{Weibel} A.,  et~al., 2024, \mn@doi [\mnras] {10.1093/mnras/stae1891}, \href {https://ui.adsabs.harvard.edu/abs/2024MNRAS.533.1808W} {533, 1808}

\bibitem[\protect\citeauthoryear{{Weinberger} et~al.,}{{Weinberger} et~al.}{2017}]{Weinberger2017}
{Weinberger} R.,  et~al., 2017, \mn@doi [\mnras] {10.1093/mnras/stw2944}, \href {https://ui.adsabs.harvard.edu/abs/2017MNRAS.465.3291W} {465, 3291}

\bibitem[\protect\citeauthoryear{{Weinberger} et~al.,}{{Weinberger} et~al.}{2018}]{Weinberger2018}
{Weinberger} R.,  et~al., 2018, \mn@doi [\mnras] {10.1093/mnras/sty1733}, \href {https://ui.adsabs.harvard.edu/abs/2018MNRAS.479.4056W} {479, 4056}

\bibitem[\protect\citeauthoryear{{Weinberger}, {Springel}  \& {Pakmor}}{{Weinberger} et~al.}{2020}]{Weinberger2020}
{Weinberger} R.,  {Springel} V.,   {Pakmor} R.,  2020, \mn@doi [\apjs] {10.3847/1538-4365/ab908c}, \href {https://ui.adsabs.harvard.edu/abs/2020ApJS..248...32W} {248, 32}

\bibitem[\protect\citeauthoryear{{Wiersma}, {Schaye}  \& {Smith}}{{Wiersma} et~al.}{2009}]{Wiersma2009}
{Wiersma} R. P.~C.,  {Schaye} J.,   {Smith} B.~D.,  2009, \mn@doi [\mnras] {10.1111/j.1365-2966.2008.14191.x}, \href {https://ui.adsabs.harvard.edu/abs/2009MNRAS.393...99W} {393, 99}

\bibitem[\protect\citeauthoryear{{Yang} et~al.,}{{Yang} et~al.}{2020a}]{Yang2020-dw}
{Yang} J.,  et~al., 2020a, \mn@doi [\apjl] {10.3847/2041-8213/ab9c26}, \href {https://ui.adsabs.harvard.edu/abs/2020ApJ...897L..14Y} {897, L14}

\bibitem[\protect\citeauthoryear{{Yang} et~al.,}{{Yang} et~al.}{2020b}]{Yang2020}
{Yang} J.,  et~al., 2020b, \mn@doi [\apj] {10.3847/1538-4357/abbc1b}, \href {https://ui.adsabs.harvard.edu/abs/2020ApJ...904...26Y} {904, 26}

\bibitem[\protect\citeauthoryear{{Yang} et~al.,}{{Yang} et~al.}{2023}]{Yang2023}
{Yang} G.,  et~al., 2023, \mn@doi [\apjl] {10.3847/2041-8213/acd639}, \href {https://ui.adsabs.harvard.edu/abs/2023ApJ...950L...5Y} {950, L5}

\bibitem[\protect\citeauthoryear{{Yeh} et~al.,}{{Yeh} et~al.}{2023}]{Yeh2023}
{Yeh} J. Y.~C.,  et~al., 2023, \mn@doi [\mnras] {10.1093/mnras/stad210}, \href {https://ui.adsabs.harvard.edu/abs/2023MNRAS.520.2757Y} {520, 2757}

\bibitem[\protect\citeauthoryear{{Yue}, {Fan}, {Yang}  \& {Wang}}{{Yue} et~al.}{2021}]{Yue2021}
{Yue} M.,  {Fan} X.,  {Yang} J.,   {Wang} F.,  2021, \mn@doi [\apjl] {10.3847/2041-8213/ac31a9}, \href {https://ui.adsabs.harvard.edu/abs/2021ApJ...921L..27Y} {921, L27}

\bibitem[\protect\citeauthoryear{{Yue} et~al.,}{{Yue} et~al.}{2024}]{Yue2023}
{Yue} M.,  et~al., 2024, \mn@doi [\apj] {10.3847/1538-4357/ad3914}, \href {https://ui.adsabs.harvard.edu/abs/2024ApJ...966..176Y} {966, 176}

\bibitem[\protect\citeauthoryear{{Zhou}, {Chen}, {Di Matteo}, {Ni}, {Croft}  \& {Bird}}{{Zhou} et~al.}{2024}]{Zhou2023}
{Zhou} Y.,  {Chen} H.,  {Di Matteo} T.,  {Ni} Y.,  {Croft} R. A.~C.,   {Bird} S.,  2024, \mn@doi [\mnras] {10.1093/mnras/stae172}, \href {https://ui.adsabs.harvard.edu/abs/2024MNRAS.528.3730Z} {528, 3730}

\bibitem[\protect\citeauthoryear{{Zier}, {Kannan}, {Smith}, {Vogelsberger}  \& {Verbeek}}{{Zier} et~al.}{2024}]{Zier2024}
{Zier} O.,  {Kannan} R.,  {Smith} A.,  {Vogelsberger} M.,   {Verbeek} E.,  2024, \mn@doi [\mnras] {10.1093/mnras/stae1837}, \href {https://ui.adsabs.harvard.edu/abs/2024MNRAS.533..268Z} {533, 268}

\bibitem[\protect\citeauthoryear{{Zier} et~al.,}{{Zier} et~al.}{2025a}]{Zier2025}
{Zier} O.,  et~al., 2025a, \mn@doi [arXiv e-prints] {10.48550/arXiv.2503.02927}, \href {https://ui.adsabs.harvard.edu/abs/2025arXiv250302927Z} {p. arXiv:2503.02927}

\bibitem[\protect\citeauthoryear{{Zier} et~al.,}{{Zier} et~al.}{2025b}]{Zier2025-PopIII}
{Zier} O.,  et~al., 2025b, \mn@doi [arXiv e-prints] {10.48550/arXiv.2503.03806}, \href {https://ui.adsabs.harvard.edu/abs/2025arXiv250303806Z} {p. arXiv:2503.03806}

\bibitem[\protect\citeauthoryear{{{\v{D}}urov{\v{c}}{\'\i}kov{\'a}}, {Katz}, {Bosman}, {Davies}, {Devriendt}  \& {Slyz}}{{{\v{D}}urov{\v{c}}{\'\i}kov{\'a}} et~al.}{2020}]{Durovcikova2020}
{{\v{D}}urov{\v{c}}{\'\i}kov{\'a}} D.,  {Katz} H.,  {Bosman} S. E.~I.,  {Davies} F.~B.,  {Devriendt} J.,   {Slyz} A.,  2020, \mn@doi [\mnras] {10.1093/mnras/staa505}, \href {https://ui.adsabs.harvard.edu/abs/2020MNRAS.493.4256D} {493, 4256}

\bibitem[\protect\citeauthoryear{{{\v{D}}urov{\v{c}}{\'\i}kov{\'a}} et~al.,}{{{\v{D}}urov{\v{c}}{\'\i}kov{\'a}} et~al.}{2024}]{Durovcikova2024}
{{\v{D}}urov{\v{c}}{\'\i}kov{\'a}} D.,  et~al., 2024, \mn@doi [\apj] {10.3847/1538-4357/ad4888}, \href {https://ui.adsabs.harvard.edu/abs/2024ApJ...969..162D} {969, 162}

\makeatother
\end{thebibliography}
\clearpage




\appendix

\section{Convergence tests}
\label{app:resolution}

In this section, we explore the resolution effects on the growth of BHs and galaxies. We consider two runs, with two different resolution levels: $z1$, corresponding to the original MTNG resolution, and same as all the runs explored in the paper (Table~\ref{tab:prop}); and $z2$, which is a factor of eight times better mass resolution (i.e., a factor of two better spatial resolution). Given the high computational costs of a $z2$ simulation, we ran a smaller zoom-in region for these tests, with a radius of $R = 10 \, \mathrm{cMpc}$, six times smaller than the runs in the main text (Table~\ref{tab:prop}). 

\begin{figure}
    \includegraphics[width=0.46\textwidth]{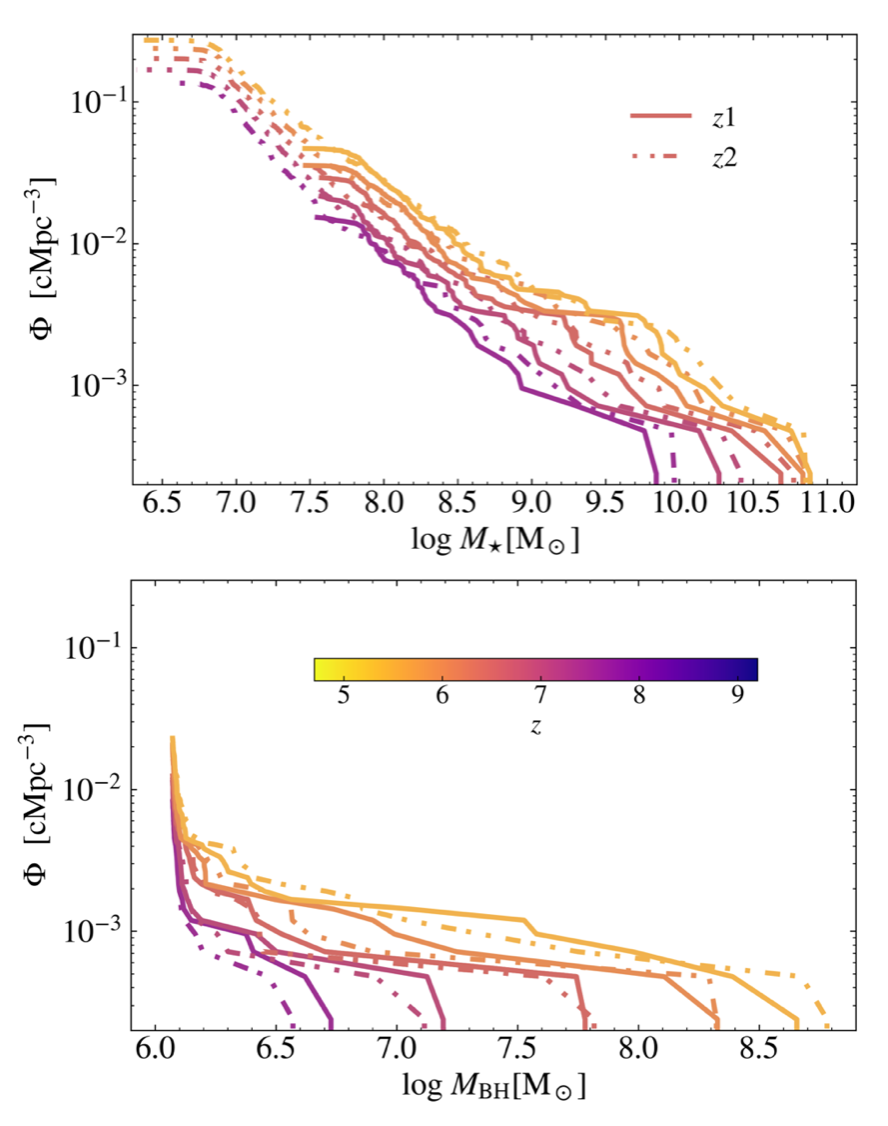}
    \centering
    \caption{\textbf{Cumulative stellar and black hole mass function, at $z = 7.5 \rightarrow 5.5$, at two different resolution levels.} The solid lines represent a zoom factor of one, the fiducial resolution employed in the study. The dashed lines represent the runs with a factor of two better spatial resolution (factor of eight better mass resolution). The colors represent the redshift, as indicated in the color bar. Overall, both stellar and black hole masses show good convergence at the population level, with the exception of low-mass galaxies ($M_\mathrm{\star} \lesssim 10^{7.5}\,\mathrm{M_\odot}$), which are only present in the $z2$ run, as the resolution in the $z1$ run is too low to capture their collapse and formation.}
    \label{fig:app_MF}
\end{figure}

We note that for a resolution level $z1$, it is impossible to form galaxies with $M_\mathrm{\star} \lesssim 10^{7.5}\,\mathrm{M_\odot}$, due to resolution limitations. Thus, they are only captured in the $z2$ simulations. For higher mass galaxies, and black holes we find good convergence for the stellar and BH mass functions (Fig.~\ref{fig:app_MF}).

\section{Calibration of escape fraction}
\label{app:CalibrationEsc}

\begin{figure}
    \includegraphics[width=0.5\textwidth]{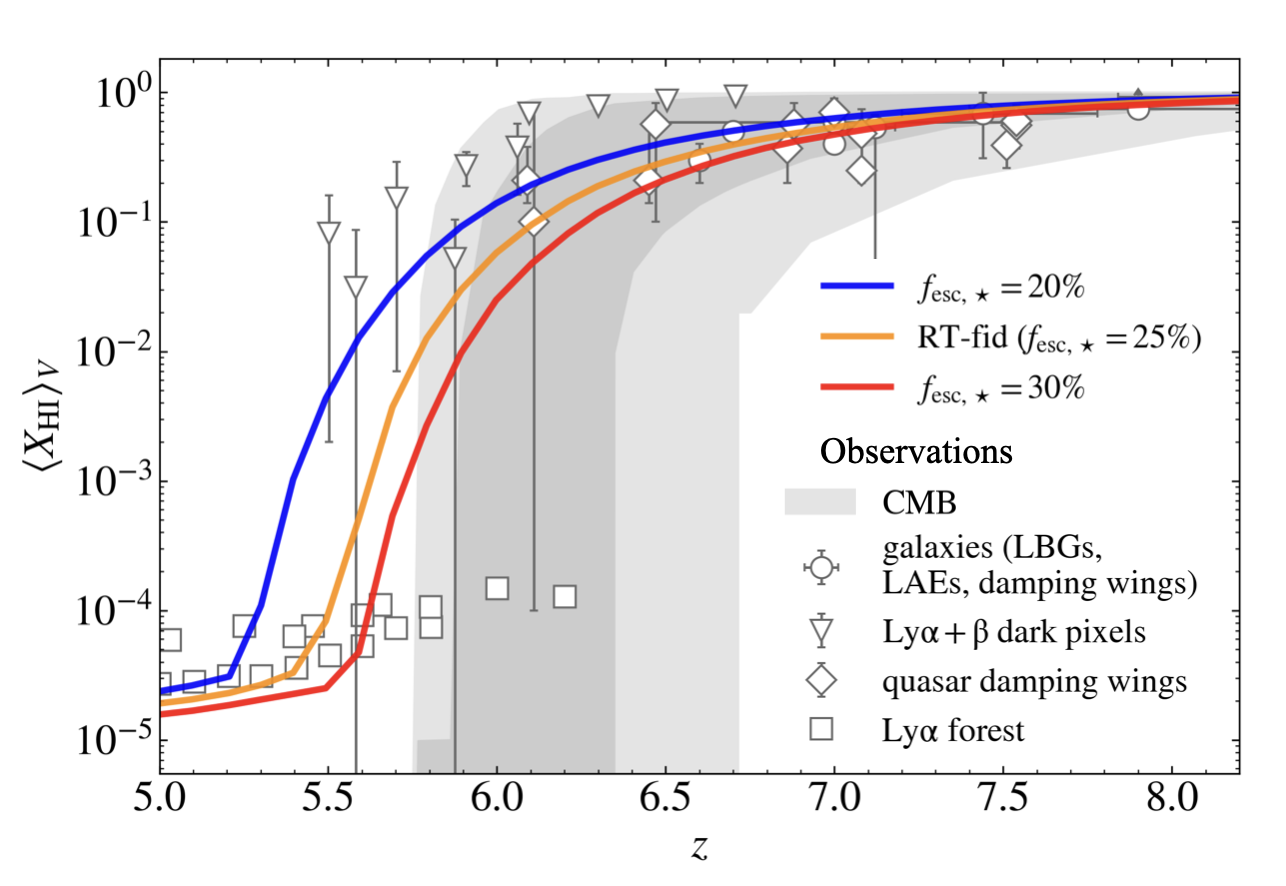}
    \caption{\textbf{Reionization history, under different stellar escape fractions.} We show the same observations as in Fig.~\ref{fig:xHIV}: quasar damping wings \citep{Banados2018,Davies2018,Yang2020-dw,Durovcikova2020,Wang2021,Durovcikova2024}; galaxies damping wings \citep{Mason2018, Ouchi2010, SobacchiMesinger2015, Mason2019, Ning2022, Umeda2023}; $\rm{Ly\alpha}$ forest \citep{Fan2006,Yang2020,Bosman2021}; $\mathrm{Ly\alpha+\beta}$ dark pixels \citep{McGreer2015,Jin2023}; and CMB \citep{Planck2020}.}
    \label{fig:app_reioniz}
\end{figure} 

Figure~\ref{fig:app_reioniz} shows the reionization history, for three different stellar escape fractions, $f_\mathrm{esc,\star}$, along with observations from quasar damping wings \citep{Banados2018,Davies2018,Yang2020-dw,Durovcikova2020,Wang2021,Durovcikova2024}; galaxies damping wings \citep{Mason2018, Ouchi2010, SobacchiMesinger2015, Mason2019, Ning2022, Umeda2023}; $\rm{Ly\alpha}$ forest \citep{Fan2006,Yang2020,Bosman2021}; $\mathrm{Ly\alpha+\beta}$ dark pixels \citep{McGreer2015,Jin2023}; and CMB \citep{Planck2020}. We use this test to calibrate the value of $f_\mathrm{esc,\star}$, which as explained in Sec.~\ref{sect:RT}, can be viewed as a resolution dependent, free parameter. The resolution and size of the Lagragian region for the simulations in Fig.~\ref{fig:app_reioniz} are the same as all the runs in the main text (Table~\ref{tab:prop}). We choose the fiducial value for $f_\mathrm{esc,\star}$ to be $f_\mathrm{esc,\star} = 25\%$.


\bsp	
\label{lastpage}
\end{document}